\documentclass[12pt]{article}

\usepackage[dvipsname]{xcolor}

\usepackage{graphicx}
\usepackage{amsmath}        
\usepackage{amssymb}        
\usepackage{slashed}
\usepackage{bm}
\usepackage{here}
\usepackage{cite}
\def\mydate{May 9, 2024}
\def\ignore#1{{}}

\def\go{\rightarrow}
\def\dd{\partial}
\def\tr{{\rm tr}\,}
\def\ep{{\epsilon}}

\def\eff{{\rm eff}}
\def\SM{{\rm SM}}
\def\KK{{\rm KK}}
\def\EM{{\rm EM}}

\def\max{{\rm max}}
\def\onehalf{\hbox{$\frac{1}{2}$}}
\def\onethird{\hbox{$\frac{1}{3}$}}
\def\twothird{\hbox{$\frac{2}{3}$}}
\def\onequarter{\hbox{$\frac{1}{4}$}}
\def\la{\langle}
\def\ra{\rangle}

\def\Tr{{\rm Tr} \,}

\def\mybig{\displaystyle \strut }

\def\myfrac#1#2{\frac{\mybig #1}{\mybig #2}}

\def\mymat#1#2{\begin{matrix}#1 \cr \noalign{\kern -2pt} #2\end{matrix}}

\def\mynoalign{\noalign{\kern 4pt}}
\def\mysnoalign{\noalign{\kern 3pt}}
\def\mytinynoalign{\noalign{\kern 2pt}}
\def\ignore#1{{}}

\newcommand{\Slash}[1]{{\ooalign{\hfil/\hfil\crcr$#1$}}}

\makeatletter
\@addtoreset{equation}{section}
\makeatother

\begin{document}

\thispagestyle{empty}

%%%%% PREPRINT NUMBERS %%%%%%
{\small \noindent \mydate ~ (typos corrected) \hfill  OU-HET-1169}

\vskip 2.5cm

%%%%%%%%%%%%%%%%%%% TITLE %%%%%%%%%%%%%%%%%%
\baselineskip=30pt plus 1pt minus 1pt

\begin{center}
{\bf \Large Coupling Sum Rules and Oblique Corrections}\\ %[12pt]
{\bf \Large  in Gauge-Higgs Unification}\\ %[12pt]
\end{center}

%%%%%%%%%%%%%%%% AUTHORS %%%%%%%%%%%%%%%%%%%%%%%

%\vspace{.0cm}
\baselineskip=22pt plus 1pt minus 1pt

\vskip 0.5cm

\begin{center}
{\large \bf  Yutaka Hosotani$^a$,  Shuichiro Funatsu$^b$, Hisaki Hatanaka$^c$}

{\large \bf  Yuta Orikasa$^d$ and Naoki Yamatsu$^e$}

\baselineskip=18pt plus 1pt minus 1pt

\vskip 10pt
{\small \it $^a$Department of Physics, Osaka University,  Toyonaka, Osaka 560-0043, 
Japan\footnote{Current address: {\it Research Center for Nuclear Physics, Osaka University, Ibaraki, Osaka 567-0047, Japan} }} \\
{\small \it $^b$Ushiku, Ibaraki 300-1234, Japan} \\
{\small \it $^c$Osaka, Osaka 536-0014, Japan} \\
{\small \it $^d$Institute of Experimental and Applied Physics, Czech Technical University in Prague,} \\
{\small \it Husova 240/5, 110 00 Prague 1, Czech Republic} \\
{\small \it $^e$Department of Physics, National Taiwan University,Taipei, Taiwan 10617, 
R.O.C.\footnote{Current address: {\it Center for Gravitational Physics and Quantum Information, Yukawa Institute for Theoretical Physics, Kyoto University, Kitashirakawa Oiwakecho, Sakyo-ku, Kyoto 606-8502, Japan} }} \\
%{\small \it Department of Physics, Osaka University,  Toyonaka, Osaka 560-0043, Japan} \\

\end{center}

\vskip 1.cm
\baselineskip=18pt plus 1pt minus 1pt

\begin{abstract}
In GUT inspired $SO(5)\times U(1)_X \times SU(3)_C$ gauge-Higgs unification (GHU) in the Randall-Sundrum 
warped  spacetime, the $W$ and $Z$ couplings
of all 4D fermion modes become nontrivial.  The $W$ and $Z$ couplings of zero-mode quarks and leptons 
slightly deviate from those in the SM, and the couplings take the matrix form in the space of Kaluza-Klein (KK) states.  
In particular, the 4D couplings and mass spectra in the KK states 
depend on the Aharonov-Bohm phase $\theta_H$ in the fifth dimension.

Nevertheless there emerge three astonishing sum rules among those coupling matrices,
which guarantees the finiteness of certain combinations of corrections to vacuum polarization tensors.  
We confirm  by numerical evaluation that the equality in the sum rules holds with 5 to 7 digits accuracy.  
Based on the sum rules we propose improved oblique parameters in GHU.  
Oblique corrections due to fermion 1-loop diagrams are found to be small.

\ignore{
For KK states $(u_n,d_n)$  ($n= 0,1,2, \cdots)$, for instance, we have $W$ and $Z$ coupling matrices 
\begin{align}
&(\hat g^W_{V/A} )_{nm} = \la  u_n | \hat g^W_{V/A}  | d_m \ra ~,~~
(\hat g^{W^\dagger}_{V/A} )_{nm} = (\hat g^W_{V/A} )_{mn}^* ~, \cr
\noalign{\kern 5pt}
&( \hat g^{Z, u}_{V/A} )_{nm} = \la  u_n | \hat g^{Z, u}_{V/A}  | u_m \ra ~,~~
( \hat g^{Z, d}_{V/A} )_{nm} = \la  d_n | \hat g^{Z, d}_{V/A}  | d_m \ra ~, \cr
\noalign{\kern 5pt}
&( \hat g^{W^3, u/d}_{V} )_{nm} =( \hat g^{Z, u/d}_{V} )_{nm} + \sin^2 \theta_W \, Q_{u/d} \, \delta_{nm} ~,
\end{align}
where $V$ and $A$ denote vector and axial vector couplings, and $(Q_u, Q_d) = (\frac{2}{3}, - \frac{1}{3})$.
They satisfy
\begin{align}
&\tr \big\{  \hat g^{W^3, u}_{V} \hat g^{W^3, u}_{V}  + \hat g^{Z, u}_{A} \hat g^{Z, u}_{A} 
+  \hat g^{W^3, d}_{V} \hat g^{W^3, d}_{V}  + \hat g^{Z, d}_{A} \hat g^{Z, d}_{A} \big\} \cr
\noalign{\kern 5pt}
&\hskip 2.cm 
= \onehalf \tr \big\{ \hat g^W_{V}  \hat g^{W^\dagger}_{V}  + \hat g^W_{A}  \hat g^{W^\dagger}_{A}  \big\}
\end{align}
and two more sum rules.
We confirmed  by numerical evaluation that the equality in the sum rules holds with 5 digits accuracy.  
Based on the sum rules we propose improved oblique parameters in GHU.  
Oblique corrections due to fermion 1-loop diagrams are found to be small.}

\end{abstract}

%\preprint{???, OU-HET/???}
%\pacs{???}

\newpage

\baselineskip=20pt plus 1pt minus 1pt
\parskip=0pt

\section{Introduction} 

Although the standard model (SM), $SU(3)_C \times SU(2)_L \times U(1)_Y$ gauge theory, has been 
successful in describing almost all of the phenomena at low energies, it has a severe gauge hierarchy 
problem when embedded in a larger theory such as grand unification.
One possible answer to this problem is  the gauge-Higgs unification  (GHU) scenario in which 
gauge symmetry is dynamically broken by an Aharonov-Bohm (AB)  phase, $\theta_H$, in the fifth dimension.
The 4D Higgs boson is identified with a 4D fluctuation mode of 
$\theta_H$.\cite{Hosotani1983, Davies1988, Hosotani1989, Davies1989,  Hatanaka1998, Hatanaka1999,  
Kubo2002, Scrucca2003, ACP2005, Cacciapaglia2006, Medina2007, HOOS2008, FHHOS2013, Yoon2018b}

Many GHU models have been proposed, among which $SO(5) \times U(1)_X \times SU(3)_C$ GHU models
in the Randall-Sundrum (RS) warped space \cite{RS1}  turn out promising candidates for describing physics beyond the SM.
The $SO(5) \times U(1)_X \times SU(3)_C$ gauge symmetry naturally incorporates the custodial symmetry
in the Higgs boson sector.\cite{ACP2005}   The orbifold boundary condition breaks $SO(5)$ to
$SO(4) \simeq SU(2)_L \times SU(2)_R$.  The $SU(2)_R \times U(1)_X$ symmetry is spontaneously broken
to $U(1)_Y$ by a brane scalar on the UV brane in the RS space.  The resultant $SU(2)_L \times U(1)_Y$ SM
symmetry is dynamically broken to $U(1)_\EM$ by the Hosotani mechanism.
The 4D Higgs boson is a zero mode of the fifth-dimensional component of gauge fields in $SO(5) / SO(4)$
which generates an AB phase in the fifth dimension.
The finite Higgs boson mass $m_H \sim 125.1\,$GeV is generated at the quantum level by the dynamics of
the AB  phase  $\theta_H$.

A realistic GHU model has been first proposed with quark-lepton multiplets in the vector ({\bf 5}) 
representation of $SO(5)$, which is referred to as the  GHU A-model.\cite{FHHOS2013}
It is recognized, however, that the A-model is difficult to be embedded in the grand unification scheme.
A natural grand-unified-theory (GUT) containing $SO(5) \times U(1)_X \times SU(3)_C$ GHU is
$SO(11)$ GHU with fermion multiplets in the spinor ({\bf 32}) and vector ({\bf 11}) 
representations.\cite{HosotaniYamatsu2015, Furui2016}
The GUT inspired $SO(5) \times U(1)_X \times SU(3)_C$ GHU model is defined with fermion multiplets
in the bulk in the $({\bf 3}, {\bf 4})$, $({\bf 1}, {\bf 4})$, $({\bf 1}, {\bf 5})$ and $({\bf 3}, {\bf 1})$
%  and $({\bf 1}, \bar{\bf 3})$
representations of $(SU(3)_C, SO(5))$.\cite{GUTinspired2019a}
$[({\bf 3}, {\bf 4}), ({\bf 1}, {\bf 4})]$ is contained in ${\bf 32}$ of $SO(11)$, whereas 
$[({\bf 1}, {\bf 5}), ({\bf 3}, {\bf 1})]$  is contained in ${\bf 11}$ of $SO(11)$.
In addition, Majorana fermions  in the singlet $({\bf 1}, {\bf 1})$ representation are introduced on the UV brane 
which provide the inverse seesaw mechanism for neutrinos.
The GUT inspired GHU model is referred to as the  GHU B-model.

The  GHU B-model is successful in many respects. It reproduces the quark, lepton, gauge and Higgs boson
spectra (except for the small mass of  the up quark), and yields nearly the same gauge couplings of the SM particles.
It can incorporate the Cabbibo-Kobayashi-Maskawa (CKM) matrix structure in the $W$ couplings 
with Flavour-Changing-Neutral-Currents (FCNC) naturally suppressed.\cite{FCNC2020a}
Many of the physical quantities at low energies are described mainly by the AB phase $\theta_H$,
being mostly independent of the parameters in the dark fermion sector.\cite{GUTinspired2020b}
Both A- and B- models predict $Z'$ particles as KK excited states of photon, $Z$ boson and $Z_R$ boson.
$Z'$ couplings of quarks and leptons exhibit large parity violation which can be explored and tested
at 250$\,$GeV $e^- e^+$ International Linear Collider (ILC).\cite{ILC2021}
By examining the dependence of event numbers on the polarization of incident electron and positron  beams
the  A- and B-models can be clearly 
distinguished.\cite{Funatsu2017a, Yoon2018a, Funatsu2019a, GUTinspired2020c, Funatsu2022b}
Effects of $Z'$ bosons can be seen in  single Higgs boson production processes as well.\cite{Yamatsu2023}

It has been established that useful and convenient quantities to investigate new physics beyond the SM are
the oblique parameters $S$, $T$ and $U$  of Peskin-Takeuchi, which represent
corrections to vacuum polarization tensors of $W$, $Z$ and 
photon.\cite{PeskinTakeuchi, Lavoura1993, PeskinSchroeder}
In the early study of gauge theory in the RS warped space it was argued that $SO(5) \times U(1)_X$ 
GHU models  may yield appreciable corrections to $S$ and $T$.\cite{Carena2006}
To evaluate oblique corrections at the one loop level in GHU, one has to know mass spectrum
and gauge couplings of all KK modes.  In Ref.\ \cite{Carena2006} $S$ and $T$ were expressed
in terms of truncated propagators of gauge bosons and fermions, by adopting a perturbative expansion in  $\theta_H$.

In this paper we use exactly determined mass spectra of fermion KK states at general $\theta_H$ at the tree level,
and evaluate $W$ and $Z$ couplings of the fermion KK states by making use of 
exactly determined wave functions of both gauge bosons and fermions in the $SO(5) \times U(1)_X$ space.
It has been known that the $W$ and $Z$ couplings of quarks and leptons are nearly the same as in the SM.
The $W$ and $Z$ couplings of the KK modes of quark and lepton multiplets become highly nontrivial, however.
They are not diagonal in the KK states at $\theta_H \not= 0$, 
taking the matrix form with nontrivial off-diagonal elements.
Further the wave functions of the $W$ and $Z$ bosons have substantial components
not only in the $SU(2)_L \times U(1)_Y$ space but also in the entire $SO(5) \times U(1)_X$ space,
which necessitates  refinement of the definition of oblique parameters.

One loop corrections to the vacuum polarization tensors of  $W$, $Z$  and photon contain 
divergences.  In the 4D $SU(2)_L \times U(1)_Y$ gauge theory certain combinations of
those vacuum polarization tensors, represented by $S$, $T$ and $U$, are finite.
In the GUT inspired $SO(5) \times U(1)_X \times SU(3)_C$ GHU various combinations of the  KK states
of fermions run along the loops in the propagators of gauge bosons.
It will be shown that there appear sum rules among the $W$ and $Z$ coupling matrices.
We shall confirm those sum rules by numerically evaluating the mass spectra of the KK modes and 
their $W$ and $Z$ coupling matrices.  The sum rules are found to hold with 5 to 7 digits accuracy.
With these sum rules certain  combinations of the 1-loop corrections to the vacuum polarization tensors
become finite, which leads to  improved oblique parameters.
It will be seen that corrections to the improved oblique parameters are small.
The total corrections are found to be $S \sim 0.01$,
$T \sim 0.12$  and  $ U \sim 0.00004$ for $\theta_H = 0.1$ and the KK mass scale $m_\KK = 13\,$TeV,
which is consistent with the current experimental data.\cite{Funatsu2022a}

Although the gauge couplings in GHU in the RS space have  highly nontrivial matrix structure, 
they satisfy remarkable identities.    The identities in the sum rules discussed in the present paper 
are associated with two-point functions  of gauge fields.   
It has been known that the exact identities hold in the combinations of the couplings appearing 
in three-point functions of gauge fields  in orbifold gauge theory.\cite{AnomalyFlow1, AnomalyFlow2}  
Triangle loop diagrams generally give rise to chiral anomalies in 4D gauge theory.
In 5D orbifold gauge theory anomaly coefficients for the three legs of various 4D KK modes of gauge fields
vary with  the AB phase in the fifth dimension.  This phenomenon is called the anomaly flow by an
AB phase.  Along triangle loops all possible KK modes of fermions run.  The sum of all those loop
contributions leads to the total anomaly coefficient which is expressed in terms of the values of 
the wave functions of gauge fields at the UV and IR branes in the RS space and orbifold boundary
conditions of the fermions.  In other words there hold sum rules for gauge coupling matrices of 
the third order.

In Section 2  the GUT inspired $SO(5) \times U(1)_X \times SU(3)_C$ GHU  is described.  
We explain how to determine the mass spectrum and wave functions of gauge bosons and  quark-lepton multiplets.
In Section 3 the $W$ and $Z$ couplings of all fermion modes are determined.  
In Section 4 fermion one-loop corrections to the vacuum polarization tensors of the $W$ boson, $Z$ boson  and 
photon  are evaluated.  In Section 5 we show that there appear three sum rules among the $W$ and $Z$ coupling matrices
of quarks, leptons and their KK modes.  The sum rules are confirmed by numerical evaluation.
Based on the coupling sum rules the improved oblique parameters are introduced in Section 6.
The finite corrections to the $S$, $T$ and $U$ parameters are evaluated, and are found to be small.
Section 7 is devoted to a summary and discussions.
In Appendix A basis functions used to express wave functions of gauge bosons and fermions are summarized.  
In Appendix B wave functions of  KK modes of down-type quarks and neutrinos are given.

\section{GUT inspired GHU} 

$SO(5) \times U(1)_X \times SU(3)_C$ GHU is defined in the RS space whose metric is given by\cite{RS1}
\begin{align}
ds^2= g_{MN} dx^M dx^N =
e^{-2\sigma(y)} \eta_{\mu\nu}dx^\mu dx^\nu+dy^2,
\label{RSmetric1}
\end{align}
where $M,N=0,1,2,3,5$, $\mu,\nu=0,1,2,3$, $y=x^5$, $\eta_{\mu\nu}=\mbox{diag}(-1,+1,+1,+1)$,
$\sigma(y)=\sigma(y+ 2L)=\sigma(-y)$, and $\sigma(y)=ky$ for $0 \le y \le L$.
In terms of the conformal coordinate $z=e^{ky}$ ($0 \le y \le L$, $1\leq z\leq z_L=e^{kL}$)
\begin{align}
ds^2=  \frac{1}{z^2} \bigg(\eta_{\mu\nu}dx^{\mu} dx^{\nu} + \frac{dz^2}{k^2}\bigg)~ .
\label{RSmetric-2}
\end{align}
The bulk region $0<y<L$ is anti-de Sitter (AdS) spacetime 
with a cosmological constant $\Lambda=-6k^2$, which is sandwiched by the
UV brane at $y=0$  and the IR brane at $y=L$.  
The KK mass scale is $m_{\rm KK}=\pi k/(z_L-1) \simeq \pi kz_L^{-1}$ for $z_L\gg 1$.

Gauge fields 
$A_M^{SO(5)}$,  $A_M^{U(1)_X}$ and $A_M^{SU(3)_C}$ of $SO(5) \times U(1)_X \times SU(3)_C$
satisfy the orbifold boundary conditions (BCs) 
\begin{align}
&\begin{pmatrix} A_\mu \cr  A_{y} \end{pmatrix} (x,y_j-y) =
P_{j} \begin{pmatrix} A_\mu \cr  - A_{y} \end{pmatrix} (x,y_j+y)P_{j}^{-1}
\label{BC-gauge1}
\end{align}
where $(y_0, y_1) = (0, L)$.   In terms of  
\begin{align}
P_{\bf 4}^{SO(5)}&=\mbox{diag}\left(I_{2},-I_{2}\right),\cr
P_{\bf 5}^{SO(5)}&=\mbox{diag}\left(I_{4},-I_{1}\right) , 
\label{BC1}
\end{align}
$P_0=P_1 = P_{\bf 5}^{SO(5)}$ for $A_M^{SO(5)}$ in the vector representation and 
$P_0=P_1= 1$ for $A_M^{U(1)_X}$ and $A_M^{SU(3)_C}$.
The 4D Higgs field is contained in the $SO(5)/SO(4)$ part of $A_y^{SO(5)}$.
The orbifold BCs break $SO(5)$ to $SO(4) \simeq SU(2)_L \times SU(2)_R$.

The matter content in the GUT inspired GHU (B-model) is summarized in Table \ref{Table:matter}.\cite{GUTinspired2019a}
Quark and lepton multiplets are introduced in three generations.  They satisfy
\begin{align}
&\Psi_{({\bf 3,4})}(x, y_j - y) = 
- P_{\bf 4}^{SO(5)} \gamma^5 \Psi_{({\bf 3,4})} (x, y_j + y) ~, \cr
&\Psi_{({\bf 3,1})^\pm}  (x, y_j - y) =
\mp \gamma^5 \Psi_{({\bf 3,1})^\pm } (x, y_j + y) ~, \cr
&\Psi_{({\bf 1,4})} (x, y_j - y) = 
- P_{\bf 4}^{SO(5)} \gamma^5 \Psi_{({\bf 1,4})} (x, y_j + y) ~.
\label{quarkleptonBC}
\end{align}
Here 5D Dirac matrices $\gamma^a$ $(a=0,1,2,3,5)$ satisfy  $\{\gamma^a,\gamma^b\}=2\eta^{ab}$
$(\eta^{ab}=\mbox{diag}(-I_1,I_4))$, and  $\gamma^5  =\mbox{diag} (1,1,-1,-1)$.
Dark fermion fields satisfy
\begin{align}
&\Psi_{({\bf 3,4})}^D (x, y_j - y) = 
(-1)^j  P_{\bf 4}^{SO(5)} \gamma^5 \Psi_{({\bf 3,4})}^D  (x, y_j + y) ~, \cr
&\Psi_{({\bf 1,5})^\pm}^D (x, y_j - y) = 
\pm P_{\bf 5}^{SO(5)} \gamma^5 \Psi_{({\bf 1,5})^\pm}^D (x, y_j + y) ~.
\label{darkFBC}
\end{align}
In addition, Majorana fermion fields ($\hat \chi (x)$) and brane scalar field ($\hat \Phi(x)$) are introduced 
on the UV brane (at $y=0$).
The brane scalar field $\hat \Phi$ spontaneously breaks $SU(2)_R \times U(1)_X$ to $U(1)_Y$ with
a vacuum expectation value (VEV) much larger than the KK mass scale $m_\KK$.
The Majorana fermion field ($\hat \chi = {\hat \chi}^c$) in each generation, combining with $\Psi_{({\bf 1,4})}$ and
$\hat \Phi$,  induces the inverse seesaw mechanism to account for a very small mass of the observed neutrino.

\begin{table}[tbh]
{%\small
\renewcommand{\arraystretch}{1.5}
\begin{center}
\caption{The matter field content in the GUT inspired  GHU model.
The $(SU(3)_C, SO(5))_{U(1)_X}$ content of each field is shown in the last column.  }
\vskip 10pt
\begin{tabular}{|c|c|c|}
%\hline
%&GUT inspired model\\
%& type B &type A\\
\hline
in the bulk &quark
&$({\bf 3}, {\bf 4})_{\frac{1}{6}} ~ ({\bf 3}, {\bf 1})_{-\frac{1}{3}}^+ 
    ~ ({\bf 3}, {\bf 1})_{-\frac{1}{3}}^-$\\
\cline{2-3}
&lepton
&$\strut ({\bf 1}, {\bf 4})_{-\frac{1}{2}}$ \\
\cline{2-3}
&dark fermion $\Psi^D$ & $({\bf 3}, {\bf 4})_{\frac{1}{6}} ~ ({\bf 1}, {\bf 5})_{0}^+ ~ ({\bf 1}, {\bf 5})_{0}^-$  \\
\hline 
on the UV brane
&Majorana fermion $\hat \chi$ &$({\bf 1}, {\bf 1})_{0} $ \\
%&{\small (Majorana)}
%&$({\bf 1}, [{\bf 1,2}])_{\frac{1}{2}, -\frac{1}{2}, -\frac{3}{2}}$ \\
\cline{2-3}
&brane scalar $\hat \Phi$ &$({\bf 1}, {\bf 4})_{\frac{1}{2}} $ \\
\hline
\end{tabular}
\label{Table:matter}
\end{center}
}
\end{table}

The action of the GUT inspired GHU has been given in Ref.\cite{GUTinspired2019a}.
Let $\Psi^J$ collectively denote all fermion fields in the bulk.   Then the action of the fermions in the bulk becomes
\begin{align}
&S_{\rm bulk}^{\rm fermion} =  \int d^5x\sqrt{-\det G} \,
\bigg\{ \sum_J  \overline{\Psi^J}  {\cal D} (c_J) \Psi^J \cr
\noalign{\kern 3pt}
&\quad 
- \sum_\alpha \Big( m_D^\alpha \overline{\Psi}{}_{({\bf 3},{\bf 1}) ^+}^{ \alpha}
\Psi_{({\bf 3},{\bf 1})^- }^{-\alpha}+ \mbox{h.c.} \Big)
- \sum_\beta \Big( m_V^\beta \overline{\Psi}{}_{({\bf 1},{\bf 5})^+ }^{\beta}
\Psi_{({\bf 1},{\bf 5})^- }^{\beta}+ \mbox{h.c.} \Big) \bigg\} ,
\label{fermionAction1}
\end{align} 
where $\overline{\Psi} = i \Psi^\dagger \gamma^0$ and
%$m_D^{\alpha}$ and  $m_V^{\beta}$  are ``pseudo-Dirac'' bulk mass terms.
\begin{align}
&{\cal D}(c)= \gamma^A {e_A}^M
\bigg( D_M+\frac{1}{8}\omega_{MBC}[\gamma^B,\gamma^C]  \bigg) -c\sigma'(y) ~, \cr
\noalign{\kern 5pt}
&D_M =  \dd_M - ig_S A_M^{SU(3)} -i g_A A_M^{SO(5)}
 -i g_B Q_X A_M ^{U(1)} ~. 
\label{covariantD}
\end{align}
The dimensionless parameter $c$ in ${\cal D}(c)$ is called the bulk mass parameter, which controls
the wave functions of the zero modes of the fermions.  
%$0 < c  < 0.5$ for the top-bottom quark multiplet and
%$0.5 < c \lesssim 1$  for other quark-lepton multiplets.
$m_D^{\alpha}$ and  $m_V^{\beta}$  are pseudo-Dirac  mass terms.
The action for the Majorana fermion field ($\hat \chi^\alpha$) is 
\begin{align}
S_{\rm brane}^{\hat \chi} = &
\int d^5x\sqrt{-\det G} \, \delta(y) \bigg\{  
\frac{1}{2}\overline{\hat \chi} {}^\alpha \gamma^\mu\partial_\mu \hat \chi^\alpha
 - \frac{1}{2} M^{\alpha \beta}  \overline{\hat \chi} {}^\alpha \hat \chi^{\beta} \bigg\} 
\label{Action-Majorana}
\end{align}
where $M^{\alpha \beta}$ represents Majorana masses.  In the present paper we take 
$M^{\alpha \beta} = M^\alpha \delta^{\alpha \beta}$ for simplicity.

In addition there are gauge-invariant brane interactions given by 
\begin{align}
&%\hspace{-3em}
 S_{\rm brane}^{\rm int}=
\int d^5x\sqrt{-\det G} \, \delta(y)
\big({\cal L}_1+{\cal L}_2+{\cal L}_3\big) ~, \cr
\noalign{\kern 5pt}
&{\cal L}_1 = 
- \Big\{ \kappa^{\alpha\beta} \,
\overline{\Psi}{}_{({\bf 3,4})}^{\alpha} \hat \Phi_{({\bf 1,4})}
\cdot \Psi_{({\bf 3,1})^+ }^{\beta}  + {\rm h.c.} \Big\} ~, \cr
\noalign{\kern 5pt}
&{\cal L}_2 =
- \Big\{ \widetilde{\kappa}^{\prime\alpha\beta} \,
\overline{\Psi}{}_{({\bf 1,4})}^{\alpha} \, \Gamma^a \,
\tilde{\hat \Phi}_{({\bf 1,4})} \cdot \big(\Psi_{({\bf 1,5})^- }^{\beta} \big)_a
 + {\rm h.c.} \Big\} ~, \cr
\noalign{\kern 5pt}
&{\cal L}_3 =
- \Big\{ \tilde{\kappa}_{\bf 1}^{\alpha \beta} \,
\overline{\chi}^\beta 
\widetilde{\hat \Phi} {}_{({\bf 1,4})}^\dag \Psi_{({\bf 1,4})}^{\alpha}   + {\rm h.c.} \Big\} ~, 
\label{BraneInteraction}
\end{align}
where $\kappa$'s are coupling constants. $\tilde{\Phi}_{({\bf 1,4})}$ denotes a conjugate field in $({\bf 1,4})$
formed from $\Phi_{({\bf 1,4})}^*$.
The brane field $\hat \Phi$ develops a nonvanishing expectation value $\la \hat \Phi \ra \not= 0$.

In the electroweak sector there are two 5D gauge couplings, $g_A$ and $g_B$, corresponding to
the gauge groups $SO(5)$ and $U(1)_X$, respectively.
The 5D gauge coupling $g_Y^{\rm 5D}$ of $U(1)_{Y}$  is given by
\begin{align}
 g_Y^{\rm 5D} =\ g_A s_\phi ~,~~
 s_\phi = \frac{g_B}{\sqrt{g_A^2+g_B^2}} ~.
\label{5DcouplingU1Y}
\end{align}
The 4D $SU(2)_L$ and $U(1)_Y$ gauge coupling constants are given  by
\begin{align}
g_w = \frac{g_A}{\sqrt{L}} ~, ~~ g_Y = \frac{ g_Y^{\rm 5D}}{\sqrt{L}} ~.
\label{gaugecoupling1}
\end{align}
The bare weak mixing angle $\theta_W^0$ is given by
\begin{align}
\sin \theta_W^0 = \frac{g_Y}{\sqrt{g_w^2 +g_Y^2}}  
= \frac{s_\phi}{\sqrt{1 + s_\phi^2}} ~.
\label{angle2}
\end{align}
As is seen below, the mixing angle determined from the ratio of $m_W$ to $m_Z$ 
slightly differs from the one in (\ref{angle2}) even at the tree level in GHU in the RS space; 
$m_W/m_Z  |_{\rm tree} \not= \cos \theta_W^0$.

The 4D Higgs boson field is a part of $A_y^{SO(5)}$.  
$A_z^{SO(5)} = (kz)^{-1} A_y^{SO(5)}$ ($1 \le z \le z_L$)  in the tensor representation is expanded as
\begin{align}
A_z^{(j5)} (x, z) &= \frac{1}{\sqrt{k}} \, \phi_j (x) u_H (z) + \cdots ,~~
u_H (z) = \sqrt{ \frac{2}{z_L^2 -1} } \, z ~, \cr
%\noalign{\kern 5pt}
%u_H (z) &= \sqrt{ \frac{2}{z_L^2 -1} } \, z ~,  \cr
\noalign{\kern 5pt}
\Phi_H (x) &= \frac{1}{\sqrt{2}} \begin{pmatrix} \phi_2 + i \phi_1 \cr \phi_4 - i\phi_3 \end{pmatrix} .
\label{4dHiggs}
\end{align}
$\Phi_H (x) $ corresponds to the doublet Higgs field in the SM.
$\Phi_H$ develops nonvanishing expectation value at the quantum level by the Hosotani mechanism.
Without loss of generality we take $\la \phi_1 \ra , \la \phi_2 \ra , \la \phi_3 \ra  =0$ and  $\la \phi_4 \ra \not= 0$.
The AB phase $\theta_H$ in the fifth dimension is given by % related to the eigenvalues of
\begin{align}
\hat W &= P \exp \bigg\{ i g_A \int_{-L}^L dy \, \la A_y^{SO(5)}  \ra \bigg\}  
=  \exp \Big\{ i  \theta_H  \cdot 2 T^{(45)} \Big\} ~ .
\label{ABphase1}
\end{align}
Here $  \la A_y^{SO(5)}  \ra = (2k)^{-1/2} \la \phi_4\ra v_H (y) T^{(45)}$, $v_H (y) = k e^{ky} u_H (z)$ 
for $0 \le y \le L$ and $v_H (-y) =  v_H (y) =  v_H (y + 2L)$.
In terms of $\theta_H$, $A_z^{(45)} $ is expanded as 
\begin{align}
&A_z^{(45)} (x, z) = \frac{1}{\sqrt{k}} \big\{ \theta_H f_H + H(x) \big\} \, u_H(z) + \cdots , \cr
\noalign{\kern 5pt}
&f_H = \frac{2}{g_A} \sqrt{ \frac{k}{z_L^2 -1}} = \frac{2}{g_w} \sqrt{ \frac{k}{L(z_L^2 -1)}} ~.
\label{ABphase3}
\end{align}
4D neutral Higgs field $H(x)$ is the fluctuation mode of the AB phase $\theta_H$.

\subsection{Spectrum and wave functions of gauge fields} 

When the VEV  $| \la \hat \Phi \ra |= w$ is sufficiently large, $w \gg m_\KK$,
the spectra of the $W$ and $Z$ towers, 
$\{ m_{W^{(n)}} = k \lambda_{W^{(n)}} , m_{Z^{(n)}} = k \lambda_{Z^{(n)}}\}$ are determined by
the zeros of
\begin{align}
 & 2S(1;\lambda_{W^{(n)}} )C'(1;\lambda_{W^{(n)}} )+\lambda_{W^{(n)}} \sin^2\theta_H =  0 ~, \cr
 \noalign{\kern 5pt}
 & 2S(1;\lambda_{Z^{(n)}} )C'(1;\lambda_{Z^{(n)}} )+(1+s_\phi^2) \lambda_{Z^{(n)}} \sin^2\theta_H =  0 ~,
 \label{WZspectrum1}
\end{align}
where  the functions $C(z;\lambda)$ and $S(z; \lambda)$ are given in Appendix A.
Note $(1+s_\phi^2)^{-1}  = \cos^2 \theta_W^0$.
The lowest modes are $W = W^{(0)}$ and $Z = Z^{(0)}$.  For $z_L \gg 1$ their masses at the tree level
are approximately given by
\begin{align}
&m_W^{\rm tree} \simeq \frac{\sin\theta_H}{\pi\sqrt{kL}} \, m_\KK ~, \cr
\noalign{\kern 5pt}
& m_Z \simeq \sqrt{1+s_\phi^2} \, \frac{\sin\theta_H}{\pi\sqrt{kL}}  \,  m_\KK 
\simeq \frac{m_W^{\rm tree}}{\cos \theta_W^0} ~.
 \label{WZmass1}
\end{align}
In this paper $m_Z = m_Z^{\rm tree} = 91.1876\,$GeV is taken as one of the input parameters.
As typical values we take $m_\KK = 13\,$TeV, $\theta_H= 0.1$, $\sin^2 \theta_W^0 =0.230634$
and $\alpha_\EM (m_Z) = 1/128$, which implies that $z_L = 3.86953 \times 10^{11}$ and $kL= 26.6816$.
The precise values determined from (\ref{WZspectrum1}) give
$m_{Z^{(0)}} \cos \theta_W^0 / m_{W^{(0)}} = 1.00002$.

Each mode of the gauge boson tower has components in the $SO(5) \times U(1)_X$ space.
Let us decompose the $SO(5)$ generators $\{ T^{jk}= - T^{kj} \, ; j,k = 1 \sim 5 \}$ 
into $SU(2)_L$ and $SU(2)_R$ generators 
$\{ T^a_{L/R}  =\onehalf ( \onehalf \ep^{ajk} T^{jk} \pm T^{a4})\,  ; a,j,k =1 \sim 3 \}$ and
$\{ \hat T^p= 2^{-1/2} T^{p5} \, ; p=1 \sim 4 \}$.
We denote the generator of $U(1)_X$ as $Q_X$.
Then 5D gauge potentials $(\sqrt{k})^{-1} \big[A_\mu^{SO(5)} (x,z) + (g_B/g_A) A_\mu^{U(1)_X} (x,z) Q_X \big]$ 
have expansion
\begin{align}
&\sum_n W_\mu^{(n)} (x) \bigg\{ h^L_{W^{(n)}} (z) \frac{T^1_L+ i T^2_L}{\sqrt{2}}
+ h^R_{W^{(n)}} (z) \frac{T^1_R+ i T^2_R}{\sqrt{2}} 
+ \hat h_{W^{(n)}} (z) \frac{\hat T^1 + i \hat T^2}{\sqrt{2}} \bigg\}
 \label{gaugeW1}
\end{align}
for the $W$ tower component,  
\begin{align}
&\sum_n Z_\mu^{(n)} (x) \Big\{ h^L_{Z^{(n)}} (z) T^3_L + h^R_{Z^{(n)}} (z)  T^3_R
+ \hat h_{Z^{(n)}} (z) \hat T^3  +  h^B_{Z^{(n)}} (z) \frac{g_B}{g_A} Q_X  \Big\}
 \label{gaugeZ1}
\end{align}
for the $Z$ tower component, and
\begin{align}
&\sum_n A_\mu^{\gamma (n)} (x) \Big\{ h^L_{\gamma^{(n)}} (z) T^3_L + h^R_{\gamma^{(n)}} (z)  T^3_R
 +  h^B_{\gamma^{(n)}} (z)  \frac{g_B}{g_A} Q_X  \Big\}
 \label{gaugegamma1}
\end{align}
for the photon tower component.  Here
\begin{align}
&\begin{pmatrix}  h^L_{W^{(n)}} (z) \cr \mynoalign
 h^R_{W^{(n)}} (z) \cr  \mynoalign
 \hat h_{W^{(n)}} (z) \end{pmatrix}
= \frac{1}{\sqrt{\, 2 \, r_{W^{(n)}} }} 
\begin{pmatrix} (1 + \cos \theta_H ) \,C(z, \lambda_{W^{(n)}} ) \cr \mynoalign
(1 - \cos \theta_H)\,C(z, \lambda_{W^{(n)}} ) \cr  \mynoalign
 - \sqrt{2} \sin \theta_H  \hat S (z, \lambda_{W^{(n)}}  ) \end{pmatrix}, \cr
 %%%%%%%%
\noalign{\kern 5pt}
&\begin{pmatrix}  h^L_{Z^{(n)}} (z) \cr \mysnoalign
 h^R_{Z^{(n)}} (z) \cr  \mysnoalign    \hat h_{Z^{(n)}} (z) \cr  \mysnoalign h^B_{Z^{(n)}} (z) \end{pmatrix} =
 \begin{pmatrix}  h^{L, su2}_{Z^{(n)}} (z) \cr \mysnoalign
 h^{R, su2}_{Z^{(n)}} (z) \cr  \mysnoalign    \hat h^{su2}_{Z^{(n)}} (z) \cr   0 \end{pmatrix}
 - \sin \theta_W^0  \begin{pmatrix} \sin \theta_W^0 \cr   \sin \theta_W^0 \cr     0 \cr 
\sqrt{1 - 2 \sin^2 \theta_W^0} \end{pmatrix} h^{Q}_{Z^{(n)}} (z) , \cr
%%%%%
 \noalign{\kern 5pt}
&\hskip 0.7cm
 \begin{pmatrix}  h^{L, su2}_{Z^{(n)}}  (z) \cr \mysnoalign
 h^{R, su2}_{Z^{(n)}}  (z) \cr  \mysnoalign   \hat h^{su2}_{Z^{(n)}} (z)  \end{pmatrix} =
\frac{1}{\sqrt{ \, 2 \, r_{Z^{(n)}} }} 
\begin{pmatrix} (1 +  \cos \theta_H) \,C(z, \lambda_{Z^{(n)}} ) \cr 
(1 -  \cos \theta_H) \,C(z, \lambda_{Z^{(n)}} ) \cr  
- \sqrt{2}  \, \sin \theta_H  \check S (z, \lambda_{Z^{(n)}}  ) \end{pmatrix} , \cr
\noalign{\kern 5pt}
&\hskip 1.cm
h^{Q}_{Z^{(n)}} (z)  = \sqrt{\frac{2}{r_{Z^{(n)}}}} \, C(z, \lambda_{Z^{(n)}} )~, \cr
\noalign{\kern 5pt}
&\begin{pmatrix}  h^L_{\gamma^{(n)}} (z) \cr \mynoalign
 h^R_{\gamma^{(n)}} (z) \cr  \mynoalign
 h^B_{\gamma^{(n)}} (z)  \end{pmatrix}
=
\begin{pmatrix} \sin \theta_W^0 \cr \mynoalign
 \sin \theta_W^0 \cr  \mynoalign
\sqrt{1 - 2 \sin^2 \theta_W^0} \end{pmatrix}
\begin{cases}
\myfrac{1}{\sqrt{ r_{\gamma^{(n)}} }} C(z, \lambda_{\gamma^{(n)}} )&{\rm for~}n \ge 1 ~, \cr
%\noalign{\kern 5pt}
\myfrac{1}{\sqrt{kL}}  &{\rm for~}n =0~, 
\end{cases}
\label{waveWZ1}
\end{align}
where the normalization factors $\{ r_{W^{(n)}} , r_{Z^{(n)}} ,r_{\gamma^{(n)}} \}$ are determined by
\begin{align}
\int_1^{z_L}  \frac{dz}{z}  \, \sum_\alpha |h_n^\alpha (z)  |^2 = 1
\end{align}
in each mode. $\hat S (z, \lambda)$is given in (\ref{functionA1}). 
The photon ($\gamma^{(0)}$) coupling is $e \, Q_\EM$ where 
$e = g_w \sin \theta_W^0$  and  $Q_\EM = T^3_L + T^3_R +Q_X$.

\subsection{Spectrum and wave functions of fermion fields} 

The spectra $\{ m_n = k \lambda_n \}$ of the KK towers of up-type quarks and charged leptons are 
determined by the zeros of
\begin{align}
S_L (1; \lambda_n, c)  S_R (1; \lambda_n, c)  + \sin^2 \frac{\theta_H}{2} = 0
\label{fermionspectrum1}
\end{align}
where $S_{L/R} (z; \lambda, c)$ is defined in (\ref{functionA2}).
The bulk mass parameter $c$ of each doublet multiplet is determined such that
the lowest value $\lambda_0$ reproduces $m_u$, $m_c$, $m_t$,  $m_e$, $m_\mu$ or $m_\tau$.
For $\theta_H=0.1$ and $m_\KK = 13\,$TeV, $(c_u, c_c, c_t) = (-0.859,  -0.719, -0.275)$
and $(c_e, c_\mu, c_\tau) = (-1.01, -0.793, - 0.675)$.
Although Eq.\ (\ref{fermionspectrum1}) is satisfied by either positive or negative $c$, the negative values
for $c$ are chosen in the B-model  to have the spectra of the KK towers of down-type quarks and 
neutrinos consistent with observation as explained below.

Wave functions of $\{ u^{(n)} \}$ and  $\{ e^{(n)} \}$ in the first generation are contained in the 
$SO(5)$ spinor multiplets $\Psi_{({\bf 3,4})}$ and $\Psi_{({\bf 1,4})}$, which are denoted as $(u,d, u',d')$ 
and $(\nu_e, e, \nu_e', e')$, respectively.  $(u,d)$  and $ (\nu_e, e)$ are $SU(2)_L$ doublets, whereas $(u',d')$  
and $ (\nu_e', e')$ are $SU(2)_R$ doublets.
5D $(u, u')(x,z)$ fields are expanded as
\begin{align}
&\frac{1}{z^2} \begin{pmatrix} u \cr u' \end{pmatrix} = 
\sqrt{k} \sum_{n=0}^\infty \bigg\{ u^{(n)}_L (x) \begin{pmatrix} f^{u^{(n)}}_L (z) \cr g^{u^{(n)}}_L (z) \end{pmatrix}
+ u^{(n)}_R (x) \begin{pmatrix} f^{u^{(n)}}_R (z) \cr g^{u^{(n)}}_R (z) \end{pmatrix} \bigg\}
\label{wave-up1}
\end{align}
where, in terms of functions defined in (\ref{functionA2}), 
\begin{align}
&\begin{pmatrix} f^{u^{(n)}}_L (z) \cr g^{u^{(n)}}_L (z) \end{pmatrix}=  \frac{1}{\sqrt{r_{u^{(n)} L}}}
\begin{pmatrix} \cos \onehalf \theta_H C_L (z, \lambda_{u^{(n)}}, c_u) \cr
\noalign{\kern 5pt}
i \sin \onehalf \theta_H \hat S_L (z, \lambda_{u^{(n)}}, c_u) \end{pmatrix}, \cr
\noalign{\kern 5pt}
&\begin{pmatrix} f^{u^{(n)}}_R (z) \cr g^{u^{(n)}}_R (z) \end{pmatrix}=  \frac{1}{\sqrt{r_{u^{(n)} R}}}
\begin{pmatrix} \cos \onehalf \theta_H S_R (z, \lambda_{u^{(n)}}, c_u) \cr
\noalign{\kern 5pt}
i \sin \onehalf \theta_H \hat C_R (z, \lambda_{u^{(n)}}, c_u) \end{pmatrix} .
\label{wave-up2}
\end{align}
The normalization factor in each mode is determined by the condition
\begin{align}
\int_1^{z_L} dz \, \Big\{ |f_n (z) |^2 + |g_n (z) |^2 \Big\} = 1 \quad {\rm for~} 
\begin{pmatrix} f_n (z) \cr g_n (z) \end{pmatrix} .
\end{align}
One can show that $r_{u^{(n)} L}= r_{u^{(n)}R} \equiv r_{u^{(n)}}$.
Note that with the use of (\ref{fermionspectrum1}) one can express $(f^{u^{(n)}}_R, g^{u^{(n)}}_R)$ as
\begin{align}
&\begin{pmatrix} f^{u^{(n)}}_R (z) \cr g^{u^{(n)}}_R (z) \end{pmatrix}=  \frac{1}{\sqrt{r_{u^{(n)} R}^\prime}}
\begin{pmatrix}   \sin  \onehalf \theta_H \hat S_R (z, \lambda_{u^{(n)}}, c_u) \cr
\noalign{\kern 5pt}
- i \cos \onehalf \theta_H  C_R (z, \lambda_{u^{(n)}}, c_u) \end{pmatrix} .
\label{wave-up3}
\end{align}
At $\theta_H = 0$, $ \lambda_{u^{(0)}} =0$ so that the zero mode $u^{(0)}$ has purely chiral structure;
$(u_L^{(0)}, d_L^{(0)})$ becomes an $SU(2)_L$ doublet, whereas $u_R^{(0)}$ and $d_R^{(0)}$ become
 $SU(2)_L$ singlets.

Wave functions of $\{ e^{(n)} \}$  have the same structure as those of $\{ u^{(n)} \}$.  
Formulas for $\{ e^{(n)} \}$ are obtained from (\ref{wave-up1})-(\ref{wave-up3}) by replacing $u^{(n)}$ and $c_u$
by $e^{(n)}$ and $c_e$.

For mass spectra and wave functions of down-type quark and neutrino multiplets 
the $SO(5)$ singlet fields $\Psi_{({\bf 3,1})^\pm}$ and Majorana brane fermions $\hat \chi$ intertwine.
In general the coupling constants $\kappa$'s in  the brane interactions given in Eq.\ (\ref{BraneInteraction}) 
are not diagonal in the generation space.  The ${\cal L}_1$ term in  (\ref{BraneInteraction}), 
with $\la \hat \Phi \ra \not= 0$, leads to the Kobayashi-Maskawa mixing matrix in the quark sector.
Further complex  $\kappa^{\alpha\beta}$'s give rise to $CP$-violation phases.
In the present paper we analyze the case in which the brane interactions given in Eq.\ (\ref{BraneInteraction}) 
are diagonal in the generation space.

In the first generation the $d$ and $d'$ components in $\Psi_{({\bf 3,4})}$ and 
$\Psi_{({\bf 3,1})^\pm} \equiv D^\pm$ intertwine with each other.  The mass spectrum $\{ m_n = k \lambda_n \}$ is
determined by the zeros of
\begin{align}
&\Big( S_L^Q S_R^Q +\sin^2\frac{\theta_H}{2} \Big)
 \big({\cal S}_{L1}^{D}{\cal S}_{R1}^{D}
 -{\cal S}_{L2}^{D}{\cal S}_{R2}^{D}\big) \cr
 \noalign{\kern 5pt}
 &\hskip 1.cm
+|\mu|^2 C_R^Q S_R^Q
 \left({\cal S}_{L1}^{D}{\cal C}_{L1}^{D} -{\cal S}_{L2}^{D}{\cal C}_{L2}^{D}\right)=0 
 \label{spectrumDown1}
\end{align}
where  $S_{L/R}^Q = S_{L/R}(1; \lambda, c_u)$, 
${\cal S}_{L j}^{D} = {\cal S}_{L j} (1; \lambda, c_{D}, \tilde m_{D})$, etc.
Functions ${\cal S}_{L/R j}$, $ {\cal C}_{L/R j}$ are given in (\ref{functionA3}).
$c_{D}$ is the bulk mass parameter of $\Psi_{({\bf 3},{\bf 1})^\pm}$ field, and
$\tilde m_{D} = m_{D}/k$ where $m_D$ is a Dirac mass connecting $D^+$ and $D^-$.
The parameter $\mu$ represents  the strength of a brane interaction among $\Psi_{({\bf 3},{\bf 4})}$, 
$\Psi_{({\bf 3},{\bf 1})^\pm}$ and $\hat \Phi$, which is necessary to reproduce a mass of
each down-type quark.
As typical values we take $(\tilde m_d, \tilde m_s, \tilde m_b) = (1,1,1)$ and 
$(\mu_{d},  \mu_{s}, \mu_{b})= (0.1, 0.1, 1)$.
We determine bulk mass parameters $c_{D}$'s  to reproduce a down-type quark mass in each
generation, finding $(c_{D_d}, c_{D_s}, c_{D_b}) = (0.6244, 0.6563,  0.8725)$.
We have chosen the negative values for $(c_u, c_c, c_t)$.  
With positive $c_u, c_c > \onehalf$ there would arise an exotic  extra light mode of charge $Q_\EM = - \frac{1}{3}$
with a mass much less than $m_\KK$ in the first and second generation from (\ref{spectrumDown1}),  
which contradicts with the observation.
One comment is in order.  Eqs.\  (\ref{fermionspectrum1}) and (\ref{spectrumDown1})  imply
that the up-type quark mass is larger than the corresponding down-type quark mass, although
in the first generation $m_u < m_d$.  The resolution of this problem is left for future investigation.
In this paper we take $(m_u,m_d) = (20, 2.9)\,$MeV at the $m_Z$ scale.  This does not affect 
gauge couplings and KK spectra of $(u,d)$ multiplets in the discussions below
as $m_u, m_d \ll m_\KK \sim 13\,$TeV.

In the first generation there are two types of series, 
$\{ d^{(n)} ; n \ge 0\}$ and $\{ D^{(n)}; n \ge 1 \}$.\footnote{In Ref.\ \cite{GHUfiniteT2021} 
the series has been decomposed into  $\{ d^{(n)} \}$, $\{ d^{\prime (n)}  \}$,  $\{ D^{+(n)} \}$ and $\{ D^{-(n)} \}$.}
For $\theta_H =0.1$ and $m_\KK = 13\,$TeV, the mass spectra of the KK excited states are 
$(m_{d^{(1)}} , m_{d^{(2)}}, m_{d^{(3)}}, \cdots ) =(12.2, 17.8, 25.1,  \cdots ) \,$TeV and
$(m_{D^{(1)}} , m_{D^{(2)}}, m_{D^{(3)}}, \cdots ) =(8.4, 16.7, 22.8,  \cdots ) \,$TeV.
The $(d, d',D^+, D^-)$ fields are expanded as
%Wave functions of each mode have four components; %$(d, d',D^+, D^-)$;
\begin{align}
\frac{1}{z^2} \begin{pmatrix} d \cr d'  \cr D^+ \cr D^- \end{pmatrix} 
&=  \sqrt{k} \sum_{n=0}^\infty \Bigg\{ d^{(n)}_L (x) 
\begin{pmatrix} f^{d^{(n)}}_L (z) \cr g^{d^{(n)}}_L (z) \cr h^{d^{(n)}}_L (z) \cr k^{d^{(n)}}_L (z)\end{pmatrix}
+ d^{(n)}_R (x) 
\begin{pmatrix} f^{d^{(n)}}_R (z) \cr g^{d^{(n)}}_R (z)  \cr h^{d^{(n)}}_R (z) \cr k^{d^{(n)}}_R (z) \end{pmatrix} \Bigg\} \cr
\noalign{\kern 5pt}
&+  \sqrt{k} \sum_{n=1}^\infty \Bigg\{ D^{(n)}_L (x) 
\begin{pmatrix} f^{D^{(n)}}_L (z) \cr g^{D^{(n)}}_L (z) \cr h^{D^{(n)}}_L (z) \cr k^{D^{(n)}}_L (z)\end{pmatrix}
+ D^{(n)}_R (x) 
\begin{pmatrix} f^{D^{(n)}}_R (z) \cr g^{D^{(n)}}_R (z)  \cr h^{D^{(n)}}_R (z) \cr k^{D^{(n)}}_R (z) \end{pmatrix} \Bigg\}.
\label{wave-down1}
\end{align}
Wave functions are normalized by
\begin{align}
&\int_1^{z_L} dz   \, \big\{ | f |^2 + |g |^2 + |h|^2 + |k |^2 \big\} = 1 
\end{align}
in each mode.
The explicit forms of the wave functions are given in (\ref{wave-down2}) and  (\ref{wave-down3}).

In the neutrino sector  the $\nu$ and $\nu'$ components in $\Psi_{({\bf 1,4})}$ and 
the brane Majorana fermion $\hat \chi$ mix with each other.
$\hat \chi (x) = \hat \chi^c (x)$ has a Majorana mass $M$.
The brane interaction ${\cal L}_3$ in (\ref{BraneInteraction}) 
% among $\Psi_{({\bf 1,4})}$, $\hat \chi$ and $\hat \Phi$  on the UV brane at $z=1$
generates a mixing brane mass term  $(m_B/\sqrt{k}) (\bar{\hat \chi} \nu'_R + \bar\nu'_R \hat \chi)$.
Because of the Majorana mass term eigen modes in the neutrino sector have both left- and right-handed
components.  The mass spectra $\{ m_{\nu^{\pm (n)}} = k \lambda_{\nu^{\pm (n)}}  \}$ are determined by  
\begin{align}
&K^\pm_\nu \equiv 
(k\lambda \mp M ) \Big\{ S_L^L S_R^L +\sin^2\frac{\theta_H}{2} \Big\}
+ \frac{m_B^2}{k} S_R^L C_R^L=0 ~.
\label{neutrinoSpectrum1}
\end{align}
where  $S_{L/R}^L = S_{L/R}(1; \lambda, c_e)$ etc.\ in the first generation.
For $c_e < - \onehalf$ and $M >0$ the gauge-Higgs seesaw mechanism, similar to the inverse seesaw mechanism,  
is at work in the $K_\nu^+$ series to generate a small neutrino mass.\cite{GHUseesaw2017}
\begin{align}
&m_{\nu_e} = k \lambda_{\nu^{+ (0)}}  \simeq
\frac{m_e^2 M}{(- 2c_e -1) m_B^2} ~.
\label{neutrinomass1}
\end{align}
There arises no light mode in the $K_\nu^-$ series.  For the KK excited modes the two series are
nearly degenerate; $ \lambda_{\nu^{- (n)}}  \simeq  \lambda_{\nu^{+ (n)}}  $ for $n \ge 1$.
We note that with $c_e > \onehalf$ the mass of the lightest mode becomes
$\sim m_e^2 M z_L^{2 c_e +1}/(2 c_e + 1)m_B^2$ so that unnecessarily large $m_B$ is required
to reproduce a small neutrino mass.  Further $c_e > \onehalf$ yields an additional exotic mode with
 a mass of $O(10\,{\rm GeV})$.  We adopt $c_e, c_\mu, c_\tau < - \onehalf$.

The Majorana field $\hat \chi$ is decomposed as
\begin{align}
&\hat \chi = \begin{pmatrix} \xi \cr \eta \end{pmatrix} , ~~
\hat \chi^c=  e^{i \delta_C} \begin{pmatrix} + \sigma^2 \eta^* \cr - \sigma^2 \xi^*  \end{pmatrix} = \hat \chi ~.
\label{Majorana1}
\end{align}
The fields are expanded as
\begin{align}
\frac{1}{z^2} \begin{pmatrix} \nu \cr \nu' \end{pmatrix}
&= \sqrt{k} \sum_{n=0}^\infty \Bigg\{ \nu^{+(n)}_{L} (x) \begin{pmatrix} f_L^{\nu^{+(n)}} (z) \cr g_L^{\nu^{+(n)}} (z) \end{pmatrix}
+ \nu^{+(n)}_{R} (x) \begin{pmatrix} f_R^{\nu^{+(n)}} (z) \cr g_R^{\nu^{+(n)}} (z) \end{pmatrix} \Bigg\} \cr
\noalign{\kern 5pt}
%&\hskip 1.5cm 
&+ \sqrt{k} \sum_{n=1}^\infty  \Bigg\{ \nu^{-(n)}_{L} (x) \begin{pmatrix} f_L^{\nu^{-(n)}} (z) \cr g_L^{\nu^{-(n)}} (z) \end{pmatrix}
+ \nu^{-(n)}_{R} (x) \begin{pmatrix} f_R^{\nu^{-(n)}} (z) \cr g_R^{\nu^{-(n)}} (z) \end{pmatrix} \Bigg\} , \cr
\noalign{\kern 5pt}
%&\hskip .5cm
\eta &= \sum_{n=0}^\infty  \nu^{+(n)}_{L} (x)  \, h^{\nu^{+(n)}} +  \sum_{n=1}^\infty  \nu^{-(n)}_{L} (x)  \, h^{\nu^{-(n)}}  ~, \cr
\noalign{\kern 5pt}
\nu^{\pm (n)}_{R}  &= \pm e^{i\delta_C} \sigma^2 ( \nu^{\pm (n)}_L )^* ~.
\label{wave-neutrino1}
\end{align}
Wave functions are normalized as  
\begin{align}
&\int_1^{z_L} dz   \, \big\{ | f_L |^2 + |g_L |^2 + | f_R |^2 + |g_R |^2 \big\}  + |h |^2 = 1 
\label{norm-neutrino1}
\end{align}
in each mode.   The explicit forms of the wave functions are given in (\ref{wave-neutrino2}).

\section{$W$ and $Z$ couplings} 

The $\gamma$, $W$ and $Z$ couplings of the fermion fields are contained in the part of the action 
\begin{align}
&\int d^4 x \int_1^{z_L} \frac{dz}{k} \sum_J \bar{\check \Psi}^J \gamma^\mu (-i)
\Big(  g_A A_\mu^{SO(5)} + g_B Q_X A_\mu^{U(1)_X} \Big)\check  \Psi^J 
\label{action1}
\end{align}
where $\check \Psi^J = z^{-2} \Psi^J  $.  
By inserting the KK expansions of the gauge and fermion fields into (\ref{action1}),
$\gamma^{(n)}$, $W^{(n)}$, and $Z^{(n)}$ couplings among the fermion KK modes are evaluated.
The photon $\gamma = \gamma^{(0)}$ couplings are universal.  
They are diagonal in the KK space,  and are  given by 
$e \, Q_\EM = g_w \sin \theta_W^0 (T^3_L + T^3_R + Q_X)$.

The $W = W^{(0)}$ couplings in the first generation of the quark multiplets are given by
\begin{align}
&\frac{g_w}{\sqrt{2}} W_\mu \bigg[ \sum_{n,m=0}^\infty \Big\{
\hat g^{Wud}_{L, nm} \bar u_L^{(n)} \gamma^\mu d_L^{(m)} + \hat g^{Wud}_{R, nm} \bar u_R^{(n)} \gamma^\mu d_R^{(m)} \Big\}  \cr
\noalign{\kern 5pt}
&\hskip 0.6cm 
+  \sum_{n=0}^\infty   \sum_{m=1}^\infty \Big\{ \hat g^{W uD}_{L, nm} 
\bar u_L^{(n)} \gamma^\mu D_L^{(m)} + \hat g^{WuD}_{R, nm} \bar u_R^{(n)} \gamma^\mu D_R^{(m)} \Big\} \bigg] 
\label{Wud1}
\end{align}
where
\begin{align}
\hat g^{Wud}_{L/R, nm} &= \sqrt{kL} \int_1^{z_L} dz \, \bigg\{ h^L_{W^{(0)}}  f^{u^{(n)}*}_{L/R}  f^{d^{(m)}}_{L/R}
+ h^R_{W^{(0)}}  g^{u^{(n)}*}_{L/R}  g^{d^{(m)}}_{L/R} \cr
\noalign{\kern 5pt}
&\hskip 2.cm
+ \frac{i}{\sqrt{2}} \,  \hat h_{W^{(0)}} \Big(  f^{u^{(n)}*}_{L/R}  g^{d^{(m)}}_{L/R} -  g^{u^{(n)}*}_{L/R} f^{d^{(m)}}_{L/R} \Big) \bigg\} .
\label{Wud2}
\end{align}
$\hat g^{WuD}_{L/R, nm}$ is obtained by replacing $d^{(m)}$ by $D^{(m)}$ in (\ref{Wud2}).
The values in the SM correspond to $\hat g^{Wud}_{L, 00} = 1$ and $\hat g^{Wud}_{R, 00} = 0$.
In the RS space off-diagonal components of $\hat g^{Wud}_{L/R, nm}$ are non-vanishing.
For $\theta_H = 0.1$ and $m_\KK = 13\,$TeV, for instance, the coupling matrices are given by
\begin{align}
\hat g^{Wud}_L &= \begin{pmatrix}
0.997645 & -0.024904 & 0.000020 & -0.002827 & 10^{-6}  & \cdots \cr
-0.024904 & 0.002498 & 0.028389 & 10^{-7} & 0.000510   &  \cr
0.000020 & 0.028389 & 0.997618 & -0.024548 & 0.000022 &  \cr
-0.002827 & 10^{-7} & -0.024548 & 0.002498 & 0.027021 & \cr
10^{-6} & 0.000510 & 0.000022 & 0.027021 & 0.997620   & \cr
 \vdots &  & & &   & \ddots
\end{pmatrix} ,  \cr
\noalign{\kern 5pt}
\hat g^{Wud}_R &= \begin{pmatrix}
10^{-12}  & 10^{-7}  & 10^{-7} & 10^{-9} & 10^{-7} &\cdots  \cr
10^{-8} & 0.002498 & 0.024145 & 10^{-8} & 10^{-6} & \cr
10^{-7} & 0.024145 & 0.997632 & -0.022564 & 0.000018 & \cr
10^{-10} & 10^{-8} & -0.022564 & 0.002498 & 0.025826 & \cr
10^{-8} & 10^{-6} & 0.000018 & 0.025826 & 0.997625 & \cr
\vdots  &  & & &   & \ddots
\end{pmatrix}
\label{Wud3}
\end{align}
where $10^{-7}$, for instance, implies $O(10^{-7})$.
Notice that the couplings for KK excited states are nearly vector-like.
However, they have very small axial-vector components; 
$\hat g^{Wud}_A = \onehalf (\hat g^{Wud}_R - \hat g^{Wud}_L)$ is $O(10^{-3})$ or less.
As is shown below, those small numbers must be properly taken into account to establish the coupling sum rules.
The $\hat g^{WuD}_{L/R, nm}$ couplings are very small; ${\rm max} \,  | \hat g^{WuD}_{L/R, nm} | \sim 2 \times 10^{-6}$.

In the lepton sector there are two types of the neutrino towers, $\{ \nu^{+(n)} \}$ ($\nu_{e1}$ series) 
and $\{ \nu^{-(n)} \}$ ($\nu_{e2}$ series), both of which have $W$ couplings.  In the first generation 
\begin{align}
&\frac{g_w}{\sqrt{2}} W_\mu \bigg[ \sum_{n,m=0}^\infty \Big\{
\hat g^{W\nu_{e1} e}_{L, nm} \,  \bar \nu_{eL}^{+(n)} \gamma^\mu e_L^{(m)} 
+ \hat g^{W\nu_{e 1} e}_{R, nm} \,   \bar \nu_{eR}^{+(n)} \gamma^\mu e_R^{(m)} \Big\}  \cr
\noalign{\kern 5pt}
&\hskip 0.6cm 
+  \sum_{n=1}^\infty   \sum_{m=0}^\infty \Big\{ 
\hat g^{W\nu_{e2} e}_{L, nm} \,   \bar \nu_{eL}^{-(n)} \gamma^\mu e_L^{(m)} 
+ \hat g^{W\nu_{e 2} e}_{R, nm} \,   \bar \nu_{eR}^{-(n)} \gamma^\mu e_R^{(m)} \Big\} \bigg] 
\label{Wnue1}
\end{align}
where
\begin{align}
\hat g^{W\nu_{e1} e}_{L, nm} 
&= \sqrt{kL} \int_1^{z_L} dz \, \bigg\{ h^L_{W^{(0)}}  f^{\nu_e^{+(n)}*}_{L/R}  f^{e^{(m)}}_{L/R}
+ h^R_{W^{(0)}}  g^{\nu_e^{+(n)}*}_{L/R}  g^{e^{(m)}}_{L/R} \cr
\noalign{\kern 5pt}
&\hskip 2.cm
+ \frac{i}{\sqrt{2}} \,  \hat h_{W^{(0)}} \Big(  f^{\nu_e^{+(n)}*}_{L/R}  g^{e^{(m)}}_{L/R} 
-  g^{\nu_e^{+(n)}*}_{L/R} f^{e^{(m)}}_{L/R} \Big) \bigg\} ~, \cr
\noalign{\kern 5pt}
\hat g^{W\nu_{e2} e}_{L, nm} 
&= \sqrt{kL} \int_1^{z_L} dz \, \bigg\{ h^L_{W^{(0)}}  f^{\nu_e^{-(n)}*}_{L/R}  f^{e^{(m)}}_{L/R}
+ h^R_{W^{(0)}}  g^{\nu_e^{-(n)}*}_{L/R}  g^{e^{(m)}}_{L/R} \cr
\noalign{\kern 5pt}
&\hskip 2.cm
+ \frac{i}{\sqrt{2}} \,  \hat h_{W^{(0)}} \Big(  f^{\nu_e^{-(n)}*}_{L/R}  g^{e^{(m)}}_{L/R} 
-  g^{\nu_e^{-(n)}*}_{L/R} f^{e^{(m)}}_{L/R} \Big) \bigg\} ~.
\label{Wnue2}
\end{align}
The values in the SM correspond to $\hat g^{W\nu_{e1} e}_{L, 00} = 1$ and $\hat g^{W\nu_{e1} e}_{R, 00} = 0$.
For $\theta_H = 0.1$, $m_\KK = 13\,$TeV and $M_e = 10^3\,$TeV, for instance, the coupling matrices are given by
\begin{align}
\hat g^{W\nu_{e1} e}_L &= \begin{pmatrix}
0.997647 & -0.023607 & 0.000018 & -0.002964 & 10^{-7}  & \cdots \cr
-0.016689& 0.001766 & 0.020087 & 10^{-8} & 0.000319   &  \cr
-0.000013 & - 0.020092 & -0.705422 & 0.017130 & 0.000015 &  \cr
0.002094 & 10^{-8} & 0.017115 & -0.001765 & -0.019154 & \cr
10^{-7} & -0.000318 & -0.000015 & -0.019170 & -0.705382   & \cr
 \vdots &  & & &   & \ddots
\end{pmatrix} ,   \cr
\noalign{\kern 5pt}
\hat g^{W\nu_{e1} e}_R &= \begin{pmatrix}
10^{-23}  & 10^{-17}  & 10^{-17} & 10^{-19} & 10^{-17} &\cdots  \cr
10^{-9} & 0001766 & 0.016811 & 10^{-8} & -0.000095   &  \cr
10^{-8} & -0.016814 & -0.705432 & 0.015656 & -0.000012 &  \cr
10^{-10} & 10^{-8} & 0.015642 & -0.001765 &-0.018204& \cr
10^{-9} & 0.000095 & -0.000012 & -0.018219 & -0.705385   & \cr
\vdots  &  & & &   & \ddots
\end{pmatrix}
\label{Wnue3}
\end{align}
and
\begin{align}
\hat g^{W\nu_{e2} e}_L &= \begin{pmatrix}
0 &0 & 0 & 0 & 0  & \cdots \cr
0.016688& -0.001766 & -0.020086 & 10^{-8} & -0.000319   &  \cr
0.000013 &  0.020092 & 0.705422 & -0.017130 & 0.000015 &  \cr
-0.002094 & 10^{-8} & -0.017107 & 0.001764 & 0.019145 & \cr
10^{-7} & 0.000319 & 0.000015 & 0.019171 & 0.705414   & \cr
 \vdots &  & & &   & \ddots
\end{pmatrix} ,   \cr
\noalign{\kern 5pt}
\hat g^{W\nu_{e2} e}_R &= \begin{pmatrix}
0  & 0 & 0 & 0 & 0 &\cdots  \cr
10^{-9} & -0001766 & -0.016810 & 10^{-8} & 0.000095   &  \cr
10^{-8} & 0.016814 & 0.705432 & -0.015656 & 0.000012 &  \cr
10^{-10} & 10^{-8} & -0.015635 & 0.001764 & 0.018196& \cr
10^{-9} & -0.000095 &  0.000012 &  0.018220 & 0.705418   & \cr
\vdots  &  & & &   & \ddots
\end{pmatrix} .
\label{Wnue4}
\end{align}
Here we have set $\hat g^{W\nu_{e2} e}_{L /R, 0n} = 0$.
It is seen that in the part of  the KK excited states, 
$\hat g^{W\nu_{e1} e}_{L, nm} \sim \hat g^{W\nu_{e1} e}_{R, nm}$
and $\hat g^{W\nu_{e2} e}_{L, nm} \sim \hat g^{W\nu_{e2} e}_{R, nm}$ for $n, m \ge 1$, 
and  $\hat g^{W\nu_{e1} e}_{L/R, nm} + \hat g^{W\nu_{e2} e}_{L/R, nm} \sim 0$ for $n \ge 1, m \ge 0$.
Those couplings are almost vector-like.
As is in the $(u,d)$ case, they have very small axial-vector components; 
$\hat g^{W\nu_{e1} e}_A $ and $\hat g^{W\nu_{e2} e}_A $ are $O(10^{-3})$ or less.

The $W$-couplings of the zero modes, namely the couplings of quarks and leptons in three generations, are
summarized in Table \ref{Table:Wcoupling1}.
Except for $(t,b)$ the $W$ couplings are universal to high accuracy; $\hat g^{W}_{L} \sim 0.997645 \equiv \hat g^{W, \rm GHU}$ 
and $\hat g^{W}_{R} \sim 0$.
The observed lepton coupling should be identified as $g_w^{\rm obs} = g_w \, \hat g^{W, \rm GHU}$.

\begin{table}[tbh]
{%\small
\renewcommand{\arraystretch}{1.2}
\begin{center}
\caption{The $W$-couplings of  quarks and leptons in units of $g_w$
for  $\theta_H = 0.1$, $m_\KK = 13\,$TeV and $M = 10^3\,$TeV.
}
\vskip 10pt
\begin{tabular}{|c|c|c|}
%\hline
%&GUT inspired model\\
%& type B &type A\\
\hline
&$\hat g^{W}_{L}$ &$\hat g^{W}_{R}$ \\
\hline
$(\nu_{e}, e)$ &$0.997647$ &$ 3 \times 10^{-23}$\\
\hline
$(\nu_{\mu}, \mu)$ &$0.997644$ &$3 \times 10^{-21}$\\
\hline 
$(\nu_{\tau}, \tau)$ &$0.997642$ &$ 4 \times 10^{-20}$\\
\hline
$(u, d)$ &$0.997645$ &$2 \times 10^{-12}$\\
\hline
$(c,s)$ &$0.997643$ &$ 8 \times 10^{-10}$\\
\hline 
$(t,b)$ &$0.997969 $ &$ 0.000011$\\
\hline 
\end{tabular}
\label{Table:Wcoupling1}
\end{center}
}
\end{table}

The $Z = Z^{(0)}$ couplings are evaluated similarly.  
The couplings in the up-type quark sector are given in the form 
\begin{align}
&\frac{g_w}{\cos \theta_W^0} Z_\mu \sum_{n,m=0}^\infty \Big\{
\hat g^{Zu}_{L, nm} \bar u_L^{(n)} \gamma^\mu u_L^{(m)} + \hat g^{Zu}_{R, nm} \bar u_R^{(n)} \gamma^\mu u_R^{(m)} \Big\} ~.
\label{Zu1}
\end{align}
As is suggested from the structure of the wave functions in (\ref{waveWZ1}), it is convenient to decompose 
the $Z$ couplings into the $U(1)_\EM$ part and the rest.  % With $h^Q_{Z^{(n)}} (z)$ in  (\ref{waveWZ1})
% \begin{align}
%h^Q_{Z^{(n)}} (z) = \sqrt{\frac{2} {r_{Z^{(n)}}}} \, C(z, \lambda_{Z^{(n)}}) 
% \label{waveZQ}
% \end{align}
We write
\begin{align}
\hat g^{Zu}_{L/R, nm} = \hat g^{Zu, su2}_{L/R, nm} - \sin^2 \theta_W^0 \,  \hat g^{Zu, \EM}_{L/R, nm} ~.  
\label{Zu2}
\end{align}
Then with $h^{L, su2}_{Z^{(0)}}$,  $h^{R, su2}_{Z^{(0)}}$,  $\hat h^{su2}_{Z^{(0)}}$ and $h^Q_{Z^{(0)}} $ 
in (\ref{waveWZ1}) the couplings are given by 
\begin{align}
\hat g^{Zu, su2}_{L/R, nm} 
&=T^3_u \cos \theta_W^0  \sqrt{kL} \int_1^{z_L} dz \, \bigg\{ h^{L, su2}_{Z^{(0)}}  f^{u^{(n)}*}_{L/R}  f^{u^{(m)}}_{L/R}
+ h^{R, su2}_{Z^{(0)}}  g^{u^{(n)}*}_{L/R}  g^{u^{(m)}}_{L/R} \cr
\noalign{\kern 5pt}
&\hskip 3.5cm
+ \frac{i}{\sqrt{2}} \,  \hat h^{su2}_{Z^{(0)}}  \Big(  f^{u^{(n)}*}_{L/R}  g^{u^{(m)}}_{L/R} -  g^{u^{(n)}*}_{L/R} f^{u^{(m)}}_{L/R} \Big) \bigg\} , \cr
\noalign{\kern 5pt}
\hat g^{Zu, \EM}_{L/R, nm}
&= Q_u  \cos \theta_W^0  \sqrt{ kL} \int_1^{z_L} dz \,  h^Q_{Z^{(0)}}  \Big\{  f^{u^{(n)}*}_{L/R}  f^{u^{(m)}}_{L/R} + 
g^{u^{(n)}*}_{L/R}  g^{u^{(m)}}_{L/R} \Big\} 
\label{Zu3}
\end{align}
where $T^3_u = \onehalf$ and $Q_u = \frac{2}{3}$. 
% With the normalization $r_{Z^{(0)}}$  in (\ref{waveWZ1}), 
One finds that 
$\hat g^{Zu, su2}_{L, 00} = 0.498844$,  $\hat g^{Zu, su2}_{R, 00} = 5 \times 10^{-12}$, 
$\hat g^{Zu, \EM}_{L, 00} = 0.666791$ and  $\hat g^{Zu, \EM}_{R, 00} = 0.666725$  
for $\theta_H = 0.1$, $m_\KK = 13\,$TeV and $\sin^2 \theta_W^0 = 0.230634$.
The $Z$ coupling matrices are given by
\begin{align}
\hat g^{Zu}_L &= \begin{pmatrix}
0.345059 & -0.012453 & 0.000009 & -0.001414 & 10^{-7}  & \cdots \cr
-0.012453  & -0.152531 & 0.014195 & -0.000004 & 0.000255   &  \cr
0.000009 & 0.014195 & 0.345047 & -0.012275 & 0.000010 &  \cr
-0.001414 &  -0.000004 & -0.012275 & -0.152531 & 0.013511 & \cr
10^{-7} &0.000255 & 0.000010 & 0.013511 & 0.345048  & \cr
 \vdots &  & & &   & \ddots
\end{pmatrix} ,  \cr
\noalign{\kern 5pt}
\hat g^{Zu}_R &= \begin{pmatrix}
-0.153769 & 0.000012  & 10^{-7} & -0.000011 & 10^{-7} &\cdots  \cr
0.000012 & -0.152536 & 0.012073 & -0.000003 & 0.000001 & \cr
10^{-7} & 0.012073 & 0.345054 & -0.011282 & 0.000008 & \cr
-0.000011 &  -0.000003 & -0.011282  & -0.152532 & 0.012913 & \cr
10^{-7} &  0.000001  & 0.000008 & 0.012913 & 0.345050 & \cr
\vdots  &  & & &   & \ddots
\end{pmatrix} .
\label{Zu4}
\end{align}
The couplings in the space of KK excited states, namely $n,m\ge 1$ elements of $\hat g^{Zu}_{nm}$,   are nearly vector-like.
The axial-vector components are small; 
$\hat g^{Zu}_A = \onehalf (\hat g^{Zu}_R - \hat g^{Zu}_L)$ is $O(10^{-3})$ or less.

In the down-type quark sector there are $\{ d^{(n)} \}$ and $\{ D^{(n)} \}$ series.
The couplings are written as
\begin{align}
&\frac{g_w}{\cos \theta_W^0} Z_\mu \bigg[ \sum_{n,m=0}^\infty \Big\{
\hat g^{Zdd}_{L, nm} \bar d_L^{(n)} \gamma^\mu d_L^{(m)} + \hat g^{Zdd}_{R, nm} \bar d_R^{(n)} \gamma^\mu d_R^{(m)} \Big\}  \cr
\noalign{\kern 5pt}
&\hskip 1.cm
+  \sum_{n=0}^\infty    \sum_{m=1}^\infty\Big\{
\hat g^{ZdD}_{L, nm} \bar d_L^{(n)} \gamma^\mu D_L^{(m)} + \hat g^{ZdD}_{R, nm} \bar d_R^{(n)} \gamma^\mu D_R^{(m)} \Big\} \cr
\noalign{\kern 5pt}
&\hskip 1.cm
+  \sum_{n=1}^\infty    \sum_{m=0}^\infty\Big\{
\hat g^{ZDd}_{L, nm} \bar D_L^{(n)} \gamma^\mu d_L^{(m)} + \hat g^{ZDd}_{R, nm} \bar D_R^{(n)} \gamma^\mu d_R^{(m)} \Big\}  \cr
\noalign{\kern 5pt}
&\hskip 1.cm
+  \sum_{n,m=1}^\infty \Big\{
\hat g^{ZDD}_{L, nm} \bar D_L^{(n)} \gamma^\mu D_L^{(m)} + \hat g^{ZDD}_{R, nm} \bar D_R^{(n)} \gamma^\mu D_R^{(m)} \Big\} 
\bigg] .
\label{Zd1}
\end{align}
With the decomposition 
$\hat g^{Zdd}_{L/R, nm} = \hat g^{Zdd, su2}_{L/R, nm} - \sin^2 \theta_W^0 \,  \hat g^{Zdd, \EM}_{L/R, nm}$ etc,
the couplings are given by
\begin{align}
\hat g^{Zdd, su2}_{L/R, nm} 
&=T^3_d \cos \theta_W^0  \sqrt{kL} \int_1^{z_L} dz \, \bigg\{  h^{L, su2}_{Z^{(0)}}  f^{d^{(n)}*}_{L/R}  f^{d^{(m)}}_{L/R}
+ h^{R,su2}_{Z^{(0)}}  g^{d^{(n)}*}_{L/R}  g^{d^{(m)}}_{L/R} \cr
\noalign{\kern 5pt}
&\hskip 3.5cm
+ \frac{i}{\sqrt{2}} \,  \hat h^{su2}_{Z^{(0)}}  \Big(  f^{d^{(n)}*}_{L/R}  g^{d^{(m)}}_{L/R} -  g^{d^{(n)}*}_{L/R} f^{d^{(m)}}_{L/R} \Big) \bigg\} , \cr
\noalign{\kern 5pt}
\hat g^{Zdd, \EM}_{L/R, nm}
&= Q_d  \cos \theta_W^0  \sqrt{ kL} \int_1^{z_L} dz \,  h^Q_{Z^{(0)}}  
\bigg(  f^{d^{(n)}*}_{L/R}  f^{d^{(m)}}_{L/R} + g^{d^{(n)}*}_{L/R}  g^{d^{(m)}}_{L/R} \cr
\noalign{\kern 5pt}
&\hskip 5.5cm
+  h^{d^{(n)}*}_{L/R}  h^{d^{(m)}}_{L/R} + k^{d^{(n)}*}_{L/R}  k^{d^{(m)}}_{L/R} \bigg) .
\label{Zd2}
\end{align}
Here $T^3_d = - \onehalf$ and $Q_d = - \frac{1}{3}$. 
The expressions for $\hat  g^{ZdD},  g^{ZDd},  g^{ZDD}$ components are obtained by the replacement
$d^{(n)} \go D^{(n)}$ and/or $d^{(m)} \go D^{(m)}$ in (\ref{Zd2}).  
Note that $\hat g^{Zdd, su2}_{L, 00} = -0.498844$,  $\hat g^{Zdd, su2}_{R, 00} = -1 \times 10^{-13}$, 
$\hat g^{Zdd, \EM}_{L, 00} = -0.333395$ and  $\hat g^{Zdd, \EM}_{R, 00} = -0.333372$  
for $\theta_H = 0.1$, $m_\KK = 13\,$TeV and $\sin^2 \theta_W^0 = 0.230634$.

The $Z$ coupling matrices are given by
\begin{align}
\hat g^{Zdd}_L &= \begin{pmatrix}
-0.421952 & 0.012453 & - 0.000011 & 0.001414 & 10^{-7}  & \cdots \cr
0.012453  & 0.075640 &  -0.014195 & 0.000002 & -0.000255   &  \cr
-0.000011 & -0.014195 & -0.421937 & 0.012275 & -0.000012 &  \cr
0.001414 &  0.000002 & 0.012275 & 0.0756407 & -0.013511 & \cr
10^{-7} &- 0.000255 & -0.000012 & -0.013511 & -0.421938 & \cr
 \vdots &  & & &   & \ddots
\end{pmatrix} ,  \cr
\noalign{\kern 5pt}
\hat g^{Zdd}_R &= \begin{pmatrix}
0.076887 & 10^{-8}   & 10^{-8} & 10^{-10}  & 10^{-8} &\cdots  \cr
10^{-8}  &0.075643 & -0.012073 & 0.000002 & -0.000001 & \cr
10^{-8} & -0.012073 & -0.421945 & 0.011282 & -0.000010 & \cr
10^{-10}  &  0.000002 & 0.011282  & 0.075641 & -0.012913 & \cr
10^{-8} &  -0.000001  & -0.000010 & -0.012913 & -0.421940 & \cr
\vdots  &  & & &   & \ddots
\end{pmatrix} .
\label{Zd3}
\end{align}
As in the case of $\hat g^{Zu}$, the couplings  in the space of the KK excited states are almost vector-like;
$\hat g^{Zdd}_A = \onehalf (\hat g^{Zdd}_R - \hat g^{Zdd}_L)$ is $O(10^{-3})$ or less.
The ${su2}$ components of $\{ D^{(n)} \}$ series are very small; $| \hat g^{ZDD, su2}_{L/R, nm} |$  is $O(10^{-12})$ or less.
Off-diagonal elements of $ \hat g^{ZDD, \EM}_{L/R} $ are $O(10^{-6})$ or less.  All diagonal elements are about
$ -0.33339$ and $| \hat g^{ZDD, \EM}_{L, nn} - \hat g^{ZDD, \EM}_{R, nn} | = O(10^{-6})$.
Also $\hat g^{ZdD}_{L/R, nm} = (\hat g^{ZDd}_{L/R, mn})^* =O(10^{-6})$ or less.

The $Z$ couplings of charged lepton multiplets have the same structure as in the up-type quark sector.
The couplings of  the electron multiplet are
\begin{align}
&\frac{g_w}{\cos \theta_W^0} Z_\mu \sum_{n,m=0}^\infty \Big\{
\hat g^{Ze}_{L, nm} \bar e_L^{(n)} \gamma^\mu e_L^{(m)} + \hat g^{Ze}_{R, nm} \bar e_R^{(n)} \gamma^\mu e_R^{(m)} \Big\} ~,
\label{Ze1}
\end{align}
where $\hat g^{Ze}_{L/R, nm}$ are given by the expressions obtained by replacing, in (\ref{Zu2}) and (\ref{Zu3}),
$u^{(n)}$ by $e^{(n)}$, $T^3_u$ by $T^3_e = - \onehalf$, and $Q_u$ by $Q_e=-1$.
$\hat g^{Ze, su2}_{L, 00} = -0.498845$,  $\hat g^{Ze, su2}_{R, 00} = -4 \times 10^{-15}$, 
$\hat g^{Ze, \EM}_{L, 00} = -1.00019$ and  $\hat g^{Ze, \EM}_{R, 00} = -1.00009$  
for $\theta_H = 0.1$, $m_\KK = 13\,$TeV and $\sin^2 \theta_W^0 = 0.230634$.
The $Z$ coupling matrices are given by
\begin{align}
\hat g^{Ze}_L &= \begin{pmatrix}
-0.268168 & 0.011804 & -0.000006 & 0.001482 & 10^{-7}  & \cdots \cr
0.011804  & 0.229421 & -0.014207 & 0.000006 & -0.000256   &  \cr
-0.000006 & -0.014207 &-0.268158 & 0.012113 & -0.000008 &  \cr
0.001482  &  0.000006 & 0.012113 & 0.229421 & -0.013556& \cr
10^{-7} &-0.000256 &  -0.000008 & -0.013556 & -0.268158  & \cr
 \vdots &  & & &   & \ddots
\end{pmatrix} ,  \cr
\noalign{\kern 5pt}
\hat g^{Ze}_R &= \begin{pmatrix}
0.230654 & 10^{-7}  & 10^{-8} &10^{-7} & 10^{-8} &\cdots  \cr
10^{-7}  & 0.229429 & -0.011890 & 0.000004 & 0.000067 & \cr
10^{-8} & -0.011890 & -0.268163 & 0.011071 & -0.000006 & \cr
10^{-7}  &  0.000004 & 0.011071  & 0.229423 & -0.012884& \cr
10^{-8} &  0.000067  &  -0.000006  &  -0.012884 & -0.268160 & \cr
\vdots  &  & & &   & \ddots
\end{pmatrix} .
\label{Ze2}
\end{align}
%The couplings in the space of KK excited states  are nearly vector-like.
The axial-vector components in the space of KK excited states are small; 
$\hat g^{Ze}_{A, nm}$ ($n,m \ge 1$) are $O(10^{-3})$ or less.

In the neutrino sector the couplings are given by
\begin{align}
&\frac{g_w}{\cos \theta_W^0} Z_\mu \bigg[ \sum_{n,m=0}^\infty \Big\{
\hat g^{Z\nu_{e 11}}_{L, nm} \bar \nu_{eL}^{+(n)} \gamma^\mu  \nu_{eL}^{+(m)} 
+ \hat g^{Z\nu_{e 11}}_{R, nm} \bar \nu_{eR}^{+(n)} \gamma^\mu  \nu_{eR}^{+(m)} \Big\}  \cr
\noalign{\kern 5pt}
&\hskip 1.cm
+  \sum_{n=0}^\infty    \sum_{m=1}^\infty\Big\{
\hat g^{Z\nu_{e 12}}_{L, nm} \bar \nu_{eL}^{+(n)} \gamma^\mu  \nu_{eL}^{-(m)}  
+ \hat g^{Z\nu_{e 12}}_{R, nm} \bar \nu_{eR}^{+(n)} \gamma^\mu  \nu_{eR}^{-(m)}  \Big\} \cr
\noalign{\kern 5pt}
&\hskip 1.cm
+  \sum_{n=1}^\infty    \sum_{m=0}^\infty\Big\{
\hat g^{Z\nu_{e 21}}_{L, nm} \bar \nu_{eL}^{-(n)}  \gamma^\mu  \nu_{eL}^{+(m)} 
+ \hat g^{Z\nu_{e 21}}_{R, nm} \bar \nu_{eR}^{-(n)} \gamma^\mu  \nu_{eR}^{+(m)} \Big\}  \cr
\noalign{\kern 5pt}
&\hskip 1.cm
+  \sum_{n,m=1}^\infty \Big\{
\hat g^{Z\nu_{e 22}}_{L, nm} \bar \nu_{eL}^{-(n)} \gamma^\mu  \nu_{eL}^{-(m)}  
+ \hat g^{Z\nu_{e 22}}_{R, nm} \bar \nu_{eR}^{-(n)} \gamma^\mu  \nu_{eR}^{-(m)}  \Big\} 
\bigg] 
\label{Znu1}
\end{align}
where
\begin{align}
\hat g^{Z\nu_{e ab}}_{L/R, nm} 
&=T^3_{\nu_e} \cos \theta_W^0  \sqrt{kL} \int_1^{z_L} dz \, 
\bigg\{  h^{L,su2}_{Z^{(0)}}  f^{\nu_{e}^{a(n)}*}_{L/R}  f^{\nu_e^{b(m)}}_{L/R}
+  h^{R,su2}_{Z^{(0)}} g^{\nu_e^{a(n)}*}_{L/R}  g^{\nu_e^{b(m)}}_{L/R} \cr
\noalign{\kern 5pt}
&\hskip 3.5cm
+ \frac{i}{\sqrt{2}} \,  \hat h^{su2}_{Z^{(0)}} \Big(  f^{\nu_e^{a(n)}*}_{L/R}  g^{\nu_e^{b(m)}}_{L/R} -  g^{\nu_e^{a(n)}*}_{L/R} f^{\nu_e^{b(m)}}_{L/R} \Big) \bigg\}  .
\label{Znu2}
\end{align}
Here $T^3_{\nu_e} = \onehalf$ and we have denoted as $(\nu_{e}^{1(n)}, \nu_{e}^{2(n)} ) = (\nu_{e}^{+(n)}, \nu_{e}^{-(n)} )$.
Note $\hat g^{Z\nu_{e 21}}_{L/R} = (\hat g^{Z\nu_{e 12}}_{L/R} )^\dagger$.
We set 
$\hat g^{Z\nu_{e 12}}_{L/R, n 0}, \hat g^{Z\nu_{e 21}}_{L/R, 0 n},  \hat g^{Z\nu_{e 22}}_{L/R, 0n}, \hat g^{Z\nu_{e 22}}_{L/R, n0} =0$.
The $Z$ coupling matrices are given by
\begin{align}
\hat g^{Z\nu_{e11}}_L &= \begin{pmatrix}
0.498845 & -0.008345 & -0.000008 & 0.001047 & 10^{-7}  & \cdots \cr
-0.008345 & 0.000624 &-0.007102 &10^{-8} & -0.000113 &  \cr
-0.000008  &-0.007102 & 0.249412 & -0.006051 & 0.000007 &  \cr
0.001047  & 10^{-8} & -0.006051 & 0.000624 & 0.006772& \cr
10^{-7} &  -0.000113 & 0.000007 &0.006772 & 0.249384 & \cr
\vdots & & & & & \ddots
\end{pmatrix} ,  \cr
\noalign{\kern 5pt}
\hat g^{Z\nu_{e11}}_R &= \begin{pmatrix}
10^{-32} & 10^{-18}  & 10^{-17} &10^{-19} & 10^{-18} &\cdots  \cr
10^{-18}  & 0.000624 &-0.005944 &10^{-8} & 0.000034 & \cr
10^{-17} & -0.005944 & 0.249417 & -0.005531 & 0.000006 & \cr
10^{-19}  &  10^{-8} &-0.005531 & 0.000624 & 0.006436 & \cr
10^{-18} &  0.000034  & 0.000006  &  0.006436 & 0.249386 & \cr
\vdots  &  & & &   & \ddots
\end{pmatrix} , \cr
\noalign{\kern 5pt}
\hat g^{Z\nu_{e22}}_L &= \begin{pmatrix}
0 & 0 & 0 & 0 & 0  & \cdots \cr
0 & 0.000624 &-0.007102 &10^{-8} & -0.000113 &  \cr
0 &-0.007102 & 0.249412 & -0.006048 & 0.000007 &  \cr
0  & 10^{-8} & -0.006048 & 0.000623 & 0.006769& \cr
0 &  -0.000113 & 0.000007 &0.006769 & 0.249407 & \cr
\vdots & & & & & \ddots
\end{pmatrix} ,  \cr
\noalign{\kern 5pt}
\hat g^{Z\nu_{e22}}_R &= \begin{pmatrix}
0 & 0  & 0 &0 & 0 &\cdots  \cr
0 & 0.000624 &-0.005943 &10^{-8} & 0.000034 & \cr
0 & -0.005943 & 0.249417 & -0.005528 & 0.000006 & \cr
0 &  10^{-8} &-0.005528 & 0.000623 & 0.006433 & \cr
0 &  0.000034  & 0.000006  &  0.006433 & 0.249409 & \cr
\vdots  &  & & &   & \ddots
\end{pmatrix} , \cr
\noalign{\kern 5pt}
%\label{Znu3}
%\end{align}
%and
%\begin{align}
\hat g^{Z\nu_{e12}}_L &= \begin{pmatrix}
0 & 0.008344 & 0.000008 & -0.001047 & 10^{-7}  & \cdots \cr
0 & -0.000624 &0.007102 &10^{-8} & 0.000113 &  \cr
0  &0.007102 & -0.249412 & 0.006048 & -0.000007 &  \cr
0  & 10^{-8} & 0.006051 & -0.000623 & -0.006772& \cr
0 &  0.000113 & -0.000007 &-0.006769 & -0.249395 & \cr
\vdots & & & & & \ddots
\end{pmatrix} ,  \cr
\noalign{\kern 5pt}
\hat g^{Z\nu_{e12}}_R &= \begin{pmatrix}
0 & 10^{-18}  & 10^{-18} &10^{-19} & 10^{-18} &\cdots  \cr
0  & -0.000624 & 0.005944 &10^{-8} & -0.000034 & \cr
0 & 0.005943 & -0.249417 & -0.005528 & -0.000006 & \cr
0  &  10^{-8} & 0.005531 & -0.000623 & -0.006436 & \cr
0 &  -0.000034  & -0.000006  &  -0.006433 & -0.249397 & \cr
\vdots  &  & & &   & \ddots
\end{pmatrix} .
\label{Znu3}
\end{align}
Notice that in the space of KK excited states
$\hat g^{Z\nu_{e11}}_{L, nm} \sim \hat g^{Z\nu_{e22}}_{L, nm} \sim - \hat g^{Z\nu_{e12}}_{L, nm} \sim - \hat g^{Z\nu_{e21}}_{L, nm}$, 
$\hat g^{Z\nu_{e11}}_{R, nm} \sim \hat g^{Z\nu_{e22}}_{R, nm} \sim - \hat g^{Z\nu_{e12}}_{R, nm} \sim - \hat g^{Z\nu_{e21}}_{R, nm}$,
and $| \hat g^{Z\nu_{eab}}_{R, nm} - \hat g^{Z\nu_{eab}}_{L, nm} |$ are
$O(10^{-3})$ or less for $n,m \ge 1$. 

The $Z$-couplings of the zero modes, namely those of leptons and quarks in three generations, are summarized in 
Table \ref{Table:Zcoupling1}.
The deviations from the SM values are tiny.

\begin{table}[tbh]
{%\small
\renewcommand{\arraystretch}{1.2}
\begin{center}
\caption{The $Z$-couplings of  leptons and quarks in units of $g_w/\cos \theta_W^0$
for  $\theta_H = 0.1$, $m_\KK = 13\,$TeV and $M = 10^3\,$TeV.
For reference the SM values $T^3_L - \sin^2 \theta_W^\SM Q_\EM$ with 
$\sin^2 \theta_W^\SM = 0.2312$ are listed as well.
}
\vskip 10pt
\begin{tabular}{|c|c|c||c|c|c|}
%\hline
%&GUT inspired model\\
%& type B &type A\\
\hline
&$\hat g^{Z}_{L}$  &$\hat g^{Z}_{R}$ &&$\hat g^{Z}_{L}$  &$\hat g^{Z}_{R}$ \\
\hline \hline
$\nu_{e}$ &$0.498845$ &$2 \times 10^{-32}$ &$ e$ &$-0.268168$ &$0.230654$\\
\hline
$\nu_{\mu}$ &$0.498843$  &$1 \times 10^{-32}$ &$\mu$ &$-0.268167$ &$0.230654$\\
\hline 
$\nu_{\tau}$ &$0.498842$  &$1 \times 10^{-32}$ &$\tau$&$-0.268166$&$0.229961$\\
\hline
\SM &0.5  & 0 &\SM &$-0.2688$& $0.2312$\\
\hline \hline
%&$\hat g^{Z u}_{L}$ &$\hat g^{Z u}_{R}$ &&$\hat g^{Z d}_{L}$  &$\hat g^{Z d}_{R}$  \\
%\hline
$u$ &$0.345059$  &$-0.153769$ &$d$ &$-0.421952$ &$0.076887$\\
\hline
$c$& $0.345058$  &$-0.153769$ &$s$& $-0.421950$ &$0.076887$\\
\hline 
$t$ &$0.345390$  &$-0.153439$ &$b$ & $-0.421945$&$0.076890$\\
\hline
\SM &0.3459  &$-0.1541$ &\SM &$-0.4229$ & $0.0771$\\
\hline 
\end{tabular}
\label{Table:Zcoupling1}
\end{center}
}
\end{table}

In observation the weak coupling constant is measured from the $W \nu_e e$ and $W \nu_\mu \mu$ couplings. 
Normalized by $\hat g^{W \nu_{e1} e}_{L,00}$, the $Z$-couplings in the first generation become
\begin{align}
\frac{1}{\hat g^{W \nu_{e1} e}_{L,00}} 
\begin{pmatrix} \hat g^{Z \nu_{e11}}_{L,00} \cr \noalign{\kern 2pt}
 \hat g^{Z \nu_{e11}}_{R,00} \cr\noalign{\kern 2pt}
 \hat g^{Z e}_{L,00} \cr \noalign{\kern 2pt} \hat g^{Z e}_{R,00}  \end{pmatrix}
&= \begin{pmatrix} 0.500022 \cr \noalign{\kern 2pt}2 \times 10^{-32} \cr \noalign{\kern 2pt}
 -0.268800 \cr  \noalign{\kern 2pt}  0.231198  \end{pmatrix}, \cr
\noalign{\kern 5pt}
\frac{1}{\hat g^{W \nu_{e1} e}_{L,00}} 
\begin{pmatrix} \hat g^{Z u}_{L,00} \cr  \noalign{\kern 2pt}
\hat g^{Z u}_{R,00} \cr  \noalign{\kern 2pt}
 \hat g^{Z dd}_{L,00} \cr  \noalign{\kern 2pt}
 \hat g^{Z dd}_{R,00}  \end{pmatrix}
&= \begin{pmatrix} 0.345873 \cr \noalign{\kern 2pt} -0.154132 \cr  \noalign{\kern 2pt}
 -0.422947 \cr \noalign{\kern 2pt}  0.077068  \end{pmatrix}.
\label{Zcoupling2}
\end{align}
The values in (\ref{Zcoupling2}) are very close to those in the SM with $\sin^2 \theta_W^\SM = 0.2312$.
For $(t,b)$ quarks
\begin{align}
\frac{1}{\hat g^{W \nu_{e1} e}_{L,00}} 
\begin{pmatrix} \hat g^{Z t}_{L,00} \cr  \noalign{\kern 2pt}
\hat g^{Z t}_{R,00} \cr  \noalign{\kern 2pt}
 \hat g^{Z bb}_{L,00} \cr  \noalign{\kern 2pt}
 \hat g^{Z bb}_{R,00}  \end{pmatrix}
&= \begin{pmatrix} 0.346205 \cr \noalign{\kern 2pt} -0.153801 \cr  \noalign{\kern 2pt}
-0.422940 \cr \noalign{\kern 2pt}   0.077071  \end{pmatrix}.
\label{Zcoupling3}
\end{align}
The deviations in the $Zb$ couplings from the SM are very small, which should be contrasted to the situation 
in some models in the RS warped space formulated in the early days.\cite{Carena2006}
The  $Zt_L$  coupling  is 0.09\% larger than that in the SM, whereas the  $Zt_R$  coupling  is
0.19\% smaller than that  in the SM.

As shown in  (\ref{Zu2}) and (\ref{Zu3}), the $Z$-coupling is decomposed into the $su2$ part and $\EM$ part.
The $su2$ part consists of three components; the $h^L_{Z^{(0)}} $ $[a]$, $h^R_{Z^{(0)}}$ $[b]$, 
$\hat h_{Z^{(0)}}$ $[ c]$ parts, 
$\hat g^{Z,  su2}_{L/R}  = \hat g^{Z, a}_{L/R}  + \hat g^{Z, b}_{L/R} + \hat g^{Z,c}_{L/R}$.
All of them are important.
For $(u,d)$ quarks, for instance, 
\begin{align}
&\begin{matrix} \cr\mynoalign \hat g^{Z, a} \cr  \hat g^{Z, b} \cr  \hat g^{Z,c} \cr \mynoalign  \hat g^{Z,  su2} \end{matrix}
~~ \begin{matrix} \cr\mynoalign : \cr   : \cr : \cr \mynoalign :  \end{matrix}
~~ \begin{matrix} u_L & d_L & u_R & d_R\cr  \mynoalign
0.49884 & -0.49884 & ~~0.0012459 & -0.000026196 \cr 
10^{-23} & 10^{-26} & ~~0.0012459 & - 0.000026196 \cr  
10^{-12} & 10^{-13} & -0.0024918 & ~~0.000052392 \cr   \mynoalign
0.49884 & -0.49884 & 10^{-11} &10^{-13}\end{matrix} ~~~ .
\label{Zcoupling4}
\end{align}
It is seen that the $SO(5)$ structure is crucial to have consistent gauge couplings of quarks and leptons.
Vanishingly small $\hat g^{Zu, su2}_{R}$ and  $\hat g^{Zd, su2}_{R}$ are due to the cancellation
among the $a$, $b$ and $c$ components, which is possible
in the $SO(5) \times U(1)_X$ gauge theory, but not in $SU(2)_L \times SU(2)_R  \times U(1)$ 
gauge theory.

The $WWZ$ coupling is evaluated similarly.  Triple couplings are written as
\begin{align}
 -i g_w \cos \theta_W^0 & \sum_{n, m, \ell} \hat g_{W^{\dagger (n)} W^{(m)} Z^{(\ell)}}
K[W^{\dagger (n)},  W^{(m)},  Z^{(\ell)}] ~, \cr
\noalign{\kern 5pt}
K[A, B, C] &= A^{\mu\nu} B_\mu C_\nu + B^{\mu\nu} C_\mu A_\nu +  C^{\mu\nu} A_\mu B_\nu ~, \cr
&A_{\mu\nu} = \dd_\mu A_\nu - \dd_\nu A_\mu ~~{\rm etc.}
\label{WWZcoupling1}
\end{align}
The SM value is $g_{WWZ}^{\SM} = g_w \cos \theta_W^\SM$.  In the current model one finds
\begin{align}
 \hat g_{W^{\dagger (n)} W^{(m)} Z^{(\ell)}} &= \frac{\sqrt{kL}}{2 \cos \theta_W^0} \int_1^{z_L} \frac{dz}{z} \,
 \Big\{ 2 h^L_{W^{(n)}}  h^L_{W^{(m)}}  h^L_{Z^{(\ell)}} + 2 h^R_{W^{(n)}}  h^R_{W^{(m)}}  h^R_{Z^{(\ell)}} \cr
 \noalign{\kern 5pt}
&\hskip 0.5cm
+  \big( h^L_{W^{(n)}} + h^R_{W^{(n)}} \big)  \hat h_{W^{(m)}}  \hat h_{Z^{(\ell)}} 
+  \hat h_{W^{(n)}}   \big(  h^L_{W^{(m)}} +  h^R_{W^{(m)}} \big)  \hat h_{Z^{(\ell)}}  \cr
\noalign{\kern 5pt}
&\hskip 5.cm
+  \hat h_{W^{(n)}}  \hat h_{W^{(m)}} \big( h^L_{Z^{(\ell)}} + h^R_{Z^{(\ell)}} \big) \Big\}
 ~.
\label{WWZcoupling2}
\end{align}
For $\theta_H = 0.1$ and $m_\KK = 13\,$TeV
\begin{align}
\hat g_{W^{\dagger (0)} W^{(0)} Z^{(0)}} - 1 &=  2.7 \times 10^{-7} ~, \cr
\noalign{\kern 5pt}
\begin{pmatrix} \hat g_{W^{\dagger (0)} W^{(0)} Z^{(1)}} \cr \hat g_{W^{\dagger (0)} W^{(0)} Z^{(2)}} \cr
\hat g_{W^{\dagger (0)} W^{(0)} Z^{(3)}} \cr \hat g_{W^{\dagger (0)} W^{(0)} Z^{(4)}} \end{pmatrix} 
&=\begin{pmatrix} -  2 \times 10^{-4} \cr 3 \times 10^{-8} \cr
 -8 \times 10^{-6} \cr -3 \times 10^{-8} \end{pmatrix}.
\label{WWZcoupling3}
\end{align}
The deviation in the $W^\dagger WZ = W^{\dagger (0)} W^{(0)} Z^{(0)}$ coupling is extremely  tiny. 

\section{Fermion 1-loop corrections} 

With the $W$ and $Z$ couplings obtained in the previous section we are going to evaluate 
oblique corrections in the GUT inspired $SO(5) \times U(1)_X \times SU(3)_C$ GHU.
%We first summarize  formulas necessary to evaluate fermion 1-loop vacuum polarization diagrams.
Let gauge fields $X_\mu$ and $Y_\mu$ couple to fermions $\psi_1$ and $\psi_2$ by
\begin{align}
&X_\mu\bar \psi_2 \gamma^\mu (g^{X, V}_{21} - g^{X, A}_{21} \gamma^5 ) \psi_1
+ Y_\mu\bar \psi_1 \gamma^\mu (g^{Y, V}_{12} - g^{Y, A}_{12} \gamma^5 ) \psi_2 ~.
\label{gaugecoupling2}
\end{align}
The vacuum polarization $\Pi_{XY}^{\mu\nu} (p)$ in which fermion $\psi_1$ and $\psi_2$ are running along the loop
is given by
\begin{align}
i \, \Pi_{XY}^{\mu\nu} (p) &= (-1) \int \frac{d^4 q}{(2\pi)^4}\, \Tr (-i \gamma^\mu) (g^{X, V}_{21} - g^{X, A}_{21} \gamma^5 )
\frac{i}{\Slash{q} - m_1 + i \ep} \cr
\noalign{\kern 5pt}
&\hskip 2.5cm
\times  (-i \gamma^\nu) (g^{Y, V}_{12} - g^{Y, A}_{12} \gamma^5 ) \frac{i}{\Slash{q} + \Slash{p}  - m_2 + i \ep} ~.
\label{loopPi1}
\end{align}
In the dimensional regularization it becomes
\begin{align}
\Pi_{XY}^{\mu\nu} (p) &=
-\frac{i}{(4\pi)^{d/2}} \, \Gamma \big( 2 - \onehalf d \big) \int_0^1 dx \, % \frac{\mu^{4-d}}{\Delta^{2 - d/2}} 
\bigg( \frac{\mu^2}{\Delta} \bigg)^{2 - d/2} \cr
\noalign{\kern 5pt}
&\hskip .3cm
\times \bigg[ (g^{X, V}_{21}  g^{Y, V}_{12}  + g^{X, A}_{21} g^{Y, A}_{12})
\Big\{ \big( x(1-x) p^2 - \Delta \big) \eta^{\mu\nu} - 2x(1-x) p^\mu p^\nu \Big\} \cr
\noalign{\kern 5pt}
&\hskip 1.cm
+ (g^{X, V}_{21}  g^{Y, V}_{12}  - g^{X, A}_{21} g^{Y, A}_{12}) \, m_1 m_2 \eta^{\mu\nu} \bigg]  \cr
\noalign{\kern 5pt}
&\equiv  \Pi_{XY}(p^2) \,  \eta^{\mu\nu}  - \Sigma_{XY} (p^2) \, p^\mu p^\nu ~, \cr
\noalign{\kern 5pt}
\Delta &= \Delta (x ; p^2, m_1, m_2) 
= - x(1-x) p^2 + (1-x) m_1^2 + x m_2^2 ~.
\label{loopPi2}
\end{align}
Expanded around $d=4$, $\Pi_{XY}(p)$ contains divergent pole terms.
\begin{align}
\Pi_{XY}(p^2) &= \Pi_{XY}^{\rm div} (p^2) + \Pi_{XY}^{\rm finite} (p^2) ~, \cr
\noalign{\kern 5pt}
\Pi_{XY}^{\rm div} (p^2) &= \frac{1}{4 \pi^2} 
\bigg[ (g^{X, V}_{21}  g^{Y, V}_{12}  + g^{X, A}_{21} g^{Y, A}_{12}) \big\{ \onethird p^2 - \onehalf (m_1^2 + m_2^2) \big\} \cr
\noalign{\kern 5pt}
&\hskip 1.5cm 
+  (g^{X, V}_{21}  g^{Y, V}_{12}  - g^{X, A}_{21} g^{Y, A}_{12}) \, m_1 m_2 \bigg] \, \hat E \equiv K^{\rm div}  \hat E ~, \cr
\noalign{\kern 5pt}
\Pi_{XY}^{\rm finite} (p^2) &= \frac{1}{4 \pi^2} \int_0^1 dx \, \ln \Delta 
\bigg[ (g^{X, V}_{21}  g^{Y, V}_{12}  + g^{X, A}_{21} g^{Y, A}_{12}) \big\{ x(1-x) p^2 - \Delta \big\} \cr
\noalign{\kern 5pt}
&\hskip 3.2cm
+  (g^{X, V}_{21}  g^{Y, V}_{12}  - g^{X, A}_{21} g^{Y, A}_{12}) \, m_1 m_2 \bigg] ~,  \cr
\noalign{\kern 5pt}
\hat E &=  - \frac{2}{4-d} + \gamma_E - \ln 4\pi \mu^2 ~.
\label{loopPi3}
\end{align}
In terms of  
\begin{align}
b_0 (s, m_1, m_2) &= \int_0^1 dx \, \ln \Delta (x; s, m_1, m_2) = b_0 (s, m_2, m_1) ~, \cr
\noalign{\kern 5pt}
b_1 (s, m_1, m_2) &= \int_0^1 dx \,  x \ln \Delta (x; s, m_1, m_2) 
= \int_0^1 dx \,  (1-x)  \ln \Delta (x; s, m_2, m_1)  ~, \cr
\noalign{\kern 5pt}
b_2 (s, m_1, m_2) &= \int_0^1 dx \,  x (1-x) \ln \Delta (x; s, m_1, m_2)  = b_2 (s, m_2, m_1 )  
\label{loopFn1}
\end{align}
$\Pi_{XY}^{\rm finite} (p^2) $ is expressed as
\begin{align}
\Pi_{XY}^{\rm finite} (p^2) &= \frac{1}{4 \pi^2} 
\bigg[ \big(g^{X, V}_{21}  g^{Y, V}_{12}  + g^{X, A}_{21} g^{Y, A}_{12} \big)
\Big\{ 2 p^2 \, b_2 (p^2, m_1, m_2) \cr
\noalign{\kern 5pt}
&\hskip 3.cm
- m_1^2 \, b_1 (p^2, m_2, m_1) - m_2^2 \, b_1 (p^2, m_1, m_2) \Big\} \cr
\noalign{\kern 5pt}
&\hskip 1.5cm
+ \big(g^{X, V}_{21}  g^{Y, V}_{12}  - g^{X, A}_{21} g^{Y, A}_{12} \big) \, m_1 m_2 \, b_0 (p^2, m_1, m_2) \bigg] .
\label{loopPi4}
\end{align}
We note that
\begin{align}
&\Pi_{XY}^{\rm finite} (p^2, m_1, m_2) 
= \mu^2 \, \Pi_{XY}^{\rm finite} \Big( \frac{p^2}{\mu^2}, \frac{m_1}{\mu}, \frac{m_2}{\mu} \Big) - K^{\rm div} \ln \mu^2 
\label{loopPi5}
\end{align}
where $K^{\rm div}$ is defined in Eq.\ (\ref{loopPi3}).
It will be seen convenient to take $\mu= m_\KK$ in (\ref{loopPi5}) to evaluate finite parts of oblique corrections 
in the current model.

%%%%%%%%%%%%%%%%%%%

%We evaluate vacuum polarization diagrams for  $W$, $Z$ and $\gamma$ generated by fermion loops.
In Section 3 we have obtained $W$ and $Z$ couplings of fermions.  We adopt the convention for vector and
axial-vector couplings given by $g_V = \onehalf (g_R + g_L)$ and $g_A = \onehalf (g_R- g_L)$ so that
$g_R g_R' + g_L g_L' = 2 ( g_V g_V' + g_A g_A')$.
To simplify expressions we introduce 
\begin{align}
&G[ \, p^2 ;  \hat g_V , \hat g_A ;  \hat g_{V'} , \hat g_{A'} ;  m_{1n}, m_{2\ell} \, ] \cr
\noalign{\kern 5pt}
&=
\sum_{n,\ell =0}^\infty \bigg[
\Big(   \hat g_{V, n\ell} \hat g_{V', \ell n} +  \hat g_{A, n\ell} \hat g_{A', \ell n}   \Big)
\Big\{ \big(\onethird p^2 - \onehalf (m_{1n}^2 + m_{2\ell}^2) \big) \hat E  \cr
\noalign{\kern 5pt}
&\hskip 0.5cm
+ 2 p^2 \, b_2 (p^2, m_{1n} , m_{2\ell} ) 
 - m_{1n}^2 \, b_1 (p^2, m_{2\ell} , m_{1n}) - m_{2\ell}^2 \, b_1 (p^2, m_{1n}, m_{2\ell})  \Big\} \cr
\noalign{\kern 5pt}
&\hskip 0.5cm
+ \Big(  \hat g_{V, n\ell} \hat g_{V', \ell n} -  \hat g_{A, n\ell} \hat g_{A', \ell n}  \Big)
m_{1n} m_{2\ell} \, \big\{ \hat E + b_0 (p^2, m_{1n}, m_{2\ell}) \big\} \bigg] .
\label{Pi0}
\end{align}
Then,  for the $WW$ vacuum polarization,  contributions from the $(u,d)$ multiplets are given by
\begin{align}
\Pi_{W W}^{ud} (p^2) &= \frac{N_C g_w^2}{8\pi^2} \, \Big\{ 
G[ \, p^2 ;  \hat g^{Wud}_V, \hat g^{Wud}_A ;  \hat g^{W^\dagger ud}_V, \hat g^{W^\dagger ud}_A ; m_{u^{(n)}}, m_{d^{(\ell)}} \, ]  \cr
\noalign{\kern 5pt}
&\hskip 1.cm
+ G[ \,p^2 ; \hat g^{W uD}_V, \hat g^{W uD}_A ;  \hat g^{W^\dagger uD}_V, \hat g^{W^\dagger uD}_A ; m_{u^{(n)}}, m_{D^{(\ell)}} \, ]\Big\} 
\label{PiWW1}
\end{align}
where $N_C=3$.
%We have omitted the contributions coming from $uD$ loops as the $ \hat g^{W uD}_{V/A, n\ell}$ couplings are extremely small.  
Note that $\hat g^{W^\dagger ud}_{V/A, \ell n}  = ( \hat g^{Wud}_{V/A, n\ell})^*$ etc.
Contributions from the $(\nu_e, e)$ multiplets are
\begin{align}
\Pi_{W W}^{\nu_e e} (p^2) &= \frac{g_w^2}{8\pi^2} \sum_{a=1}^2
G[ \, p^2; \hat g^{W\nu_{ea} e}_V, \hat g^{W\nu_{ea} e}_A ;  \hat g^{W^\dagger \nu_{ea} e}_V, \hat g^{W^\dagger\nu_{ea} e}_A ; 
m_{\nu_{ea}^{(n)}}, m_{e^{(\ell)}} \, ] ~.
\label{PiWW2}
\end{align}
For the $ZZ$ vacuum polarization,  contributions from the $(u,d)$ multiplets are 
\begin{align}
\Pi_{ZZ}^{ud} (p^2) &= \frac{N_C g_w^2}{4\pi^2 \cos^2 \theta_W^0} \, \Big\{
G[ \, p^2;  \hat g^{Z u}_V, \hat g^{Zu}_A ;  \hat g^{Z u}_V, \hat g^{Z u}_A ;  m_{u^{(n)}}, m_{u^{(\ell)}} \, ]  \cr
\noalign{\kern 5pt}
&\hskip 1.cm
+ G[ \, p^2; \hat g^{Z dd}_V, \hat g^{Zdd}_A ;  \hat g^{Z dd}_V, \hat g^{Z dd}_A ;  m_{d^{(n)}}, m_{d^{(\ell)}} \, ]  \cr
\noalign{\kern 5pt}
&\hskip 1.cm
+ G[ \, p^2; \hat g^{Z DD}_V, \hat g^{ZDD}_A ;  \hat g^{Z DD}_V, \hat g^{Z DD}_A ;  m_{D^{(n)}}, m_{D^{(\ell)}} \, ]  \cr
\noalign{\kern 5pt}
&\hskip 1.cm
+ 2 \, G[ \, p^2; \hat g^{Z dD}_V, \hat g^{ZdD}_A ;  \hat g^{Z Dd}_V, \hat g^{Z Dd}_A ;  m_{d^{(n)}}, m_{D^{(\ell)}} \, ]  \Big\} .
\label{PiZZ1}
\end{align}
Here we have set $\hat g^{Z dD}_{V/A, n0} = \hat g^{Z Dd}_{V/A, 0n} = \hat g^{Z DD}_{V/A, n0}  = \hat g^{Z DD}_{V/A, 0n} = 0$.
Contributions from the $(\nu_e, e)$ multiplets are
\begin{align}
\Pi_{ZZ}^{\nu_e e} (p^2) &= \frac{g_w^2}{4\pi^2 \cos^2 \theta_W^0} \, \Big\{
G[ \, p^2; \hat g^{Z e}_V, \hat g^{Ze}_A ;  \hat g^{Z e}_V, \hat g^{Z e}_A ;  m_{e^{(n)}}, m_{e^{(\ell)}} \, ]  \cr
\noalign{\kern 5pt}
&\hskip 0.5cm
+ \sum_{a=1}^2  \sum_{b=1}^2 
G[ \, p^2; \hat g^{Z \nu_{e a b}}_V, \hat g^{Z\nu_{e a b}}_A ;  \hat g^{Z \nu_{e ba}}_V, \hat g^{Z \nu_{e ba}}_A ;  
m_{\nu_{ea}^{(n)}}, m_{\nu_{eb}^{(\ell)}} \, ]    \Big\} .
\label{PiZZ2}
\end{align}

The photon  couplings are universal.  They are  diagonal and vectorlike.  
Noting that $b_1 (p^2, m, m) = \onehalf b_0 (p^2, m, m)$, one finds for the $\gamma \gamma$ vacuum polarization that
\begin{align}
\Pi_{\gamma \gamma}^{ud} (p^2) &=  \frac{N_C g_w^2 \sin^2 \theta_W^0}{4\pi^2} \, p^2 \, 
\bigg\{ \sum_{n=0}^\infty Q_u^2 \Big( \onethird   \hat E + 2 b_2(p^2, m_{u^{(n)}}, m_{u^{(n)}})  \Big) \cr
\noalign{\kern 5pt}
&\hskip 3.cm
+  \sum_{n=0}^\infty  Q_d^2  \Big(  \onethird  \hat E +  2  b_2(p^2, m_{d^{(n)}}, m_{d^{(n)}}) \Big) \cr
\noalign{\kern 5pt}
&\hskip 3.cm
+ \sum_{n=1 }^\infty  Q_D^2  \Big(  \onethird  \hat E +  2  b_2(p^2, m_{D^{(n)}}, m_{D^{(n)}}) \Big) \bigg\} , \cr
\noalign{\kern 5pt}
\Pi_{\gamma \gamma}^{\nu_e e} (p^2) &=  \frac{g_w^2 \sin^2 \theta_W^0}{4\pi^2} \, p^2 \, 
\sum_{n=0}^\infty Q_e^2 \Big( \onethird   \hat E + 2 b_2(p^2, m_{e^{(n)}}, m_{e^{(n)}})  \Big) 
\label{Pigamma1}
\end{align}
where $Q_u = \twothird$, $Q_d= Q_D = - \onethird$ and $Q_e = -1$.
For the $Z \gamma$ vacuum polarization one finds that
\begin{align}
\Pi_{Z \gamma}^{ud} (p^2) &=  \frac{N_C g_w^2 \sin \theta_W^0}{4\pi^2 \cos \theta_W^0} \, p^2 \, 
\bigg\{ \sum_{n=0}^\infty Q_u \hat g^{Zu}_{V, nn} \Big( \onethird   \hat E + 2 b_2(p^2, m_{u^{(n)}}, m_{u^{(n)}})  \Big) \cr
\noalign{\kern 5pt}
&\hskip 3.cm
+  \sum_{n=0}^\infty  Q_d  \hat g^{Zdd}_{V, nn} \Big(  \onethird  \hat E +  2  b_2(p^2, m_{d^{(n)}}, m_{d^{(n)}}) \Big) \cr
\noalign{\kern 5pt}
&\hskip 3.cm
+ \sum_{n=1 }^\infty  Q_D  \hat g^{ZDD}_{V, nn}  \Big(  \onethird  \hat E +  2  b_2(p^2, m_{D^{(n)}}, m_{D^{(n)}}) \Big) \bigg\} , \cr
\noalign{\kern 5pt}
\Pi_{Z \gamma}^{\nu_e e} (p^2) &=  \frac{g_w^2 \sin \theta_W^0}{4\pi^2 \cos \theta_W^0} \, p^2 \, 
\sum_{n=0}^\infty Q_e  \hat g^{Ze}_{V, nn}  \Big( \onethird   \hat E + 2 b_2(p^2, m_{e^{(n)}}, m_{e^{(n)}})  \Big) ~.
\label{PiZgamma1}
\end{align}
Note that $\Pi_{\gamma \gamma} (0)  = \Pi_{Z \gamma} (0)  = 0$ as a consequence of the Ward-Takahashi identity
in $U(1)_\EM$.

Expressions for $\Pi( p^2)$ for the second and third generations are obtained similarly.

\section{Coupling sum rules} 

Each $\Pi( p^2)$ in the previous section contains divergent terms proportional to $\hat E$.  
In the SM some of them are absorbed by renormalization constants, and specific combinations of 
the  $\Pi( p^2)$'s, namely the $S$, $T$ and $U$ combinations,  remain finite.\cite{PeskinTakeuchi, Lavoura1993, PeskinSchroeder}
In GHU all KK modes of fermions contribute to $\Pi( p^2)$, and their couplings $\hat g^W_{V/A}$ and $\hat g^Z_{V/A}$ 
are highly nontrivial.   The couplings $\hat g^W_{V/A}$ and $\hat g^Z_{V/A}$ take the matrix form with nonvanishing
off-diagonal elements.  Further even in the subspace of the KK excited states the axial vector couplings are nonvanishing.

In this section we show that there exist three identities among $W$ and $Z$ coupling matrices in each fermion
doublet-multiplet, which are associated with the divergent terms in $\Pi_{WW} (p^2)$, $\Pi_{ZZ} (p^2)$ and
$\Pi_{Z \gamma} (p^2)$.
We define the $W^3$ coupling matrix $\hat g^{W^3}_V$, say for the $(u,d)$ multiplet, by
\begin{align}
\hat g^{Zu}_{V, n \ell}~ &=  ~\hat g^{W^3 u}_{V, n \ell} - \sin^2 \theta_W^0 \, Q_u \, \delta_{n\ell} ~, \cr
\noalign{\kern 5pt}
\begin{pmatrix} \hat g^{Zdd}_{V, n \ell} \cr \mynoalign  \hat g^{ZDD}_{V, n \ell}    \end{pmatrix}
&= \begin{pmatrix} \hat g^{W^3 dd}_{V, n \ell} \cr \mysnoalign  \hat g^{W^3 DD}_{V, n \ell}   \end{pmatrix}
- \sin^2 \theta_W^0 \, Q_d \, \delta_{n\ell} ~.
\label{W3coupling1}
\end{align}
We stress that $\hat g^{W^3 u}_{V}$ and $\hat g^{W^3 dd}_{V}$ slightly differ from $\hat g^{Zu, su2}_{V}$ and
 $\hat g^{Zdd, su2}_{V}$ defined in (\ref{Zu2}) and (\ref{Zd2}).
 Numerically all elements of $\hat g^{W^3 DD}_{V}$, $\hat g^{Z DD}_{A}$,  $\hat g^{Z dD}_{V/A}$ and $\hat g^{Z Dd}_{V/A}$
 are $O(10^{-6})$ or less.
 In the following we safely omit the contributions coming from the $D$ modes in the expressions for the coupling sum rules.
 
 We define
 \begin{align}
 A_0^{ud} &= \Tr \Big\{ \hat g^{W^3 u}_{V} \hat g^{W^3 u}_{V} + \hat g^{Z u}_{A}  \hat g^{Z u}_{A} 
 +  \hat g^{W^3 dd}_{V} \hat g^{W^3 dd}_{V} + \hat g^{Z dd}_{A}  \hat g^{Zdd}_{A} \Big\} ~, \cr
 \noalign{\kern 5pt}
  A_1^{ud} &= \sum_{n, \ell}
  \Big\{ \hat g^{W^3 u}_{V, n\ell} \hat g^{W^3 u}_{V, \ell n} (m_{u^{(n)} } - m_{u^{(\ell )}} )^2
  + \hat g^{Z u}_{A, n\ell}  \hat g^{Z u}_{A, \ell n}  ( m_{u^{(n)} } + m_{u^{(\ell )}} )^2 \cr
 \noalign{\kern 0pt}
 &\hskip 1.cm
 +  \hat g^{W^3 dd}_{V, n\ell} \hat g^{W^3 dd}_{V, \ell n}  (m_{d^{(n)} } - m_{d^{(\ell )}} )^2
 + \hat g^{Z dd}_{A, n\ell}  \hat g^{Zdd}_{A, \ell n}  (m_{d^{(n)} } + m_{d^{(\ell )}} )^2  \Big\} ~,  \cr
\noalign{\kern 5pt}
B^{ud} &=  Q_u \Tr  \hat g^{W^3 u}_{V} + Q_d  \Tr  \hat g^{W^3 dd}_{V} ~, \cr
\noalign{\kern 5pt}
C^{ud} &=  Q_u^2 \Tr I+ Q_d^2  \Tr  I ~, \cr
\noalign{\kern 5pt}
D_0^{ud} &= \Tr \Big\{ \hat g^{W ud}_{V} \hat g^{W^\dagger ud}_{V} + \hat g^{W ud}_{A}  \hat g^{W^\dagger ud}_{A}  \Big\} ~, \cr
\noalign{\kern 5pt}
D_1^{ud} &= \sum_{n, \ell}  \Big\{ \hat g^{W ud}_{V, n\ell} \hat g^{W^\dagger ud}_{V, \ell n}  (m_{u^{(n)} } - m_{d^{(\ell )}} )^2
+ \hat g^{W ud}_{A, n\ell}  \hat g^{W^\dagger ud}_{A, \ell n}  (m_{u^{(n)} } + m_{d^{(\ell )}} )^2  \Big\} ~.
\label{coeff1}
\end{align}
Here $\Tr$ in $Q_u^2 \Tr I$ implies the trace over the $u^{(n)}$ states.
 Then the divergent parts of $\Pi^{ud} (p^2)$ are expressed as 
 \begin{align}
& \Pi^{ud}_{ZZ} (p^2)^{\rm div} =  \frac{N_C g_w^2}{4\pi^2 \cos^2 \theta_W^0} \, \Big\{
\frac{1}{3}  \big( A_0^{ud} - 2 \sin^2 \theta_W^0 B^{ud} + \sin^4 \theta_W^0 C^{ud} \big) p^2 -\frac{1}{2} A_1^{ud} \Big\} \, \hat E ~, \cr
\noalign{\kern 5pt}
&\Pi^{ud}_{Z\gamma} (p^2)^{\rm div} =  \frac{N_C g_w^2 \sin \theta_W^0}{4\pi^2 \cos \theta_W^0} \, 
 \frac{1}{3}  \big( B^{ud} -  \sin^2 \theta_W^0 C^{ud} \big) p^2 \, \hat E ~, \cr
 \noalign{\kern 5pt}
&\Pi^{ud}_{\gamma\gamma} (p^2)^{\rm div} =  \frac{N_C g_w^2 \sin^2 \theta_W^0}{4\pi^2} \, 
 \frac{1}{3}  \, C^{ud}  p^2 \, \hat E ~, \cr
 \noalign{\kern 5pt}
&\Pi^{ud}_{WW} (p^2)^{\rm div} =  \frac{N_C g_w^2}{8\pi^2} \, 
\Big\{  \frac{1}{3}  \, D_0^{ud}  p^2  - \frac{1}{2} D_1^{ud} \Big\}  \, \hat E ~.
\label{PiDiv1}
 \end{align}
For the $(\nu_e, e)$ doublet-multiplet 
\begin{align}
\hat g^{Ze}_{V, n \ell}  &=  \hat g^{W^3 e}_{V, n \ell} - \sin^2 \theta_W^0 \, Q_e  \delta_{n\ell}
\label{W3coupling2}
\end{align}
and we define
 \begin{align}
 A_0^{\nu_e e} &= \Tr \Big\{ \hat g^{W^3 e}_{V} \hat g^{W^3 e}_{V} + \hat g^{Z e}_{A}  \hat g^{Z e}_{A} 
 +  \sum_{a=1}^2 \sum_{b=1}^2 \big( \hat g^{Z \nu_{eab}}_{V} \hat g^{Z \nu_{eba}}_{V} 
 + \hat g^{Z \nu_{eab}}_{A}  \hat g^{Z \nu_{eba}}_{A}  \big) \Big\} ~, \cr
 \noalign{\kern 5pt}
  A_1^{\nu_e e} &= \sum_{n, \ell}
  \bigg( \hat g^{W^3 e}_{V, n\ell} \hat g^{W^3 e}_{V, \ell n} (m_{e^{(n)} } - m_{e^{(\ell )}} )^2
  + \hat g^{Z e}_{A, n\ell}  \hat g^{Z e}_{A, \ell n}  ( m_{e^{(n)} } + m_{e^{(\ell )}} )^2 \cr
 \noalign{\kern 0pt}
 &\hskip 0.cm
 +  \sum_{a=1}^2\sum_{b=1}^2 \Big\{ \hat g^{Z \nu_{eab}}_{V, n\ell} \hat g^{Z \nu_{eba}}_{V, \ell n}  (m_{\nu_{ea}^{(n)} } - m_{\nu_{eb}^{(\ell )}} )^2
 + \hat g^{Z \nu_{eab}}_{A, n\ell}  \hat g^{Z \nu_{eba}}_{A, \ell n}  (m_{\nu_{ea}^{(n)} } + m_{\nu_{eb}^{(\ell )}} )^2  \Big\}  \bigg)~,  \cr
\noalign{\kern 5pt}
B^{\nu_e e} &=  Q_e \Tr  \hat g^{W^3 e}_{V}  ~, \cr
\noalign{\kern 5pt}
C^{\nu_e e} &=  Q_e^2 \Tr I~, \cr
\noalign{\kern 5pt}
D_0^{\nu_e e} &= \sum_{a=1}^2  \Tr \Big\{ \hat g^{W \nu_{ea} e}_{V} \hat g^{W^\dagger \nu_{ea} e}_{V} 
+ \hat g^{W \nu_{ea} e}_{A}  \hat g^{W^\dagger \nu_{ea} e}_{A}  \Big\} ~, \cr
\noalign{\kern 5pt}
D_1^{\nu_e e} &=  \sum_{a=1}^2   \sum_{n, \ell}  
\Big\{ \hat g^{W \nu_{ea} e}_{V, n\ell} \hat g^{W^\dagger \nu_{ea} e}_{V, \ell n}  (m_{\nu_{ea}^{(n)} } - m_{e^{(\ell )}} )^2
+ \hat g^{W \nu_{ea} e}_{A, n\ell}  \hat g^{W^\dagger\nu_{ea} e}_{A, \ell n}  (m_{\nu_{ea}^{(n)} } + m_{e^{(\ell )}} )^2  \Big\} .
\label{coeff2}
\end{align}
The divergent parts of $\Pi^{\nu_e e}  (p^2)$ are given by the expressions in (\ref{PiDiv1}) where
$N_C=1$ and the superscript `$ud$' is replaced by `$\nu_e e$'.
Note that these coefficients depend on the fermion doublet; $A_0^{ud} \not= A_0^{\nu_e e}$ etc.

Although all of the coupling matrices $\hat g_{V/A}$ are rather nontrivial as shown in Section 3, there appear astonishing
relations among $A_0$, $A_1$, $B$, $D_0$ and $D_1$.
We are going to establish, by numerical evaluation from  the coupling matrices, the following coupling sum rules
\begin{align}
&\begin{cases} A_0^{ud} = h^{ud} B^{ud}  \cr \noalign{\kern 4pt}
D_0^{ud} = 2 A_0^{ud}  \cr \noalign{\kern 4pt}
D_1^{ud} = 2 A_1^{ud}  \end{cases} 
, \hskip .5cm
\begin{cases} A_0^{\nu_e e} = h^{\nu_e e} B^{\nu_e e}  \cr \noalign{\kern 4pt}
D_0^{\nu_e e} =2 A_0^{\nu_e e} \cr \noalign{\kern 4pt}
D_1^{\nu_e e} =2 A_1^{\nu_e e}  \end{cases} 
\label{SumRule1}
\end{align}
to high accuracy, where
\begin{align}
&h^{ud} =\hat g^{Zu, su2}_{L,00} -  \hat g^{Zdd, su2}_{L,00} ~,~~~
h^{\nu_e e} = \hat g^{Z \nu_{e11}}_{L,00} - \hat g^{Ze, su2}_{L,00}~.
\label{SumRule2}
\end{align}
Similar relations hold for the second and third generations.  For the $(t,b)$ doublet, we use 
$h^{tb} = -  2\hat g^{Zbb, su2}_{L,00}$.
Numerical values of $\hat g^{Z, su2}_{L,00}$ and $\hat g^W_{L,00}$ for $\theta_H=0.1$, $m_\KK=13\,$TeV
and $M=10^3\,$TeV are summarized in Table \ref{Table:hfactor}.
The factors $h$ are close to, but not exactly 1.

\begin{table}[tbh]
{%\small
\renewcommand{\arraystretch}{1.3}
\begin{center}
\caption{The couplings $\hat g^{Z, su2}_{L,00}$ and $\hat g^W_{L,00}$ for $\theta_H=0.1$, $m_\KK=13\,$TeV
and $M=10^3\,$TeV.  
}
\vskip 10pt
\begin{tabular}{|c|c|c|c|}
%\hline
%&GUT inspired model\\
%& type B &type A\\
\hline
&$2 \hat g^{Z \nu_{e11}}_{L,00}$  &$-2 \hat g^{Ze, su2}_{L,00}$ &$\hat g^{W \nu_{e1} e}_{L,00}$   \\
\hline %\hline
$(\nu_{e}, e)$ &$0.997690$ &$0.997691$  &$0.997647$ \\
\hline
$(\nu_{\mu}, \mu)$ &$0.997686$  &$0.997687$  &$0.997644$ \\
\hline 
$(\nu_{\tau}, \tau)$ &$0.997684$  &$0.997684$ &$0.997642$\\
\hline
\noalign{\kern 3pt}
\hline
&$2 \hat g^{Z u, su2}_{L,00}$  &$-2 \hat g^{Zdd, su2}_{L,00}$ &$\hat g^{W ud}_{L,00}$   \\
\hline
$(u,d)$ &$0.997688$  &$0.997688$  &$0.997645$ \\
\hline
$(c,s)$& $0.997685$  &$0.997685$  & $0.997643$ \\
\hline 
$(t,b)$ &$0.998344$  &$0.997671$  & $0.997969$\\
\hline
\end{tabular}
\label{Table:hfactor}
\end{center}
}
\end{table}

In the SM $h=1$, $A_0 = B = \onehalf D_0 = \onequarter$, $A_1^{ud} = \onehalf D_1^{ud} = \onequarter (m_u^2 + m_d^2)$ 
etc.\ so that the relations in (\ref{SumRule1}) are satisfied for each doublet.
In the current GHU model the relations are highly nontrivial.  We have included contributions coming from the KK modes
$n=0$ to $n=12$. 
The mass spectrum of the KK states and the 13-by-13 coupling matrices are determined with double precision.
To confirm the accuracy of the coupling sum rules we introduce
\begin{align}
\Delta_S &= \frac{A_0 - h B}{A_0} ~, \cr
\noalign{\kern 5pt}
\Delta_T &= \frac{A_1 - \onehalf D_1}{A_1} ~, \cr
\noalign{\kern 5pt}
\Delta_U &= \frac{A_0 - \onehalf D_0}{A_0} ~.
\label{SumRule3}
\end{align}
Obtained results for $A_0$, $A_1$, $\Delta_S$, $\Delta_T$ and $\Delta_U$ are summarized in Table \ref{Table:sumrule1}.

\begin{table}[tbh]
{%\small
\renewcommand{\arraystretch}{1.3}
\begin{center}
\caption{The couplings sum rules.  
The values of  $A_0$, $A_1$, $\Delta_S$, $\Delta_T$, $\Delta_U$ and $h$ in (\ref{SumRule1})-(\ref{SumRule3}) 
are tabulated for each doublet for $\theta_H=0.1$, $m_\KK=13\,$TeV and $M=10^3\,$TeV. 
The numerical values are evaluated by including the contributions coming from the KK towers of
fermions up to the $n=12$ level.
}
\vskip 10pt
\begin{tabular}{|c|c|c|c|c|c|c|}
%\hline
%&GUT inspired model\\
%& type B &type A\\
\hline
&$A_0$  &$A_1$ &$\Delta_S$  &$\Delta_T$ &$\Delta_U$ &$h$  \\
\hline
\noalign{\kern 1pt}
\hline 
$(\nu_{e}, e)$ &$3.24204$ &$44.1410$  &$3.6 \times 10^{-5}$ &$5.3 \times 10^{-5}$  &$-3.4 \times 10^{-7}$ &$0.997690$\\
\hline
$(\nu_{\mu}, \mu)$ &$3.24214$  &$43.5089$  &$7.1\times 10^{-5}$ &$5.3 \times 10^{-5}$ &$-3.4 \times 10^{-7}$ &$0.997687$ \\
\hline 
$(\nu_{\tau}, \tau)$ &$3.24219$  &$43.1535$ &$- 1.5 \times 10^{-5}$ &$5.3 \times 10^{-5}$ &$-2.7 \times 10^{-7}$ &$0.997684$\\
\hline
\noalign{\kern 1pt}
\hline
$(u,d)$ &$3.24210$  &$43.7063$  &$- 4.0 \times 10^{-6}$ &$5.5 \times 10^{-5}$ &$-3.4 \times 10^{-7}$ &$0.997688$\\
\hline
$(c,s)$& $3.24217$  &$43.2862$  &$2.0 \times 10^{-5}$ &$5.3 \times 10^{-5}$ &$-3.4 \times 10^{-7}$ &$0.997685$\\
\hline 
$(t,b)$ &$3.24262$  &$42.5613$  & $1.0 \times 10^{-4}$ &$4.1 \times 10^{-5}$ &$-2.8 \times 10^{-7}$ &$0.997671$\\
\hline
\end{tabular}
\label{Table:sumrule1}
\end{center}
}
\end{table}

It is seen that the coupling sum rules (\ref{SumRule1}) are valid with 5 to 7 digits accuracy, at least numerically.
In view of the nontrivial matrix structure of the gauge couplings the coupling sum rules (\ref{SumRule1}) are
highly nontrivial.  The relations are expected as the consequences of the 5D gauge invariance in GHU.
Although the values of $A_0$ and $A_1$ increase with $n$,  $\Delta_S$, $\Delta_T$ and $\Delta_U$ remain small.
For the $(\nu_e, e)$ multiplet, for instance, as $n$ increases from 12  to 16, the variations are
$A_0: 3.24 \go 4.24$, $A_1: 44.1 \go 57.9$, $\Delta_S : 3.6 \times 10^{-5} \go 7.4 \times 10^{-5}$,
$\Delta_T : 5.34 \times 10^{-5} \go 5.34 \times 10^{-5}$,  $\Delta_U : - 3.44 \times 10^{-7} \go - 3.41 \times 10^{-7}$.

In this paper we have considered the vacuum polarization tensors of the photon, $W$ and $Z$ bosons only.
Rigorous theoretical derivation of the coupling sum rules would require treating the whole
KK towers of the $SO(5) \times U(1)_X$ gauge bosons, which is beyond the scope of the current paper.
It is expected that similar coupling sum rules hold even in each sector of KK excited modes of the gauge fields.

One comment is in order about the appearance of the $h$ factor in the relation $A_0 = h B$.
The relation involves the $Z$ and $U(1)_\EM$ couplings.
As emphasized in Section 3, around (\ref{Zcoupling4}), the $Z$ couplings have the effective
$SU(2)_\eff  \times U(1)_\EM$ structure.
In the SM the $Z$ couplings to quarks and leptons are given by 
$(g_w/\cos \theta_W^\SM) (T^3_L - \sin^2 \theta_W^\SM Q_\EM)$.
In GHU the $W$ and $Z$ couplings to a doublet $\beta = [(u,d), (\nu_e, e) , \cdots]$ are given 
approximately by
\begin{align}
W:  &\quad \frac{g_w  \,\hat g^{W \beta}_{L, 00} }{\sqrt{2}} \,  \big( T^1_\eff + i \, T^2_\eff  \big) ~, \cr
\noalign{\kern 5pt}
Z ~: &\quad \frac{g_w \,  h^{\beta}}{\cos \theta_W^0} \, \Big(  T^3_\eff - \frac{\sin^2 \theta_W^0}{h^\beta} \, Q_\EM \Big) ~, 
\label{Zcoupling5h}
\end{align}
where  left-handed (right-handed) quarks and leptons are $SU(2)_\eff$ doublets (singlets), and
$h^{\beta} =  \hat g^{Z\beta_u, su2}_{L,00} -  \hat g^{Z\beta_d, su2}_{L,00} $ for $\beta = (\beta_u, \beta_d)$.
The factor $h^{\beta}$ is not equal to 1 in GHU even at the tree level. 
It is very close to $\hat g^{W \beta}_{L, 00}$.
The relevant quantity for the forward-backward asymmetry in $e^- e^+\go f \bar f$ at the $Z$ pole, for instance,  
is $\sin^2 \theta_W^0 / h^{ \beta} $ which is about $ \sin^2 \theta_W^\SM$.
The $h$ factor in (\ref{Zcoupling5h}) effectively appears in the relation $A_0 = h B$, which affects the definition of
the $S$ parameter in GHU as discussed below.

\section{Improved oblique parameters} 

The oblique parameters $S$, $T$ and $U$  of Peskin-Takeuchi are useful to investigate new physics 
beyond the SM.   %\cite{PeskinTakeuchi, Lavoura1993, PeskinSchroeder}
These parameters are expressed in terms of the vacuum polarization tensors of $W$, $Z$ and photon.
Certain combinations of those vacuum polarization tensors are finite, and are expected to represent 
important parts of the corrections to physical quantities.

In GHU some improvement is necessary.  In the most general situation the $S$, $T$ and $U$
parameters should be defined as certain combinations of  the vacuum polarization tensors of
all $SO(5) \times U(1)_X$ gauge fields including the KK excited modes.  Only the combinations
which are finite at the quantum level could serve as quantities measuring corrections to 
physical quantities.  
In this section we examine the finite corrections to the $S$, $T$ and $U$ parameters associated to
the vacuum polarization tensors of $W$, $Z$ and photon.  We should remember that these quantities
are not directly-measured physical quantities.  Directly-measured physical quantities expressed
in terms of four-fermi vertices, for instance, involve contributions coming from the KK modes of
the gauge bosons in GHU.  

In the previous section we have established  three coupling sum rules to high accuracy.
For each fermion doublet $\beta$ the sum rules are
\begin{align}
\begin{cases}
A_0^\beta = h^\beta B^\beta  &{\rm for~} S~, \cr \noalign{\kern 5pt}
A_1^\beta = \onehalf D_1^\beta &{\rm for~} T~, \cr \noalign{\kern 5pt}
A_0^\beta = \onehalf D_0^\beta &{\rm for~} U~.     \end{cases}
\label{Sumrule4}
\end{align}
With these sum rules at hand we propose the following  $S$, $T$ and $U$ for each fermion doublet $\beta$
at the one loop level;
\begin{align}
\alpha_* S^\beta &= \frac{\sin^2 2 \theta_W^0}{m_Z^2}  \bigg\{ \Pi_{ZZ}^\beta (m_Z^2) -  \Pi_{ZZ}^\beta (0) 
-\frac{\cos 2\theta_W^0 + h^\beta - 1}{\sin \theta_W^0 \cos \theta_W^0} \Pi_{Z\gamma}^\beta (m_Z^2) \cr
\noalign{\kern 5pt}
&\hskip 5.cm
- \Big( 1 + \frac{h^\beta - 1}{\cos^2 \theta_W^0} \Big)  \Pi_{\gamma \gamma}^\beta (m_Z^2) \bigg\} ~, \cr
\noalign{\kern 5pt}
\alpha_* T^\beta &= \frac{1}{m_Z^2 \cos^2 \theta_W^0}  \Pi_{WW}^\beta (0) - \frac{1}{m_Z^2}  \Pi_{ZZ}^\beta (0) ~, \cr
\noalign{\kern 5pt}
\alpha_* U^\beta &= \frac{1}{m_Z^2 \cos^2 \theta_W^0}  \Big\{ \Pi_{WW}^\beta (m_W^2) - \Pi_{WW}^\beta (0)  \Big\} \cr
%- \frac{\cos^2 \theta_W^0 }{m_Z^2}  \Big\{ \Pi_{ZZ}^\beta (m_Z^2) - \Pi_{ZZ}^\beta (0)  \Big\} \cr
\noalign{\kern 5pt}
&\hskip -0.5cm
- \frac{\cos^2 \theta_W^0 }{m_Z^2}  \Big\{ \Pi_{ZZ}^\beta (m_Z^2) - \Pi_{ZZ}^\beta (0)  \Big\}
- \frac{\sin 2 \theta_W^0 }{m_Z^2} \Pi_{Z\gamma}^\beta (m_Z^2) 
- \frac{\sin^2 \theta_W^0 }{m_Z^2} \Pi_{\gamma\gamma}^\beta (m_Z^2)
\label{newSTU1}
\end{align}
where $\alpha_* = \alpha_\EM (m_Z^2)$.
In GHU $m_Z \cos \theta_W^0 \not= m_W^{\rm tree}$.
The terms proportional to $h^\beta -1$ in $\alpha_* S^\beta $ represent the improvement from the standard
expression for $S$.  It is straightforward to confirm that $S, T, U$ defined by (\ref{newSTU1}) are finite
as a consequence of the sum rules in (\ref{Sumrule4}).  
In the numerical evaluation of finite $S^\beta$,  $T^\beta$,  $U^\beta$ by using the gauge coupling matrices 
of finite-dimensional rows and columns,  one has to use the $h$ factor defined by $h^\beta = A_0^\beta/ B^\beta$,
otherwise the result would be afflicted  with the uncertainty associated with the divergence.
Also notice that the weak mixing angle $\theta_W^0$ entering in (\ref{newSTU1}) is the angle defined in (\ref{angle2}).

To explicitly express $S^{\beta}$,  $T^{\beta}$ and $U^{\beta}$ in terms of the gauge couplings and mass spectra,
it is convenient to introduce
\begin{align}
\check b_0 (s, m_1, m_2) &= b_0  (s, m_1, m_2) - b_0 (0, m_1, m_2) ~, \cr
\noalign{\kern 5pt}
\check b_1 (s, m_1, m_2) &= b_1  (s, m_1, m_2) - b_1(0, m_1, m_2) ~, \cr
\noalign{\kern 5pt}
J_\pm (m_V, m_1, m_2) 
&= - b_2 (m_V^2, m_1, m_2)   \pm  \frac{m_1 m_2}{2 m_V^2} \, \check b_0 (m_V^2, m_2, m_1)  \cr
\noalign{\kern 5pt}
&+ \frac{m_1^2}{2 m_V^2} \, \check b_1 (m_V^2, m_2, m_1)
+ \frac{m_2^2}{2 m_V^2} \, \check b_1 (m_V^2, m_1, m_2) ~,
\label{loopFn2}
\end{align}
and 
\begin{align}
&H [ \,  \hat g_V , \hat g_A ;  \hat g_{V'} , \hat g_{A'} ;  m_V, m_{1n}, m_{2\ell} \, ] \cr
\noalign{\kern 5pt}
&=
\sum_{n,\ell =0}^\infty \Big\{
\hat g_{V, n\ell} \hat g_{V', \ell n}  J_- (m_V, m_{1n}, m_{2\ell}) 
+ \hat g_{A, n\ell} \hat g_{A', \ell n} J_+ (m_V, m_{1n}, m_{2\ell}) \Big\} ~ , \cr
\noalign{\kern 5pt}
&K [ \,  \hat g_V , \hat g_A ;  \hat g_{V'} , \hat g_{A'} ;  m_{1n}, m_{2\ell} ] \cr
\noalign{\kern 5pt}
&=
\sum_{n,\ell =0}^\infty \Big\{
\big( \hat g_{V, n\ell} \hat g_{V', \ell n}  +  \hat g_{A, n\ell} \hat g_{A', \ell n} \big) 
\big[ m_{1n}^2 b_1 (0, m_{2\ell}, m_{1n} ) + m_{2\ell}^2 b_1 (0, m_{1n} , m_{2\ell} ) \big] \cr
&\hskip 1.3cm
- \big( \hat g_{V, n\ell} \hat g_{V', \ell n}  -  \hat g_{A, n\ell} \hat g_{A', \ell n} \big) 
m_{1n} m_{2\ell} \,  b_0 (0, m_{1n} , m_{2\ell} ) \Big\}  ~ .
\label{finitePiFn}
\end{align}
We note that $b_0 (s, m, m) = 2 b_1 (s, m, m)$.
Then $S^{(u,d)}$ is given by
\begin{align}
\alpha_* S^{(u,d)} &= - \frac{2 N_C \sin^2 \theta_W^0}{\pi^2}\bigg\{
H [ \,  \hat g_V^{W^3 u} , \hat g_A^{Z u} ;  \hat g_{V}^{W^3 u} , \hat g_{A}^{Z u} ;  m_Z, m_{u^{(n)}}, m_{u^{(\ell)}} \, ] \cr
\noalign{\kern 5pt}
&\hskip 2.8cm
+ H [ \,  \hat g_V^{W^3 dd} , \hat g_A^{Z dd} ;  \hat g_{V}^{W^3 dd} , \hat g_{A}^{Z dd} ;  m_Z, m_{d^{(n)}}, m_{d^{(\ell)}} \, ] \cr
\noalign{\kern 5pt}
&\hskip -1.0cm
+ h^{(u,d)}  \sum_{n=0}^\infty \Big( Q_u  \hat g_{V,nn}^{W^3 u} b_2(m_Z^2, m_{u^{(n)}}, m_{u^{(n)}})
+ Q_d  \hat g_{V,nn}^{W^3 dd} b_2(m_Z^2, m_{d^{(n)}}, m_{d^{(n)}}) \Big) \bigg\} .
\label{newS1}
\end{align}
$T^{(u,d)}$ is given by
\begin{align}
\alpha_* T^{(u,d)} &= \frac{N_C}{4\pi^2 \cos^2 \theta_W^0 m_Z^2}  \bigg\{
-\frac{1}{2} K[\,  \hat g_V^{W ud} ,  \hat g_A^{W ud};  \hat g_V^{W^\dagger  ud} ,  \hat g_A^{W^\dagger  ud} ; m_{u^{(n)}},m_{d^{(\ell)}} ]\cr
\noalign{\kern 5pt}
&\hskip 3.cm 
+  K[\,  \hat g_V^{W^3 u} ,  \hat g_A^{Z u};  \hat g_V^{W^3  u} ,  \hat g_A^{Z  u} ; m_{u^{(n)}},m_{u^{(\ell)}}] \cr
\noalign{\kern 5pt}
&\hskip 3.cm 
+  K[\,  \hat g_V^{W^3 dd} ,  \hat g_A^{Z dd};  \hat g_V^{W^3  dd} ,  \hat g_A^{Z dd} ; m_{d^{(n)}},m_{d^{(\ell)}}] \bigg\} .
\label{newT1}
\end{align}
$U^{(u,d)}$ is given by
\begin{align}
\alpha_*U^{(u,d)} &= \frac{N_C}{2\pi^2 }  \bigg\{
-\frac{1}{2} H[\,  \hat g_V^{W ud} ,  \hat g_A^{W ud};  \hat g_V^{W^\dagger  ud} ,  \hat g_A^{W^\dagger  ud} ; 
m_W, m_{u^{(n)}},m_{d^{(\ell)}} ]\cr
\noalign{\kern 5pt}
&\hskip 2.cm 
+  H[\,  \hat g_V^{W^3 u} ,  \hat g_A^{Z u};  \hat g_V^{W^3  u} ,  \hat g_A^{Z  u} ; m_Z,  m_{u^{(n)}},m_{u^{(\ell)}}] \cr
\noalign{\kern 5pt}
&\hskip 2.cm 
+  H[\,  \hat g_V^{W^3 dd} ,  \hat g_A^{Z dd};  \hat g_V^{W^3  dd} ,  \hat g_A^{Z dd} ; m_Z, m_{d^{(n)}},m_{d^{(\ell)}}] \bigg\} .
\label{newU1}
\end{align}
For the lepton doublet $(\nu_e, e)$, $S^{(\nu_e, e)} $ is given by
\begin{align}
\alpha_* S^{(\nu_e, e)} &= - \frac{2 \sin^2 \theta_W^0}{\pi^2}\bigg\{
H [ \,  \hat g_V^{W^3 e} , \hat g_A^{Z e} ;  \hat g_{V}^{W^3 e} , \hat g_{A}^{Z e} ;  m_Z, m_{e^{(n)}}, m_{e^{(\ell)}} \, ] \cr
\noalign{\kern 5pt}
&\hskip 1.cm
+ \sum_{a=1}^2  \sum_{b=1}^2
H [ \,  \hat g_V^{Z \nu_{eab}} , \hat g_A^{Z \nu_{eab}} ;  \hat g_{V}^{Z \nu_{eba}} , \hat g_{A}^{Z  \nu_{eba}} 
;  m_Z, m_{\nu_{ea}^{(n)}}, m_{\nu_{eb}^{(\ell)}} \, ] \cr
\noalign{\kern 5pt}
&\hskip 1.0cm
+ h^{(\nu_e, e)}  \sum_{n=0}^\infty  Q_e  \hat g_{V,nn}^{W^3 e} b_2(m_Z^2, m_{e^{(n)}}, m_{e^{(n)}})
 \bigg\} .
\label{newS2}
\end{align}
$T^{(\nu_e, e)}$ is given by
\begin{align}
\alpha_* T^{(\nu_e, e)} &= \frac{1}{4\pi^2 \cos^2 \theta_W^0 m_Z^2}  \bigg\{
-\frac{1}{2} \sum_{a=1}^2
K[\,  \hat g_V^{W \nu_{ea} e} ,  \hat g_A^{W \nu_{ea} e};  \hat g_V^{W^\dagger  \nu_{ea} e} ,  \hat g_A^{W^\dagger  \nu_{ea} e} 
; m_{\nu_{ea}^{(n)}},m_{e^{(\ell)}} ]\cr
\noalign{\kern 5pt}
&\hskip 3.cm 
+  K[\,  \hat g_V^{W^3 e} ,  \hat g_A^{Z e};  \hat g_V^{W^3  e} ,  \hat g_A^{Z  e} ; m_{e^{(n)}},m_{e^{(\ell)}}] \cr
\noalign{\kern 5pt}
&\hskip 3.cm 
+   \sum_{a=1}^2  \sum_{b=1}^2
K[\,  \hat g_V^{Z \nu_{eab}} ,  \hat g_A^{Z \nu_{eab}};  \hat g_V^{Z \nu_{eba}} ,  \hat g_A^{Z \nu_{eba}} 
; m_{\nu_{ea}^{(n)}},m_{\nu_{eb}^{(\ell)}}] \bigg\} .
\label{newT2}
\end{align}
$U^{(\nu_e, e)}$ is given by
\begin{align}
\alpha_*U^{(\nu_e, e)} &= \frac{1}{2\pi^2 }  \bigg\{
-\frac{1}{2}  \sum_{a=1}^2 
H[\,  \hat g_V^{W \nu_{ea} e} ,  \hat g_A^{W \nu_{ea} e};  \hat g_V^{W^\dagger \nu_{ea} e} ,  \hat g_A^{W^\dagger \nu_{ea} e} ; 
m_W, m_{\nu_{ea}^{(n)}},m_{e^{(\ell)}} ]\cr
\noalign{\kern 5pt}
&\hskip 2.cm 
+  H[\,  \hat g_V^{W^3 e} ,  \hat g_A^{Z e};  \hat g_V^{W^3  e} ,  \hat g_A^{Z  e} ; m_Z,  m_{e^{(n)}},m_{e^{(\ell)}}] \cr
\noalign{\kern 5pt}
&\hskip 2.cm 
+  \sum_{a=1}^2  \sum_{b=1}^2
H[\,  \hat g_V^{Z \nu_{eab}} ,  \hat g_A^{Z \nu_{eab}};  \hat g_V^{Z \nu_{eba}} ,  \hat g_A^{Z \nu_{eba}} 
; m_Z, m_{\nu_{ea}^{(n)}},m_{\nu_{eb}^{(\ell)}}] \bigg\} .
\label{newU2}
\end{align}
Formulas for the quark-lepton multiplets in the second and third generations are obtained similarly.

We have evaluated the improved $S$, $T$, $U$ described above from the gauge coupling matrices 
determined in the space of the KK modes of $n=0$ to $n_{\rm max} =12$ levels.  In the evaluation the usage of the 
identity (\ref{loopPi5}) reduces numerical errors.  In the combinations of the above $S^\beta$, $T^\beta$, $U^\beta$  
the sum of the $K^{\rm div} \ln \mu^2$ part in (\ref{loopPi5}) vanishes thanks to the coupling sum rules in (\ref{Sumrule4}).
In Table \ref{Table:STU1} we have tabulated the values of $S^\beta$, $T^\beta$, $U^\beta$ beyond the SM contributions
for $\theta_H=0.1$, $m_\KK = 13\,$TeV  and $M=10^3\,$TeV. 
% It is seen that the corrections are rather small in the GUT inspired GHU.  
The total values are $S \sim 0.01$, $T \sim 0.12$ and $U \sim 0.00004$ when the contributions of fermions up to 
the $n_{\rm max} =12$ level are taken into account.

\begin{table}[tbh]
{%\small
\renewcommand{\arraystretch}{1.3}
\begin{center}
\caption{Corrections to $S^\beta$, $T^\beta$, $U^\beta$ for $\theta_H=0.1$ and $m_\KK = 13\,$TeV.
The numerical values are evaluated by including the contributions coming from the KK towers of
fermions up to the $n=n_{\rm max} =12$ level.
The values in the neutrino sector are obtained by setting $m_{\nu_e} = m_{\nu_\mu}  = m_{\nu_\tau}  = 10^{-12}\,$GeV
and Majorana masses $M_e = M_\mu = M_\tau = 10^6\,$GeV.  
In the last row the average increments per level, namely $({\rm total})/n_{\rm max}$, are listed.
% The value $S^{(\nu_\tau ,\tau})$  depends on  $m_{\nu_\tau}$ and $M_\tau$.  
% The value can be lowered without affecting $T^{(\nu_\tau, \tau)}$ and $U^{(\nu_\tau , \tau)}$ very much.
}
\vskip 10pt
\begin{tabular}{|c|c|c|c|}
%\hline
%&GUT inspired model\\
%& type B &type A\\
\hline
$\beta$ &$S^\beta$  &$T^\beta$ &$U^\beta$ \\
\hline
\noalign{\kern 1pt}
\hline 
$(\nu_{e}, e)$ &$0.0010$ &$0.0126$  &$3.7 \times 10^{-6}$ \\
\hline
$(\nu_{\mu}, \mu)$ &$0.0009$  &$0.0122$  &$3.7 \times 10^{-6}$ \\
\hline 
$(\nu_{\tau}, \tau)$ &$0.0016$  &$0.0129$ &$3.7 \times 10^{-6}$ \\
\hline
% \noalign{\kern 1pt}
% \hline
$(u,d)$ &$0.0028$  &$0.0382$  &$1.1 \times 10^{-5}$ \\
\hline
$(c,s)$& $0.0026$  &$0.0360$  &$1.1 \times 10^{-5}$ \\
\hline 
$(t,b)$ &$0.0013$  &$0.0058$  & $7.9 \times 10^{-6}$\\
\hline
\hline 
total &$0.010$  &$0.12$  & $0.00004$\\
\hline
\hline 
per level&$8.4 \times 10^{-4}$  &$9.7 \times 10^{-3}$  & $3.4 \times 10^{-6}$\\
\hline
\end{tabular}
\label{Table:STU1}
\end{center}
}
\end{table}

Unlike the case of the coupling sum rules, however,  the parameters $S$, $T$, $U$ evaluated in this manner 
seem to increase as $n_{\rm max} $ is increased.  Let us denote these parameters as $S^{(0)}$, $T^{(0)}$, $U^{(0)}$ 
to stress that they are oblique parameters associated with the zero modes $W^{(0)}$, $Z^{(0)}$ and $\gamma^{(0)}$ of the
gauge fields.   The average increments in  $S^{(0)}$, $T^{(0)}$, $U^{(0)}$  per level are small (as listed in the last row
in Table \ref{Table:STU1}),  but do not vary very much in the range  $12 \le n_{\rm max} \le 16$.
This does not necessarily mean that oblique corrections to physical quantities become large in GHU.
The current estimates of $S, T, U$ from experimental data in the SM framework  are $S^{\rm RPP} = - 0.02 \pm 0.10$, 
$T^{\rm RPP}= 0.03 \pm 0.12$ and  $U^{\rm RPP} = 0.01 \pm 0.11$
where the superscript RPP stands for Review of Particle Physics.\cite{pdg2022chap10}
As emphasized in the beginning of this section, oblique corrections associated not only with $W^{(0)}$, $Z^{(0)}$ and
$\gamma^{(0)}$, but also with the KK excited modes $W^{(n)}$, $Z^{(n)}$ and $\gamma^{(n)}$ become
important for physical observable quantities in GHU.  In particular, the  couplings of left-handed  quarks and leptons to
$W^{(1)}$, $Z^{(1)}$ and $\gamma^{(1)}$ are large in the GUT inspired GHU.  
To compare with  $S^{\rm RPP}$, $T^{\rm RPP}$ and $U^{\rm RPP}$, one needs to include,
in addition to $S^{(0)}$, $T^{(0)}$ and $U^{(0)}$, oblique corrections to the propagators of  the KK gauge bosons.
Contributions coming from internal fermions at the high KK levels equally affect the oblique corrections
to the KK gauge boson propagators.
To have definitive understanding of the contributions of KK fermions at the one loop level in GHU, 
it is necessary to directly evaluate observable quantities, which is left for future investigation.

\begin{figure}[tbh]
\centering
\includegraphics[width=110mm]{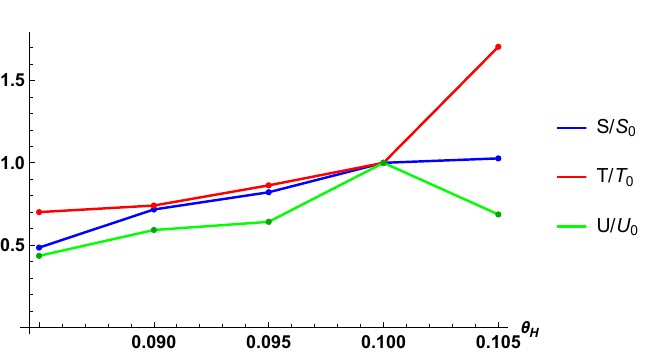}
\caption{The $\theta_H$-dependence of the $S$, $T$, $U$ parameters is plotted for $m_\KK = 13\,$TeV and
$n_\max = 12$.  $S_0$, $T_0$ and $U_0$ are the values of  $S$, $T$ and $U$ at $\theta_H = 0.1$, respectively.
}   
\label{fig:STU-theta-dep}
\end{figure}

So far we have presented the results for $\theta_H = 0.1$ with $m_\KK = 13\,$TeV.  The $\theta_H$-dependence of 
the oblique parameters is explored similarly.  It should be noted that with  $m_\KK = 13\,$TeV fixed, the value of 
$\theta_H$ can be lowered up to $\theta_H^{\rm min} \sim 0.08$ to reproduce the mass of the top quark.
To realize smaller values of $\theta_H$, one needs to increase $m_\KK$.
One expects that the oblique corrections should get smaller as $\theta_H$ gets smaller.
Indeed this is the case.  The $\theta_H$-dependence of $S$, $T$ and $U$ are depicted in the range
$0.085 \le \theta_H \le 0.105$  in Fig.\ \ref{fig:STU-theta-dep}  for  $m_\KK = 13\,$TeV and $n_{\rm max} = 12$.
For larger values of $\theta_H$ the oblique corrections get larger.  It is anticipated from  the viewpoint
that as  $\theta_H$ varies from 0 to $ \pi$,  the gauge symmetry changes from
$SU(2)_L \times U(1)_Y$ to  $SU(2)_R \times U(1)_{Y'}$.
The reliable numerical evaluation of the $Z$ couplings, particularly  in the bottom quark tower, 
becomes harder for larger values of $\theta_H$ because of the singular behavior of the $h_{L/R}^{b^{(n)}}$ and 
$k_{L/R}^{b^{(n)}}$ components of the wave functions near the UV brane at $z=1$.

%%%%%%%%%%%%%%%%%%%%
%%%%%%%%%%%%%%%%%%%

\section{Summary and discussions} 

In this paper we have examined the GUT inspired $SO(5) \times U(1)_X \times SU(3)_C$ GHU model in the RS warped space.
The $W$ and $Z$ couplings of quarks, leptons and their KK excited modes take the matrix form in the KK space.  
These coupling matrices have nontrivial off-diagonal elements, and have both vector and axial-vector components.
Nevertheless these coupling matrices satisfy three sum rules (\ref{Sumrule4}).  We have confirmed these coupling sum rules
numerically from the evaluated  $W$ and $Z$ coupling matrices.
The rigorous derivation of the coupling sum rules would require the full treatment of the gauge bosons in the $SO(5) \times U(1)_X$
theory.  It is noteworthy that the sum rules hold even in the subspace of the $W$, $Z$ and photon vacuum polarization tensors
to very high accuracy.
The appearance of the $h^\beta \not= 1$ factor in the relation $A_0^\beta = h^\beta B^\beta$ in  (\ref{Sumrule4}) is anticipated 
from the vertex correction in the $Z$ couplings at the tree level as exhibited in the approximate formula in (\ref{Zcoupling5h}).

With the coupling sum rules at hand, one can evaluate the finite oblique corrections unambiguously.  
The corrections are evaluated by using the mass spectrum and gauge coupling matrices determined numerically.
We have found for $\theta_H=0.1$ and $m_\KK = 13\,$TeV that $ S \sim 0.01$, $ T \sim 0.12$ 
and $ U \sim 0.00004$ when the contributions of the fermion loops up to the $n_{\rm max}=12$ level are taken into account.
It was argued in the very early stage of the investigation \cite{Carena2006} that there may arise a large correction to $S$ 
 in gauge theory in the RS space.   We have found that the corrections in the GUT inspired GHU are small 
 by direct evaluation of one loop diagrams.
We note that Yoon and Peskin have evaluated the oblique corrections in a different $SO(5) \times U(1)_X$ GHU model
in a different method.\cite{Yoon2018b}  Their result also indicates small corrections for $m_\KK = 13\,$TeV.
However, to have definitive understanding of the contributions of KK fermions at the one loop level, 
it is necessary to evaluate observable quantities, by taking account of oblique corrections to the KK modes of the
gauge fields.

The coupling sum rules presented in this paper are highly nontrivial.  There must be some reason behind them, possibly
originating from the 5D gauge invariance in the GHU scheme.  Further investigation is necessary.

\section*{Acknowledgements}

One of the authors (Y.H.) would like to thank Masashi Aiko for many deep comments on the
oblique corrections.  This work is supported in part  by Japan Society for the Promotion of Science, Grants-in-Aid for Scientific 
Research, Grant No. JP19K03873 (Y.H.), 
European Regional Development Fund-Project Engineering Applications of Microworld Physics 
(No.\ CZ.02.1.01/0.0/0.0/16 019/0000766) (Y.O.),
and  the Ministry of Science and Technology of Taiwan under Grant No.\ MOST-111-2811-M-002-047-MY2 (N.Y.).

\vskip 1.cm

%%%%%%%%%%%%%%%%%%%%%%%%%%%%%

\appendix

\section{Basis functions} 

We  summarize the basis functions used for wave functions of gauge and fermion fields.
For gauge fields we introduce
\begin{align}
 F_{\alpha, \beta}(u, v) &\equiv J_\alpha(u) Y_\beta(v) - Y_\alpha(u) J_\beta(v) ~, \cr
\noalign{\kern 5pt}
 C(z; \lambda) &= \frac{\pi}{2} \lambda z z_L F_{1,0}(\lambda z, \lambda z_L) ~,  \cr
 S(z; \lambda) &= -\frac{\pi}{2} \lambda  z F_{1,1}(\lambda z, \lambda z_L) ~, \cr
 C^\prime (z; \lambda) &= \frac{\pi}{2} \lambda^2 z z_L F_{0,0}(\lambda z, \lambda z_L) ~,  \cr
S^\prime (z; \lambda) &= -\frac{\pi}{2} \lambda^2 z  F_{0,1}(\lambda z, \lambda z_L)~, \cr
\noalign{\kern 5pt}
\hat  S(z; \lambda) & =  \frac{C(1; \lambda)}{S(1; \lambda)} \,  S(z; \lambda) ~,
\label{functionA1}
\end{align}
where $J_\alpha (u)$ and $Y_\alpha (u)$ are Bessel functions of  the first and second kind.
They satisfy
\begin{align}
&- z \frac{d}{dz} \frac{1}{z} \frac{d}{dz} \begin{pmatrix} C \cr S \end{pmatrix} 
= \lambda^{2} \begin{pmatrix} C \cr S \end{pmatrix} ~,  
\label{relationA1}
\end{align}
with the boundary conditions $C(z_{L} ; \lambda)  = z_{L}$, $C' (z_{L} ; \lambda)  =S(z_{L} ; \lambda)  = 0 $, 
$S' (z_{L} ; \lambda)  = \lambda$, and $CS' - S C' = \lambda z$.
%We also make use of $\hat  S(z; \lambda) =  C(1; \lambda) S(z; \lambda) /S(1; \lambda)$.

For fermion fields with a bulk mass parameter $c$, we define 
\begin{align}
\begin{pmatrix} C_L \cr S_L \end{pmatrix} (z; \lambda,c)
&= \pm \frac{\pi}{2} \lambda \sqrt{z z_L} F_{c+\frac12, c\mp\frac12}(\lambda z, \lambda z_L) ~, \cr
\begin{pmatrix} C_R \cr S_R \end{pmatrix} (z; \lambda,c)
&= \mp \frac{\pi}{2} \lambda \sqrt{z z_L} F_{c- \frac12, c\pm\frac12}(\lambda z, \lambda z_L) ~, \cr
%\noalign{\kern 5pt}
\begin{pmatrix} \hat S_L \cr \hat C_R \end{pmatrix} (z; \lambda,c)
&= \frac{C_L (1; \lambda,c)}{S_L (1; \lambda,c)}  \begin{pmatrix} S_L \cr C_R \end{pmatrix} (z; \lambda,c)~ , \cr
%\noalign{\kern 5pt}
\begin{pmatrix} \hat S_R \cr \hat C_L \end{pmatrix} (z; \lambda,c)
&= \frac{C_R (1; \lambda,c)}{S_R (1; \lambda,c)}  \begin{pmatrix} S_R \cr C_L \end{pmatrix} (z; \lambda,c)~ .
\label{functionA2}
\end{align}
These functions satisfy 
\begin{align}
&D_{+} (c) \begin{pmatrix} C_{L} \cr S_{L} \end{pmatrix} = \lambda  \begin{pmatrix} S_{R} \cr C_{R} \end{pmatrix}, \cr
\noalign{\kern 5pt}
&D_{-} (c) \begin{pmatrix} C_{R} \cr S_{R} \end{pmatrix} = \lambda  \begin{pmatrix} S_{L} \cr C_{L} \end{pmatrix}, ~~
D_{\pm} (c) = \pm \frac{d}{dz} + \frac{c}{z} ~, 
\label{relationA2}
\end{align}
with the boundary conditions $C_{R/L} =1$, $S_{R/L} = 0$ at $z=z_{L} $, and 
$C_L C_R - S_L S_R=1$.
We also use
\begin{align}
{\cal C}_{L1}(z; \lambda, c, \tilde m) &= C_L(z; \lambda, c+\tilde{m})+C_L(z; \lambda, c-\tilde{m}) ~, \cr
{\cal C}_{L2}(z; \lambda, c, \tilde m) &= S_L(z; \lambda, c+\tilde{m})-S_L(z; \lambda, c-\tilde{m}) ~, \cr
{\cal S}_{L1}(z; \lambda, c, \tilde m) &= S_L(z; \lambda, c+\tilde{m})+S_L(z; \lambda,c-\tilde{m}) ~, \cr
{\cal S}_{L2}(z; \lambda, c, \tilde m) &= C_L(z; \lambda, c+\tilde{m})-C_L(z; \lambda, c-\tilde{m}) ~, \cr
{\cal C}_{R1}(z; \lambda, c, \tilde m) &= C_R(z; \lambda, c+\tilde{m})+C_R(z; \lambda, c-\tilde{m}) ~, \cr
{\cal C}_{R2}(z; \lambda, c, \tilde m) &= S_R(z; \lambda, c+\tilde{m})-S_R(z; \lambda,c-\tilde{m}) ~, \cr
{\cal S}_{R1}(z; \lambda, c, \tilde m) &= S_R(z; \lambda, c+\tilde{m})+S_R(z; \lambda, c-\tilde{m}) ~, \cr
{\cal S}_{R2}(z; \lambda, c, \tilde m) &= C_R(z; \lambda, c+\tilde{m})-C_R(z; \lambda, c-\tilde{m}) ~.
\label{functionA3}
\end{align}

\section{Wave functions} 

Wave functions of down-type quark and neutrino multiplets are given below.

\subsection{Down-type quarks} 

The $(d, d', D^+, D^-)$ fields are expanded as in (\ref{wave-down1}). The wave functions are given by
\begin{align}
&\begin{pmatrix} f^{d^{(n)}}_L (z) \cr \mynoalign g^{d^{(n)}}_L (z) \cr \mynoalign 
h^{d^{(n)}}_L (z) \cr \mynoalign k^{d^{(n)}}_L (z)\end{pmatrix} =
\frac{1}{ \sqrt{r_{d^{(n)}}} }
\begin{pmatrix}  \alpha_{d^{(n)}} C_L(z; \lambda_{d^{(n)}} , c_u)\cr
\mynoalign
\beta_{d^{(n)}}S_L(z; \lambda_{d^{(n)}} , c_u)\cr
\mynoalign
a_{d^{(n)}}{\cal C}_{L2}(z;\lambda_{d^{(n)}} , c_{D_d}, \tilde m_{D_d})
+b_{d^{(n)}}{\cal C}_{L1}(z;\lambda_{d^{(n)}}, c_{D_d}, \tilde m_{D_d}) \cr
\mynoalign
a_{d^{(n)}}{\cal S}_{L1}(z;\lambda_{d^{(n)}}, c_{D_d}, \tilde m_{D_d})
+b_{d^{(n)}}{\cal S}_{L2}(z;\lambda_{d^{(n)}}, c_{D_d}, \tilde m_{D_d}) \cr  \end{pmatrix} \cr
\noalign{\kern 5pt}
&\begin{pmatrix} f^{d^{(n)}}_R (z) \cr \mynoalign g^{d^{(n)}}_R (z) \cr \mynoalign 
h^{d^{(n)}}_R (z) \cr \mynoalign k^{d^{(n)}}_R (z)\end{pmatrix} =
\frac{1}{ \sqrt{r_{d^{(n)}}} }
\begin{pmatrix}  \alpha_{d^{(n)}} S_R(z; \lambda_{d^{(n)}} , c_u)\cr
\mynoalign
\beta_{d^{(n)}}C_R(z; \lambda_{d^{(n)}} , c_u)\cr
\mynoalign
a_{d^{(n)}}{\cal S}_{R2}(z;\lambda_{d^{(n)}} , c_{D_d}, \tilde m_{D_d})
+b_{d^{(n)}}{\cal S}_{R1}(z;\lambda_{d^{(n)}}, c_{D_d}, \tilde m_{D_d}) \cr
\mynoalign
a_{d^{(n)}}{\cal C}_{R1}(z;\lambda_{d^{(n)}}, c_{D_d}, \tilde m_{D_d})
+b_{d^{(n)}}{\cal C}_{R2}(z;\lambda_{d^{(n)}}, c_{D_d}, \tilde m_{D_d}) \cr  \end{pmatrix}
\label{wave-down2}
\end{align}
for the $d^{(n)}$ mode where
\begin{align}
\beta_{d^{(n)}} &= 
- i \frac{\cos\onehalf \theta_H S_R(1,\lambda_{d^{(n)}} , c_u) }{\sin \onehalf \theta_H C_R(1,\lambda_{d^{(n)}} , c_u)} \, 
 \alpha_{d^{(n)}} ~, \cr
 \noalign{\kern 5pt}
a_{d^{(n)}} &= i \, \frac{\mu_d S_R(1,\lambda_{d^{(n)}} , c_u)}{\sin \onehalf \theta_H} \,
\frac{{\cal S}_{L2}(1 ;\lambda_{d^{(n)}}, c_{D_d}, \tilde m_{D_d})}{{\cal F}_1 (1 ;\lambda_{d^{(n)}}, c_{D_d}, \tilde m_{D_d})} \, 
 \alpha_{d^{(n)}} ~, \cr
\noalign{\kern 5pt}
b_{d^{(n)}} &= - 
\frac{ {\cal S}_{L1}(1 ;\lambda_{d^{(n)}}, c_{D_d}, \tilde m_{D_d})} {{\cal S}_{L2}(1 ;\lambda_{d^{(n)}}, c_{D_d}, \tilde m_{D_d})} \,
a_{d^{(n)}} ~, \cr
\noalign{\kern 5pt}
{\cal F}_1 &=  {\cal S}_{L1}  {\cal S}_{R1} - {\cal S}_{L2}  {\cal S}_{R2} ~.
\label{wave-down3}
\end{align}
For the $D^{(n)}$ mode the formulas are obtained by replacing $d^{(n)}$ by $D^{(n)}$ in (\ref{wave-down2}) and 
(\ref{wave-down3}).

Except for the $d^{(0)}$ mode, namely $d$-quark, the $d^{(n)}$ ($n \ge 1$) modes are mostly contained in $(d, d')$ fields,
whereas the $D^{(n)}$ ($n \ge 1$) modes are mostly contained in $(D^+, D^-)$ fields. 
In Table \ref{Table:downnorm1} the norm of each component ($N_f = \int_1^{z_L}  dz \, |f|^2$ etc.) is tabulated.
For comparison we list the norms of $u$ and $u'$ components of the $u^{(n)}$ modes in Table \ref{Table:upnorm1}.

\begin{table}[tbh]
{%\small
\renewcommand{\arraystretch}{1.2}
\begin{center}
\caption{The norm of each component for the $d^{(n)}$ and $D^{(n)}$ modes.
$N_f = \int_1^{z_L}  dz \, |f|^2$, etc.
}
\vskip 10pt
\begin{tabular}{|c|c|c|c|c|}
%\hline
%&GUT inspired model\\
%& type B &type A\\
\hline
&$N_f \,(d)$ &$N_g \,(d')$ &$N_h  \,(D^+)$ &$N_k  \,(D^-)$ \\
\hline
\hline
$d_L^{(0)}$ &$1. $ &$ 1 \times 10^{-23}$ &$3 \times 10^{-14}$ &$2 \times 10^{-14}$\\
\hline 
$d_L^{(1)}$ &$2 \times 10^{-22} $ &$ 1.$ &$1 \times 10^{-10}$ &$1 \times 10^{-11}$\\
\hline
$d_L^{(2)}$ &$1. $ &$ 7 \times 10^{-22}$ &$1 \times 10^{-12}$ &$1 \times 10^{-12}$\\
\hline 
$d_L^{(3)}$ &$8 \times 10^{-22} $ &$ 1.$ &$3 \times 10^{-10}$ &$2 \times 10^{-10}$\\
\hline
$d_L^{(4)}$ &$1. $ &$ 2 \times 10^{-21}$ &$2 \times 10^{-12}$ &$1 \times 10^{-12}$\\
\hline
\hline
$d_R^{(0)}$ &$5 \times 10^{-5} $ &$ 0.021$ &$0.387$ &$0.592$\\
\hline 
$d_R^{(1)}$ &$1 \times 10^{-14} $ &$ 1.$ &$2  \times 10^{-10}$ &$1 \times 10^{-10}$\\
\hline
$d_R^{(2)}$ &$1. $ &$ 3 \times 10^{-14}$ &$2 \times 10^{-12}$ &$8 \times 10^{-13}$\\
\hline 
$d_R^{(3)}$ &$2 \times 10^{-14} $ &$ 1.$ &$3 \times 10^{-10}$ &$4 \times 10^{-10}$\\
\hline
$d_R^{(4)}$ &$1. $ &$ 3 \times 10^{-14}$ &$3 \times 10^{-12}$ &$8 \times 10^{-13}$\\
\hline
\hline
$D_L^{(1)}$ &$9 \times 10^{-14} $ &$ 8 \times 10^{-11}$ &$0.697$ &$0.303$\\
\hline
$D_L^{(2)}$ &$2 \times 10^{-12} $ &$ 7 \times 10^{-11}$ &$0.633$ &$0.367$\\
\hline
$D_L^{(3)}$ &$2  \times 10^{-13} $ &$ 3  \times 10^{-10}$ &$0.284$ &$0.716$\\
\hline
$D_L^{(4)}$ &$3 \times 10^{-12} $ &$ 1 \times 10^{-10}$ &$0.781$ &$0.219$\\
\hline
\hline
$D_R^{(1)}$ &$2 \times 10^{-5} $ &$ 7  \times 10^{-3}$ &$0.125$ &$0.868$\\
\hline
$D_R^{(2)}$ &$7  \times 10^{-6} $ &$ 3 \times 10^{-3}$ &$0.952$ &$0.045$\\
\hline
$D_R^{(3)}$ &$6 \times 10^{-6} $ &$ 2  \times 10^{-3}$ &$0.034$ &$0.964$\\
\hline
$D_R^{(4)}$ &$5 \times 10^{-6} $ &$ 2 \times 10^{-3}$ &$0.973$ &$0.025$\\
\hline
\end{tabular}
\label{Table:downnorm1}
\end{center}
}
\end{table}

\begin{table}[tbh]
{%\small
\renewcommand{\arraystretch}{1.2}
\begin{center}
\caption{The norm of  the $u^{(n)}$ modes.
$N_f = \int_1^{z_L}  dz \, |f|^2$,  etc.
}
\vskip 10pt
\begin{tabular}{|c|c|c|}
%\hline
%&GUT inspired model\\
%& type B &type A\\
\hline
&$N_f \,(u)$ &$N_g \,(u')$ \\
\hline
$u_L^{(0)}$ &$1. $ &$ 2  \times 10^{-20}$\\
\hline
$u_L^{(1)}$ &$2 \times 10^{-19} $ &$1.$\\
\hline 
$u_L^{(2)}$ &$1. $ &$ 3  \times 10^{-19}$\\
\hline
$u_L^{(3)}$ &$6 \times 10^{-19} $ &$1.$\\
\hline
$u_L^{(4)}$ &$1. $ &$ 8 \times 10^{-19}$\\
\hline 
\end{tabular}
\hskip 5pt %\quad
\begin{tabular}{|c|c|c|}
%\hline
%&GUT inspired model\\
%& type B &type A\\
\hline
 &$N_f  \,(u)$ &$N_g \,(u')$ \\
\hline
$u_R^{(0)}$ &$0.002$ &$0.998$\\
\hline
$u_R^{(1)}$ &$2 \times 10^{-11} $ &$1.$\\
\hline 
$u_R^{(2)}$ &$1. $ &$ 1 \times 10^{-11}$\\
\hline
$u_R^{(3)}$ &$1 \times 10^{-11} $ &$1.$\\
\hline
$u_R^{(4)}$ &$1. $ &$ 1 \times 10^{-11}$\\
\hline 
\end{tabular}
\label{Table:upnorm1}
\end{center}
}
\end{table}

One can see that $(u_L^{(0)}, d_L^{(0)})$ is an $SU(2)_L$ doublet.
On the other hand $u_R^{(0)}$ and $d_R^{(0)})$ are nearly $SU(2)_L$ singlets.   
Further $d_R^{(0)})$ has major components in the $D^\pm$ fields. Its $SU(2)_R$ portion is small. %only 2.1\%.
Although the $W$ boson acquires a small $SU(2)_R$ portion %(about $6.2 \times 10^{-6}$) 
at $\theta_H =0.1$,  its coupling to $(u_R^{(0)}, d_R^{(0)})$ is suppressed significantly.

\subsection{Neutrinos} 

The $(\nu, \nu', \hat \chi)$ fields are expanded as in (\ref{wave-neutrino1}).   The wave functions are given by
\begin{align}
\begin{pmatrix} f_L^{\nu^{\pm (n)}} (z) \cr \mynoalign g_L^{\nu^{\pm(n)}} (z)  \cr\mynoalign
f_R^{\nu^{\pm (n)}} (z) \cr \mynoalign g_R^{\nu^{\pm(n)}} (z)\end{pmatrix} 
&= \frac{1}{\sqrt{ r_{\nu^{\pm (n)}}}}
\begin{pmatrix} 
\sin \onehalf \theta_H C_L(z;\lambda_{\nu^{\pm (n)}}, c_e)/ S_R(1; \lambda_{\nu^{\pm (n)}}, c_e) \cr \mynoalign
- i \cos \onehalf \theta_H S_L(z;\lambda_{\nu^{\pm (n)}}, c_e)/ C_R(1; \lambda_{\nu^{\pm (n)}}, c_e) \cr \mynoalign
\sin \onehalf \theta_H S_R(z;\lambda_{\nu^{\pm (n)}}, c_e)/ S_R(1; \lambda_{\nu^{\pm (n)}}, c_e) \cr \mynoalign
-i \cos \onehalf \theta_H C_L(z;\lambda_{\nu^{\pm (n)}}, c_e)/ C_R(1; \lambda_{\nu^{\pm (n)}}, c_e) 
\end{pmatrix} , \cr
\noalign{\kern 5pt}
h^{\nu^{\pm (n)}} &=  \frac{1}{\sqrt{ r_{\nu^{\pm (n)}}}} \frac{- i m_B}{k \lambda_{\nu^{\pm (n)}} \mp M} ~.
\label{wave-neutrino2}
\end{align}
The normalization factor $r_{\nu^{\pm (n)}}$ is determined by (\ref{norm-neutrino1}).

The $\nu^{+ (0)}$ mode is nearly left-handed,  saturated with $\nu_L$.  
$\nu^{\pm (n)}$ ($n \ge 1$) modes are almost vector-like.
$\nu^{\pm (2\ell -1)}$ ($\ell \ge 1$) modes are saturated by $\nu_L$ and $\nu_R$,
whereas $\nu^{\pm (2\ell )}$ ($\ell \ge 1$) modes are saturated by $\nu_L'$ and $\nu_R'$.
As an example we take $m_{\nu_e} = 10^{-3}\,$eV and $M_e = 10^6\,$GeV, which gives $m_{Be} = 4.8 \times 10^5\,$GeV.
The norm of each component is tabulated in Table \ref{Table:nu-norm1}.

\begin{table}[tbh]
{%\small
\renewcommand{\arraystretch}{1.2}
\begin{center}
\caption{The norm of  the $\nu^{\pm (n)}$ modes.
$N_f = \int_1^{z_L}  dz \, |f|^2$,  etc.
}
\vskip 10pt
\begin{tabular}{|c|c|c|c|c|c|}
%\hline
%&GUT inspired model\\
%& type B &type A\\
\hline
&$N_{f_L} \,(\nu)$ &$N_{g_L} \,(\nu')$ &$N_{f_R} \,(\nu)$ &$N_{g_R} \,(\nu')$ &$N_{h} $\\
\hline
\hline
$\nu^{+ (0)}$ &$1. $ &$ 6 \times 10^{-61}$ &$1 \times 10^{-20}$ &$4 \times 10^{-18}$ &$1 \times 10^{-18}$ \\
\hline
$\nu^{+ (1)}$ &$1 \times 10^{-28} $ &$0.5$ &$1 \times 10^{-17} $ &$0.5$  &$1 \times 10^{-15} $\\
\hline 
$\nu^{+ (2)}$ &$0.5$ &$ 9 \times 10^{-28}$ &$0.5$ &$ 5 \times 10^{-17}$ &$ 1 \times 10^{-17}$\\
\hline
$\nu^{+ (3)}$ &$2 \times 10^{-27} $ &$0.5$ &$4 \times 10^{-17} $ &$0.5$  &$4  \times 10^{-15} $\\
\hline 
$\nu^{+ (4)}$ &$0.5$ &$7 \times 10^{-26}$ &$0.5$ &$ 1 \times 10^{-15}$ &$ 3 \times 10^{-16}$\\
\hline 
\hline
$\nu^{- (1)}$ &$2 \times 10^{-28} $ &$0.5$ &$2 \times 10^{-17} $ &$0.5$  &$2 \times 10^{-15} $\\
\hline 
$\nu^{- (2)}$ &$0.5$ &$ 1 \times 10^{-27}$ &$0.5$ &$ 5 \times 10^{-17}$ &$ 1 \times 10^{-17}$\\
\hline
$\nu^{- (3)}$ &$3 \times 10^{-27} $ &$0.5$ &$8 \times 10^{-17} $ &$0.5$  &$7 \times 10^{-15} $\\
\hline 
$\nu^{- (4)}$ &$0.5$ &$ 2 \times 10^{-26}$ &$0.5$ &$ 3 \times 10^{-16}$ &$ 7 \times 10^{-17}$\\
\hline 
\end{tabular}
\label{Table:nu-norm1}
\end{center}
}
\end{table}

\vskip 1.cm

% A useful Journal macro
%\def\jnl#1#2#3#4{{#1}{\bf #2} (#4) #3}
\def\jnl#1#2#3#4{{#1}{\bf #2},  #3 (#4)}

\def\Zphys{{\em Z.\ Phys.} }
\def\jssc{{\em J.\ Solid State Chem.\ }}
\def\jpsJ{{\em J.\ Phys.\ Soc.\ Japan }}
\def\ptps{{\em Prog.\ Theoret.\ Phys.\ Suppl.\ }}
\def\PTP{{\em Prog.\ Theoret.\ Phys.\  }}
\def\PTEP{{\em Prog.\ Theoret.\ Exp.\  Phys.\  }}
\def\JMP{{\em J. Math.\ Phys.} }
\def\NPB{{\em Nucl.\ Phys.} B}
\def\NP{{\em Nucl.\ Phys.} }
\def\PLB{{\it Phys.\ Lett.} B}
\def\PL{{\em Phys.\ Lett.} }
\def\PRL{\em Phys.\ Rev.\ Lett. }
\def\PRB{{\em Phys.\ Rev.} B}
\def\PRD{{\em Phys.\ Rev.} D}
\def\PRe{{\em Phys.\ Rep.} }
\def\AP{{\em Ann.\ Phys.\ (N.Y.)} }
\def\RMP{{\em Rev.\ Mod.\ Phys.} }
\def\ZPC{{\em Z.\ Phys.} C}
\def\SCI{\em Science}
\def\CMP{\em Comm.\ Math.\ Phys. }
\def\MPLA{{\em Mod.\ Phys.\ Lett.} A}
\def\IJMPA{{\em Int.\ J.\ Mod.\ Phys.} A}
\def\IJMPB{{\em Int.\ J.\ Mod.\ Phys.} B}
\def\EPJC{{\em Eur.\ Phys.\ J.} C}
\def\PR{{\em Phys.\ Rev.} }
\def\JHEP{{\em JHEP} }
\def\JCAP{{\em JCAP} }
\def\cmp{{\em Com.\ Math.\ Phys.}}
\def\JPA{{\em J.\  Phys.} A}
\def\JPG{{\em J.\  Phys.} G}
\def\NJP{{\em New.\ J.\  Phys.} }
\def\CQG{\em Class.\ Quant.\ Grav. }
\def\ATMP{{\em Adv.\ Theoret.\ Math.\ Phys.} }
\def\ibid{{\em ibid.} }
\def\ChP{{\em Chin.Phys.}C}
\def\NCA{{\it Nuovo Cim.} A}

%%%%%%%%%%5

\renewenvironment{thebibliography}[1]
         {\begin{list}{[$\,$\arabic{enumi}$\,$]}  % {\arabic{enumi}.}
         {\usecounter{enumi}\setlength{\parsep}{0pt}
          \setlength{\itemsep}{0pt}  \renewcommand{\baselinestretch}{1.2}
          \settowidth
         {\labelwidth}{#1 ~ ~}\sloppy}}{\end{list}}

\vskip 1.cm

\leftline{\Large \bf References}

%%%%%%%%%%%%% BIBLIOGRAPHY  %%%%%%%%%%%%%%%%%%%%


\begin{thebibliography}{99}
%%%%%%%%%%%%%%%%%%%%%%%%%%%%%%%%%%%%%%%%%%%%%%%

%%%%%%%%
\bibitem{Hosotani1983}
Y.\ Hosotani, 
{\it ``Dynamical mass generation by compact extra dimensions''}, 
\jnl{\PLB}{126}{309}{1983}.

\bibitem{Davies1988}
A.~T.~Davies and A.~McLachlan,
{\it ``Gauge group breaking by Wilson loops''},
\jnl{\PLB}{200}{305}{1988}.

\bibitem{Hosotani1989}
Y.\ Hosotani,  
{\it ``Dynamics of nonintegrable phases and gauge symmetry breaking''}, 
\jnl{\AP}{190}{233}{1989}.

\bibitem{Davies1989}
A.~T.~Davies and A.~McLachlan,
{\it ``Congruency class effects in the Hosotani model''},
\jnl{\NPB}{317}{237}{1989}.

\bibitem{Hatanaka1998}
H.\ Hatanaka, T.\ Inami, C.S.\ Lim,
{\it ``The gauge hierarchy problem and higher dimensional gauge theories''}, 
\jnl{\MPLA}{13}{2601}{1998}.

\bibitem{Hatanaka1999}
H.\ Hatanaka,
{\it ``Matter representations and gauge symmetry breaking via compactified space''}, 
\jnl{\PTP}{102}{407}{1999}.

\bibitem{Kubo2002}
M.\ Kubo, C.S.\ Lim and H.\ Yamashita, 
{\it ``The Hosotani mechanism in bulk gauge theories with an orbifold extra space $S^1/Z_2$''}, 
\jnl{\MPLA}{17}{2249}{2002}.



%%  SU(3) or SO(5) x U(1) gauge-Higgs %%%

\bibitem{Scrucca2003}
C.A.~Scrucca, M.~Serone,  L.~Silvestrini,
{\it ``Electroweak symmetry breaking and fermion masses from extra dimensions''}, 
\jnl{\NPB}{669}{128}{2003}.

\bibitem{ACP2005}
K.~Agashe, R.~Contino and A.~Pomarol,
{\it ``The minimal composite Higgs model''}, 
\jnl{\NPB}{719}{165}{2005}.
%Nucl.\ Phys.\ {\bf B719}, 165 (2005).

\bibitem{Cacciapaglia2006}
G.~Cacciapaglia, C.~Csaki, S.C.~Park,
{\it ``Fully radiative electroweak symmetry breaking''}, 
\jnl{\JHEP}{0603}{099}{2006}.

\bibitem{Medina2007}
A.~D.~Medina, N.~R.~Shah and C.~E.~M.~Wagner, 
{\it ``Gauge-Higgs unification and radiative electroweak symmetry breaking in warped 
extra dimensions''}, 
\jnl{\PRD}{76}{095010}{2007}.

\bibitem{HOOS2008} 
Y.~Hosotani, K.~Oda, T.~Ohnuma and Y.~Sakamura,
{\it ``Dynamical electroweak symmetry breaking in $SO(5) \times U(1)$ gauge-Higgs 
unification with top and bottom quarks''}, 
\jnl{\PRD}{78}{096002}{2008}; 
%Phys.\ Rev.\  {\bf D78}, 096002 (2008); 
%{\it Erratum ibid.\ }{\bf D79}, 079902 (2009).
{\it Erratum}-\jnl{\ibid}{{\rm D}79}{079902}{2009}.

\bibitem{FHHOS2013} 
 S.~Funatsu, H.~Hatanaka, Y.~Hosotani, Y.~Orikasa and T.~Shimotani,
{\it ``Novel universality and Higgs decay $H\to \gamma\gamma, gg$ in the 
$SO(5) \times U(1)$ gauge-Higgs unification''}, 
\jnl{\PLB}{722}{94}{2013}.


\bibitem{Yoon2018b}
J.~Yoon and M.~E. Peskin, 
{\it ``Dissection of an $SO(5) \times U(1)$ gauge-Higgs  unification model''},
\jnl{\PRD}{100}{015001}{2019}. 




%%%%%%%%%%%
\bibitem{RS1}
L.\ Randall and R.\ Sundrum,
{\it ``A large mass hierarchy from a small extra dimension''}, 
\jnl{\PRL}{83}{3370}{1999}.



%%%%%%%%%%

\bibitem{HosotaniYamatsu2015}
Y.~Hosotani and N.~Yamatsu, 
{\it ``Gauge-Higgs grand unification''}, 
\jnl{\PTEP}{2015}{111B01}{2015}, arXiv:1504.03817 [hep-ph].

\bibitem{Furui2016}
A.~Furui, Y.~Hosotani, and N.~Yamatsu, 
{\it ``Toward realistic gauge-Higgs grand  unification''}, 
\jnl{\PTEP}{2016}{093B01}{2016}, arXiv:1606.07222 [hep-ph].


%%%% GUT inspired GHU %%%%%%

\bibitem{GUTinspired2019a}
 S.~Funatsu, H.~Hatanaka, Y.~Hosotani, Y.~Orikasa and N.~Yamatsu,
{\it ``GUT inspired $SO(5)\times U(1) \times SU(3)$ gauge-Higgs unification''}, 
\jnl{\PRD}{99}{095010}{2019}.   

\bibitem{FCNC2020a}
 S.~Funatsu, H.~Hatanaka, Y.~Hosotani, Y.~Orikasa and N.~Yamatsu,
{\it ``CKM matrix and FCNC suppression in $SO(5)\times U(1) \times SU(3)$ gauge-Higgs unification''}, 
\jnl{\PRD}{101}{055016}{2020}.   

\bibitem{GUTinspired2020b}
 S.~Funatsu, H.~Hatanaka, Y.~Hosotani, Y.~Orikasa and N.~Yamatsu,
{\it ``Effective potential and universality in GUT-inspired gauge-Higgs unification''}, 
\jnl{\PRD}{102}{015005}{2020}.   

%%%%%%%%%%%%%%%%%%
\bibitem{ILC2021}
The International Linear Collider: Report to Snowmass 2021 ILC International Development Team, 
A.Aryshev et al.
arXiv:2203.07622 [physics.acc-ph].



%%%%%%
\bibitem{Funatsu2017a}
S.~Funatsu, H.~Hatanaka, Y.~Hosotani, and Y.~Orikasa, 
{\it ``Distinct signals of  the gauge-Higgs unification in $e^+e^-$ collider experiments''}, 
\jnl{\PLB}{775}{297}{2017}.   

\bibitem{Yoon2018a}
J.~Yoon and M.~E. Peskin, 
{\it ``Fermion pair production in $SO(5) \times U(1)$  gauge-Higgs unification models''}, 
arXiv:1811.07877 [hep-ph].

\bibitem{Funatsu2019a}
S.~Funatsu, 
{\it ``Forward-backward asymmetry in the gauge-Higgs unification at the International Linear Collider''},
\jnl{\EPJC}{79}{854}{2019}.   

\bibitem{Funatsu2022b}
S.~Funatsu, H.~Hatanaka, Y.~Hosotani, Y.~Orikasa, and N.~Yamatsu, 
{\it ``Bhabha scattering in the gauge-Higgs unification''}, 
\jnl{\PRD}{106}{015010}{2022}.
% 2203.16030 [hep-ph]


\bibitem{GUTinspired2020c}
 S.~Funatsu, H.~Hatanaka, Y.~Hosotani, Y.~Orikasa and N.~Yamatsu,
{\it ``Fermion pair production at $e^- e^+$ linear collider experiments
in GUT inspired gauge-Higgs unification''}, 
\jnl{\PRD}{102}{015029}{2020}.

\bibitem{Yamatsu2023}
 S.~Funatsu, H.~Hatanaka, Y.~Orikasa and N.~Yamatsu,
{\it ``Single Higgs boson production at electron-positron colliders in gauge-Higgs unification''},
arXiv:2301.07833 [hep-ph].

%%%%%%%%%
\bibitem{PeskinTakeuchi}
M.E.\ Peskin and T.\ Takeuchi,
{\it ``New constraint on a strongly interacting Higgs sector''},
\jnl{\PRL}{65}{964}{1990};
{\it ``Estimation of oblique electroweak corrections''},
\jnl{\PRD}{46}{381}{1992}.

\bibitem{Lavoura1993}
L.\  Lavoura and J.P.\  Silva,
{\it ``Oblique corrections  from vectorlike singlet and doublet quarks''},
\jnl{\PRD}{47}{2046}{1993}.

\bibitem{PeskinSchroeder}
M.E.\ Peskin and D.V. Schroeder,
{\it ``An Introduction to Quantum Field Theory''},  Chap. 21, Addison-Wesley Publishing Company (1995).

\bibitem{Carena2006}
M.\ Carena, E.\ Ponton, J.\ Santiago and C.E.M.\ Wagner,
{\it ``Light Kaluza-Klein states in Randall-Sundrum models with custodial $SU(2)$''},
\jnl{\NPB}{759}{202}{2006}.

%%%%%%%%%%%%%%%%%%%%%%%

\bibitem{Funatsu2022a}
S.~Funatsu, H.~Hatanaka, Y.~Hosotani, Y.~Orikasa, and N.~Yamatsu, 
{\it ``Signals of $W'$  and $Z'$ bosons at the LHC in the $SU(3) \times SO(5) \times U(1)$ gauge-Higgs unification''}, 
\jnl{\PRD}{105}{055015}{2022}.
% 2111.05624 [hep-ph]


%%%%%%%%%%%%%%%%%%%%%%
\bibitem{AnomalyFlow1}
S.~Funatsu, H.~Hatanaka, Y.~Hosotani, Y.~Orikasa and N.~Yamatsu,
{\it ``Anomaly flow by an Aharonov-Bohm phase''}, 
\jnl{\PTEP}{2022}{043B04}{2022}, arXiv:2202.01393 [hep-ph].


\bibitem{AnomalyFlow2}
Y.~Hosotani,
{\it ``Universality in anomaly flow''}, 
\jnl{\PTEP}{2022}{073B01}{2022}, arXiv:2205.00154 [hep-th].



\bibitem{GHUfiniteT2021}
 S.~Funatsu, H.~Hatanaka, Y.~Hosotani, Y.~Orikasa and N.~Yamatsu,
{\it ``Electroweak and left-right phase transitions in $SO(5) \times U(1) \times  SU(3)$  gauge-Higgs unification''}, 
\jnl{\PRD}{104}{115018}{2021}.   

\bibitem{GHUseesaw2017}
Y.~Hosotani and N.~Yamatsu, 
{\it ``Gauge-Higgs seesaw mechanism in 6-dimensional grand unification''}, 
\jnl{\PTEP}{2017}{091B01}{2017}, arXiv:1706.03503 [hep-ph].


\bibitem{pdg2022chap10}
Particle Data Group Collaboration,  R. L. Workman et al., 
{\it ``Review of Particle Physics''},
\jnl{\PTEP}{2022}{083C01}{2022}, Chap.\ 10.

\end{thebibliography}
\end{document}